\documentclass[preprint]{elsarticle}

\journal{Computer Physics Communications}

\usepackage{algorithm}
\usepackage{algorithmic}
\usepackage{amssymb}
\usepackage{amsmath}
\usepackage{amsthm}
\usepackage{caption}
\usepackage{comment}
\usepackage{float}
\usepackage[latin1]{inputenc}
\usepackage[pdfborder={0 0 0}, colorlinks=true, linkcolor=blue, citecolor=red]{hyperref}\hypersetup{pdfborder=0 0 0}
\usepackage{mathrsfs}
\usepackage{multicol}
\usepackage{multirow}
\usepackage{subfigure}
\usepackage{wrapfig}
\usepackage{ulem}
\usepackage{upgreek}
\usepackage{url}
\usepackage[capitalise]{cleveref}
\usepackage{listings}

\crefname{lstlisting}{lst}{lsts}
\Crefname{lstlisting}{Lst}{Lsts}

\usepackage{soul,xargs}
\usepackage[pdftex,dvipsnames]{xcolor}

\usepackage[colorinlistoftodos,prependcaption,textsize=tiny]{todonotes}
\newcommandx{\question}[2][1=]{\todo[linecolor=red,backgroundcolor=red!20,bordercolor=red,#1]{#2}}
\newcommandx{\change}[2][1=]{\todo[linecolor=blue,backgroundcolor=blue!25,bordercolor=blue,#1]{#2}}
\newcommandx{\info}[2][1=]{\todo[linecolor=OliveGreen,backgroundcolor=OliveGreen!25,bordercolor=OliveGreen,#1]{#2}}
\newcommandx{\improve}[2][1=]{\todo[linecolor=Plum,backgroundcolor=Plum!25,bordercolor=Plum,#1]{#2}}
\newcommandx{\thiswillnotshow}[2][1=]{\todo[disable,#1]{#2}}
\newcommandx{\answer}[2][1=]{\todo[linecolor=blue,backgroundcolor=White!25,bordercolor=Plum,#1]{#2}}

\graphicspath{{./figures/}}

\begin{document}

\newcommand{\TODO}[1]{ \fbox{\parbox{3in}{\bf TODO: #1}}}

\newcommand{\grbf}[1] {\mbox{\boldmath${#1}$\unboldmath}}
\newcommand{\gbf}[1] {\mathbf{#1}}


\newcommand{\CPP}{C\nolinebreak\hspace{-.05em}\raisebox{.4ex}{\tiny\bf +}\nolinebreak\hspace{-.10em}\raisebox{.4ex}{\tiny\bf +}}

\newcommand{\beq} {\begin{equation}}
\newcommand{\eeq} {\end{equation}}
\newcommand{\bdm} {\begin{displaymath}}
\newcommand{\edm} {\end{displaymath}}
\newcommand{\bit}{\begin{itemize}}
\newcommand{\eit}{\end{itemize}}
\newcommand{\bde}{\begin{description}}
\newcommand{\ede}{\end{description}}
\newcommand{\bce}{\begin{center}}
\newcommand{\ece}{\end{center}}
\newcommand{\ben} {\begin{enumerate}}
\newcommand{\een} {\end{enumerate}}
\newcommand{\bea} {\begin{eqnarray}}
\newcommand{\eea} {\end{eqnarray}}
\newcommand{\barr} {\begin{array}}
\newcommand{\earr} {\end{array}}
\newcommand{\bean} {\begin{eqnarray*}}
\newcommand{\eean} {\end{eqnarray*}}
\newcommand{\edoc} {\end{document}}
\newcommand{\On}{\hspace{0.2cm} \mbox{on} \hspace{0.2cm}}
\newcommand{\In}{\hspace{0.2cm} \mbox{in} \hspace{0.2cm}}
\newcommand{\Minimize}{\hspace{0.2cm} \mbox{minimize} \hspace{0.2cm}}
\newcommand{\Maximize}{\hspace{0.2cm} \mbox{maximize} \hspace{0.2cm}}
\newcommand{\SubjectTo}{\hspace{0.2cm} \mbox{subject} \hspace{0.15cm}
            \mbox{to} \hspace{0.2cm}}

\def\etal{{\it et al.~}}

\newcommand{\point}[1] {\ensuremath{\boldsymbol{#1}}}
\newcommand{\vvec}[1] {\ensuremath{\boldsymbol{#1}}}
\renewcommand{\vec}[1] {\ensuremath{\boldsymbol{#1}}}
\newcommand{\cvec}[1] {\ensuremath{\boldsymbol{{#1}}}}
\newcommand{\ten}[1] {\ensuremath{\boldsymbol{#1}}}
\newcommand{\tenfour}[1] {\ensuremath{\boldsymbol{\mathsf{#1}}}}
\newcommand{\comp}[1]{\begin{pmatrix}{ #1 }\end{pmatrix}}
\newcommand{\vv}{\ensuremath{\vec{v}}}
\newcommand{\vr}{\left(\ensuremath{\vec{r}}\right)}
\newcommand{\att}{\left(t\right)}
\newcommand{\tvv}{\ensuremath{\tilde{\vec{v}}}}
\newcommand{\ww}{\ensuremath{\vec{w}}}
\newcommand{\GG}{\ensuremath{\ten{G}}}
\newcommand{\FF}{\ensuremath{\boldsymbol{\mathfrak{F}}}}
\newcommand{\tEE}{\ensuremath{\tilde{\vec{E}}}}
\newcommand{\nn}{\ensuremath{\vec{n}}}
\renewcommand{\SS}{\ensuremath{\ten{S}}}
\newcommand{\tSS}{\ensuremath{\tilde{\ten{S}}}}
\newcommand\tK{{\bf K}}
\newcommand{\cp}{\ensuremath{{c_p}}}
\newcommand{\cs}{\ensuremath{{c_s}}}
\newcommand{\tcs}{\ensuremath{{\tilde{c}_s}}}
\newcommand{\Vp}{\ensuremath{{V^e_p}}}
\newcommand{\Vs}{\ensuremath{{V^e_s}}}
\newcommand{\Ss}{\ensuremath{{S^e_s}}}
\newcommand{\nxnx}[1] {\ensuremath{\nn\times\left(\nn\times{#1}\right)}}

\newcommand{\dd}[2] {\ensuremath{\frac{\partial {#1}}{\partial {#2}}}}
\newcommand{\DD}[2] {\ensuremath{\frac{d {#1}}{d {#2}}}}
\newcommand{\Grad} {\ensuremath{\nabla}}
\newcommand{\Gradv}[1]{\ensuremath{\nabla_{#1}}}
\newcommand{\Div} {\ensuremath{\nabla\cdot}}
\newcommand{\Divv}[1] {\ensuremath{\nabla_{#1}\cdot}}
\newcommand{\Curl} {\ensuremath{\nabla\times}}
\newcommand{\Cten}{\ensuremath{\tenfour{C}}}

\newcommand{\eval}[2][\right]{\relax
  \ifx#1\right\relax \left.\fi#2#1\rvert}

\providecommand{\abs}[1]{\left\lvert#1\right\rvert}
\providecommand{\norm}[1]{\left\Vert#1\right\Vert}

\newcommand{\Nel} {\ensuremath{{N_T}}}
\newcommand{\Ned} {\ensuremath{{N_\text{ed}}}}
\newcommand{\De} {\ensuremath{{\mathsf{D}^e}}}
\newcommand{\Dep} {\ensuremath{{\mathsf{D}^{e'}}}}
\newcommand{\Dhat} {\ensuremath{\hat{\mathsf{D}}}}
\newcommand{\Dehat} {\ensuremath{{\hat{\mathsf{D}}^e}}}
\newcommand{\half} {\ensuremath{\frac{1}{2}}}
\newcommand{\IN} {\ensuremath{{\mathcal{I}_N}}}
\newcommand{\INe} {\ensuremath{{\mathcal{I}_{N_e}}}}

\newcommand{\mc}[1]{\mathcal{#1}}
\newcommand{\mcC}[1]{\mathcal{#1}}
\newcommand{\mb}[1]{\mathbf{#1}}
\newcommand{\mbb}[1]{\mathbb{#1}}
\newcommand{\mi}[1]{\mathit{#1}}
\newcommand{\mf}[1]{\mathfrak{#1}}
\newcommand{\Oi}[1]{\int_{\mc{D}} #1 \, d\vec{x}}
\newcommand{\TOi}[1]{\int_{0}^T \int_{\mc{D}} #1 \, d\vec{x}dt}
\newcommand{\Ti}[1]{\int_{0}^T #1 \, dt}
\newcommand{\TDei}[1]{\int_{0}^T\int_{\De} #1 \, d\vec{x}dt}
\newcommand{\TDeid}[1]{\int_{0}^T\int_{\De,N} #1 \, d\vec{x}dt}
\newcommand{\Dei}[1]{\int_{\De} #1 \, d\vec{x}}
\newcommand{\DeI}[1]{\int_{D} #1 \, d\vec{x}}
\newcommand{\DeiM}[1]{\int_{\Dhat} #1 \, d\vec{r}}
\newcommand{\DeMd}[1]{\int_{\De,N_e} #1 \, d\vec{x}}
\newcommand{\DeiMd}[1]{\int_{\Dhat,N_e} #1 \, d\vec{r}}
\newcommand{\DeMdN}[1]{\int_{D,N} #1 \, d\vec{x}}
\newcommand{\DeiMdN}[1]{\int_{\Dhat,N} #1 \, d\vec{r}}
\newcommand{\DeiMdNC}[2]{\int_{\Dhat,N_{#2}} #1 \, d\vec{r}}
\newcommand{\pDei}[1]{\int_{\partial \De} #1 \, d\vec{x}}
\newcommand{\pDeiM}[1]{\int_{\partial \Dhat} #1 \, d\vec{r}}
\newcommand{\pDeiMd}[1]{\int_{\partial \Dhat,N_e} #1 \, d\vec{r}}
\newcommand{\pDeiMdN}[1]{\int_{\partial \Dhat,N} #1 \, d\vec{r}}
\newcommand{\pDeiMdNC}[2]{\int_{\partial \Dhat,N_{#2}} #1 \, d\vec{r}}
\newcommand{\pMortar}[2]{\int_{\mc{M}_{#2}} #1 \, d\vec{r}}
\newcommand{\pMortard}[2]{\int_{\mc{M}^{#2},N_{#2}} #1 \, d\vec{r}}
\newcommand{\pMortardd}[3]{\int_{\mc{M}^{#2},N_{#3}} #1 \, d\vec{r}}
\newcommand{\pMortardh}[3]{\int_{\mc{M}_{#2},N_{#3}} #1 \, d\vec{r}}
\newcommand{\pMortardhn}[4]{\int_{{#2}_{#3},N_{#4}} #1 \, d\vec{r}}
\newcommand{\TpDei}[1]{\int_{0}^T\int_{\partial \De} #1 \, d\vec{x}}
\newcommand{\TpDeid}[1]{\int_{0}^T\int_{\partial \De,N} #1 \, d\vec{x}}
\newcommand{\Deid}[1]{\int_{\De,N_e} #1 \, d\vec{x}}
\newcommand{\DeidN}[1]{\int_{\De,N} #1 \, d\vec{x}}
\newcommand{\pDeidN}[1]{\int_{\partial \De,N} #1 \, d\vec{x}}
\newcommand{\pDeid}[1]{\int_{\partial \De,N_e} #1 \, d\vec{x}}
\newcommand{\nor}[1]{\left\| #1 \right\|}
\newcommand{\snor}[1]{\left| #1 \right|}
\newcommand{\LRp}[1]{\left( #1 \right)}
\newcommand{\LRs}[1]{\left[ #1 \right]}
\newcommand{\LRa}[1]{\left< #1 \right>}
\newcommand{\LRc}[1]{\left\{ #1 \right\}}
\newcommand{\pp}[2]{\frac{\partial #1}{\partial #2}}
\newcommand{\ppcp}[1]{\frac{\partial #1}{\partial \vec{c_p}}}
\newcommand{\ppcpj}[1]{\frac{\partial #1}{\partial c_p}}
\newcommand{\ppcpmf}[1]{\frac{\partial #1}{\partial \vec{\mf{c_p}}}}
\newcommand{\ppm}[1]{\frac{\partial #1}{\partial \vec{m}}}
\newcommand{\ppho}[3]{\frac{\partial^{#3} #1}{{\partial #2}^{#3}}}
\newcommand{\ppmi}[1]{\frac{\partial #1}{\partial m_i}}
\newcommand{\ppmj}[1]{\frac{\partial #1}{\partial m_j}}
\newcommand{\ppcpm}[1]{\frac{\partial #1}{\partial \vec{c_p}^-}}
\newcommand{\ppcpp}[1]{\frac{\partial #1}{\partial \vec{c_p}^+}}
\newcommand{\ppcs}[1]{\frac{\partial #1}{\partial \vec{c_s}}}
\newcommand{\ppcsm}[1]{\frac{\partial #1}{\partial \vec{c_s}^-}}
\newcommand{\ppcsp}[1]{\frac{\partial #1}{\partial \vec{c_s}^+}}
\newcommand{\td}[2]{\frac{d #1}{d #2}}
\newcommand{\Rvert}[1]{\left. #1 \right|}
\newcommand{\bs}[1]{\boldsymbol{#1}}
\newcommand{\ipDed}[3]{\LRp{ #1, #2}_{\De,N}^{#3}}

\newcommand{\bigO}{\mathcal{O}}

\newcommand{\iOm}[1]{\int_{\Omega} #1 \, d\Omega}
\newcommand{\iGb}[1]{\int_{\Gamma_{b}} #1 \, ds}
\newcommand{\iGs}[1]{\int_{\Gamma_{s}} #1 \, ds}
\newcommand{\iGsy}[1]{\int_{\Gamma_{s}} #1 \, ds(\mb{y})}
\newcommand{\RE}{\mbox{{I}}\hspace{-0.5ex}\mbox{{R}}}
\newcommand{\iC}[1]{\int_{0}^{2\pi} #1 \, d\theta}
\newcommand{\iGr}[1]{\int_{\Gamma_{\infty}} #1 \, ds}

\newcommand{\jump}[1] {\ensuremath{[\![{#1}]\!]}}
\newcommand{\diff}[1] {\ensuremath{\LRs{#1}}}
\newcommand{\average}[1] {\ensuremath{\LRc{\!\!\LRc{#1}\!\!}}}

\newcounter{tablecell}

\newcommand*{\numbercell}{%
  \stepcounter{tablecell}%
  \thetablecell. 
}

\newtheorem{propo}{Proposition}
\newtheorem{coro}{Corollary}
\newtheorem{theo}{Theorem}
\newtheorem{lem}{Lemma}

\newtheorem{defi}{definition}

\newtheorem{rema}{remark}

\newcommand{\figlab}[1]{\label{fig:#1}}
\newcommand{\eqnlab}[1]{\label{eq:#1}}
\newcommand{\theolab}[1]{\label{theo:#1}}
\newcommand{\corolab}[1]{\label{coro:#1}}
\newcommand{\propolab}[1]{\label{propo:#1}}
\newcommand{\lemlab}[1]{\label{lem:#1}}
\newcommand{\defilab}[1]{\label{defi:#1}}
\newcommand{\remalab}[1]{\label{rema:#1}}
\newcommand{\tablab}[1]{\label{tab:#1}}

\newcommand{\figref}[1]{\ref{fig:#1}}
\newcommand{\theoref}[1]{\ref{theo:#1}}
\newcommand{\defiref}[1]{\ref{defi:#1}}
\newcommand{\remaref}[1]{\ref{rema:#1}}
\newcommand{\cororef}[1]{\ref{coro:#1}}
\newcommand{\proporef}[1]{\ref{propo:#1}}
\newcommand{\lemref}[1]{\ref{lem:#1}}
\newcommand{\eqnref}[1]{\eqref{eq:#1}}
\newcommand{\alglab}[1]{\label{alg:#1}}
\newcommand{\algref}[1]{\ref{alg:#1}}
\newcommand{\seclab}[1]{\label{sect:#1}}
\newcommand{\secref}[1]{\ref{sect:#1}}
\newcommand{\tabref}[1]{\ref{tab:#1}}

\renewcommand{\u}{u}
\newcommand{\uk}{\u^k}
\newcommand{\upk}{\u^{+k}}
\newcommand{\ukp}{\u^{k+1}}
\newcommand{\umkp}{\u^{-k+1}}
\newcommand{\umk}{\u^{-k}}
\renewcommand{\v}{v}
\newcommand{\vk}{{\v}^k}
\newcommand{\vkp}{{\v}^{k+1}}
\newcommand{\un}{\u_h}
\newcommand{\e}{e}
\newcommand{\w}{w}
\newcommand{\wb}{\mb{\w}}
\newcommand{\J}{J}
\newcommand{\Jb}{\mb{J}}
\newcommand{\q}{q}
\renewcommand{\r}{r}
\newcommand{\qh}{\hat{\q}}
\newcommand{\qs}{\q^*}
\newcommand{\qht}{\tilde{\qh}}
\newcommand{\qbar}{\overline{\q}}
\newcommand{\qb}{{\mb{\q}}}
\newcommand{\sig}{{\sigma}}
\newcommand{\thetah}{{\hat{\theta}}}
\newcommand{\thetas}{{\theta^*}}
\newcommand{\thetahn}{{\thetah_h}}
\newcommand{\sign}[1]{\text{ sgn}\!\LRp{#1}}
\newcommand{\sigh}{\hat{\sig}}
\newcommand{\sigs}{\sig^*}
\newcommand{\sigk}{\sig^{k}}
\newcommand{\sigpk}{\sig^{+k}}
\newcommand{\sigkp}{\sig^{k+1}}
\newcommand{\sigmkp}{\sig^{-k+1}}
\newcommand{\sigmk}{\sig^{-k}}
\newcommand{\sigb}{{\bs{\sig}}}
\newcommand{\sigbk}{\sigb^k}
\newcommand{\sigbkp}{\sigb^{k+1}}
\newcommand{\sigbn}{\sigb_h}
\newcommand{\sigbs}{\sigb^*}
\newcommand{\taub}{{\bs{\tau}}}
\newcommand{\taustar}{\tau_*}
\newcommand{\kappab}{{\bs{\kappa}}}
\newcommand{\gamb}{{\bs{\gamma}}}
\newcommand{\ut}{\tilde{\u}}
\newcommand{\sigbt}{\tilde{\sigb}}
\newcommand{\vt}{\tilde{\v}}
\newcommand{\taubt}{\tilde{\taub}}
\newcommand{\ub}{\mb{\u}}
\newcommand{\ubl}{\mb{\u}_{\lambda}}
\newcommand{\ubf}{\mb{\u}_{f}}
\newcommand{\ubp}{\mb{\u}^+}
\newcommand{\ubm}{\mb{\u}^-}
\newcommand{\vb}{\mb{\v}}
\newcommand{\xb}{\mb{x}}
\newcommand{\um}{\u^-}
\newcommand{\up}{\u^+}
\newcommand{\ubk}{\ub^k}
\newcommand{\ubkp}{\ub^{k+1}}

\newcommand{\m}{{m}}

\newcommand{\ud}{{\dot{\u}}}
\newcommand{\vd}{{\dot{\v}}}
\newcommand{\sigbd}{{\dot{\sigb}}}
\newcommand{\taubd}{\dot{\taub}}

\renewcommand{\d}{{d}}
\newcommand{\R}{{\mathbb{R}}}
\newcommand{\kap}{\kappa}
\newcommand{\F}{\mb{F}}
\newcommand{\f}{f}
\newcommand{\g}{g}
\newcommand{\fb}{\mb{\f}}
\newcommand{\gb}{\mb{\g}}
\newcommand{\gD}{g_D}
\newcommand{\pOmega}{\partial \Omega}
\newcommand{\K}{T}
\newcommand{\Kp}{\K^+}
\newcommand{\Km}{\K^-}
\newcommand{\pK}{{\partial \K}}
\newcommand{\pKin}{{\pK^{\text{in}}}}
\newcommand{\pKout}{{\pK^{\text{out}}}}
\newcommand{\Kj}{{\K_j}}
\newcommand{\Gh}{{\mc{E}_h}}
\newcommand{\Gho}{{\mc{E}_h^o}}
\newcommand{\Ghb}{{\mc{E}_h^{\partial}}}
\newcommand{\Poly}{{\mc{P}}}
\newcommand{\D}{{D}}
\newcommand{\p}{{p}}
\newcommand{\px}{\partial_x}
\newcommand{\ppx}{\partial^2_x}
\newcommand{\pres}{{q}}
\newcommand{\presl}{{q}_{\lambda}}
\newcommand{\presf}{{q}_{f}}
\newcommand{\Ts}{{\mb{T}}}
\newcommand{\Vb}{{\mb{V}}}
\newcommand{\V}{{V}}
\newcommand{\Lam}{{\Lambda}}
\newcommand{\Lamb}{\bs{\Lambda}}
\newcommand{\mub}{\bs{\mu}}
\newcommand{\lambdab}{\bs{\lambda}}
\newcommand{\Thn}{\mathcal{T}_h}
\newcommand{\Th}{{\Ts_h}}
\newcommand{\Vh}{{\V_h}}
\newcommand{\Vbh}{{\Vb_h}}
\newcommand{\Lamh}{{\Lam_h}}
\newcommand{\Lambh}{{\Lamb_h}}
\newcommand{\Lamho}{{\overline{\Lam}_h}}
\newcommand{\Lambht}{{\Lamb_h^t}}
\newcommand{\Ltwo}{{L^2\LRp{\Omega}}}
\newcommand{\Ltwoe}{{L^2\LRp{\e}}}
\newcommand{\Ltw}{{L^2\LRp{\K}}}
\newcommand{\VhK}{{\V_h\LRp{\K}}}
\newcommand{\ThK}{{\Ts_h\LRp{\K}}}
\newcommand{\VbhK}{{\Vb_h\LRp{\K}}}
\newcommand{\Lamhe}{{\Lam_h\LRp{\e}}}
\newcommand{\Lambhe}{{\Lamb_h\LRp{\e}}}
\renewcommand{\L}{\mc{L}}
\renewcommand{\D}{\mc{D}}
\newcommand{\T}{\mc{T}}
\newcommand{\Tt}{\tilde{\T}}
\newcommand{\A}{\mb{A}}
\newcommand{\B}{\mb{B}}
\renewcommand{\D}{\mb{D}}
\newcommand{\Am}{\A^-}
\newcommand{\Ap}{\A^+}
\newcommand{\Aa}{\snor{\A}}
\newcommand{\Rb}{\mb{R}}
\newcommand{\C}{\mb{C}}
\renewcommand{\S}{\mb{S}}
\newcommand{\Sm}{\S^{-}}
\newcommand{\Sp}{\S^{+}}
\newcommand{\Spm}{\S^{\pm}}

\newcommand{\Lte}{{L^2\LRp{\Gh}}}
\newcommand{\n}{\mb{n}}
\newcommand{\nm}{\n^-}
\newcommand{\np}{\n^+}
\newcommand{\uh}{{\hat{\u}}}
\newcommand{\uhk}{{\uh}^k}
\newcommand{\uhkp}{{\uh}^{k+1}}
\newcommand{\us}{{\u^*}}
\newcommand{\uhn}{{\uh_h}}
\newcommand{\ubh}{{\hat{\ub}}}
\newcommand{\ubs}{{\ub^*}}
\newcommand{\ubht}{{\tilde{\ubh}}}
\newcommand{\uht}{{\tilde{\uh}}}
\newcommand{\uhtn}{{\uht_h}}
\newcommand{\uhd}{{\dot{\uh}}}
\newcommand{\ubhk}{\ubh^k}
\newcommand{\ubhkp}{\ubh^{k+1}}
\newcommand{\sigbh}{{\hat{\sigb}}}
\newcommand{\Fh}{{\hat{\F}}}
\newcommand{\fh}{{\bar{\f}}}
\newcommand{\Fs}{\F^*}
\renewcommand{\c}{{c}}

\newcommand{\zb}{\mb{z}}
\renewcommand{\sb}{\mb{s}}
\newcommand{\rb}{\mb{r}}
\newcommand{\betab}{\bs{\beta}}
\newcommand{\veps}{{\varepsilon}}
\newcommand{\vepsb}{\bs{\veps}}
\newcommand{\Hb}{\mb{H}}
\newcommand{\Hbt}{\Hb^t}
\newcommand\E{\mathcal{E}}
\newcommand\Eh{\mathcal{E}_h}
\newcommand\Ehi{\mathcal{E}_{h,i}}
\newcommand\Ehiint{\mathcal{E}_{h,i}^\mathrm{int}}
\newcommand\EhG{\mathcal{E}_{h}^{\Gamma}}
\newcommand\EhiG{\mathcal{E}_{h,i}^{\Gamma}}
\newcommand\EhijG{\mathcal{E}_{h,i,j}^{\Gamma}}
\newcommand\EhjiG{\mathcal{E}_{h,j,i}^{\Gamma}}
\newcommand{\Eb}{\mb{E}}
\newcommand{\Ebb}{\mbb{E}}
\newcommand{\Ebt}{\Eb^t}
\newcommand{\hb}{\mb{h}}
\newcommand{\eb}{\mb{e}}
\newcommand{\Hbh}{\hat{\mb{H}}}
\newcommand{\Ebh}{\hat{\mb{E}}}
\newcommand{\Hbs}{{\mb{H}^*}}
\newcommand{\Ebs}{{\mb{E}^*}}
\newcommand{\Ebht}{\Ebh^t}
\newcommand{\Hbht}{\Hbh^t}
\newcommand{\Lb}{\mb{L}}
\newcommand{\Gb}{\mb{G}}
\newcommand{\Lbh}{\hat{\Lb}}
\newcommand{\Lbs}{\Lb^*}
\newcommand{\Ib}{\mb{I}}
\newcommand{\I}{I}
\newcommand{\Q}{Q}
\newcommand{\Sigb}{\bs{\Sigma}}
\newcommand{\Piu}{\Pi_\u}
\newcommand{\Pisigb}{\Pi_\sigb}
\newcommand{\Z}{Z}
\newcommand{\Y}{Y}
\newcommand{\rhou}{\rho\u}
\newcommand{\rhov}{\rho\v}
\newcommand{\rhoe}{\rho_e}
\newcommand{\rhoa}{\tilde\rho}
\newcommand{\rhoaa}{\hat\rho}
\renewcommand{\P}{P}
\newcommand{\h}{H}

\newcommand{\mygreen}[1]{{\color[rgb]{0,.65,0} #1}}
\newcommand{\mywhite}[1]{{\color[rgb]{1.0,1.0,1.0} #1}}
\newcommand{\myred}[1]{{\color[rgb]{0.65,0.0,0.0} #1}}
\newcommand{\myblue}[1]{{\color[rgb]{0.0,0.0,1.0} #1}}
\newcommand{\mygray}[1]{{\color[rgb]{.15,.35,.60} #1}}
\newcommand{\mythemec}[1]{{\color[RGB]{51,108,121} #1}}
\newcommand{\mydarkred}[1]{{\color[rgb]{.6,.1,.1} #1}}
\newcommand{\mydarkblue}[1]{{\color[rgb]{.1,.1,.9} #1}}
\newcommand{\mygreenback}[1]{{\color[rgb]{.75,1,.75} #1}}
\newcommand{\myorange}[1]{{\color[rgb]{.76,.39,.13} #1}}
\newcommand{\mygrass}[1]{{\color[rgb]{.19,.64,.13} #1}}
\newcommand{\mysierp}[1]{{\color[RGB]{209,28,209} #1}}

\newcommand{\tanbui}[2]{\textcolor{blue}{#1} \textcolor{red}{#2}}

\newcommand{\vel}{\bs{\vartheta}}
\newcommand{\vels}{\vel^*}
\newcommand{\vele}{\vel^e}
\newcommand{\velh}{\hat{\vel}}
\newcommand{\velk}{\vel^k}
\newcommand{\velkp}{\vel^{k+1}}
\newcommand{\vhk}{\hat\v^{k}}

\newcommand{\phis}{\phi^*}
\newcommand{\phie}{\phi^e}
\newcommand{\phih}{\hat{\phi}}
\newcommand{\phihk}{\phih^k}
\newcommand{\phik}{\phi^k}
\newcommand{\phikp}{\phi^{k+1}}

\begin{frontmatter}

\title{Scaling and performance portability of the particle-in-cell scheme for plasma physics applications 
    through mini-apps targeting exascale architectures}

    \author[srk]{Sriramkrishnan Muralikrishnan}
    \author[Matthias]{Matthias Frey}
    \author[Alex]{Alessandro Vinciguerra}
    \author[Alex]{Michael Ligotino}
    \author[Antoine]{Antoine J. Cerfon}
    \author[Miroslav]{Miroslav Stoyanov}
    \author[Rahul]{Rahulkumar Gayatri}
    \author[srk]{Andreas Adelmann}
    \address[srk]{Paul Scherrer Institut, Forschungsstrasse 111, 5232 Villigen, Switzerland.}
    \address[Matthias]{Mathematical Institute, University of St Andrews, KY16 9SS, UK.}
    \address[Alex]{ETH Zurich, Switzerland.}
    \address[Antoine]{Courant Institute of Mathematical Sciences, New York University, New York NY 10012, USA.}
    \address[Miroslav]{Oak Ridge National Laboratory, Oak Ridge, USA.}
    \address[Rahul]{National Energy Research Scientific Computing Center, Berkeley, CA, USA.}

\begin{abstract}
We perform a scaling and performance portability study of the particle-in-cell scheme for plasma physics applications through a set of mini-apps we name ``Alpine", which can make use of exascale computing capabilities. 
The mini-apps are based on Independent Parallel Particle Layer, a framework that is designed around performance portable and 
dimension independent particles and fields.

We benchmark the simulations with varying parameters such as grid resolutions ($512^3$ to $2048^3$) and number of simulation particles ($10^9$ to $10^{11}$) with the following mini-apps: weak and strong 
Landau damping, bump-on-tail and two-stream instabilities, and the dynamics of an electron bunch in a charge-neutral Penning trap. We show strong and weak scaling and analyze the performance of different components on 
several pre-exascale architectures such as Piz-Daint, Cori, Summit and Perlmutter. While the scaling and portability study helps identify the performance critical components of the particle-in-cell scheme in the current state-of-the-art
 computing architectures, the mini-apps by themselves can be used to develop new algorithms and optimize their high performance implementations targeting exascale architectures.   
\end{abstract}


\end{frontmatter}

\section{Introduction}
    Heterogeneous computing architectures are unavoidable as scientific computing moves towards the era of exascale computing. This is already evident from some of the existing and future supercomputers listed in Table \ref{tab:supercomputers} which consist of 
    different types of CPUs and GPUs. 

        \begin{table}[h!b!t!]
        \centering
        \begin{tabular}{|r|c|c|c|}
        \hline
            \!\!\! Name \!\!\!\! & \!\!\! Location \!\!\!\! & \!\!\! CPUs \!\!\!\! & \!\!\! GPUs \!\!\!\!\\
        \hline
            \myred{Perlmutter} & NERSC, USA  & AMD Milan & NVIDIA A100 \\
            \myred{Summit} & ORNL, USA  & IBM POWER9 & NVIDIA V100 \\
            \myred{Sunway TaihuLight} & NSCC-Wuxi, China  & Sunway SW26010 & - \\
            \myred{Fugaku} & RIKEN, Japan & ARM A64FX & - \\
            \myblue{Frontier} & ORNL, USA & AMD EPYC & AMD Instinct \\
            \myblue{Aurora} & ANL, USA & Intel Xeon & Intel Xe \\
        \hline
        \end{tabular}
        \caption{\label{tab:supercomputers}An incomplete list of some of the existing (in red) and future (in blue) computing systems showing the diversity of HPC landscape.}
        \end{table}

    The programming paradigms or languages used with these
    architectures can differ significantly from one another. As a result, naively written codes could require rewrites of considerable portions of the codes simply to achieve compatibility with a given architecture, to say nothing of efficiency. In that context, performance portability is a key criteria for current and future simulation codes, if they are to be able to take advantage of the newest developments in computing architectures with minimal maintenance.

    In the plasma physics community, particle-in-cell (PIC) schemes have been widely used for the simulation of kinetic plasmas since their inception \cite{hockney1988computer,birdsall2004plasma,dawson1983particle}. The attractive features of PIC schemes include simplicity, ease of parallelization and robustness for a wide variety of 
    physical scenarios \cite{ricketson2016sparse}. 
    Because of their flexibility and versatility, PIC schemes are employed 
    in many production level plasma simulation and particle accelerator codes such as TRISTAN-MP \cite{spitkovsky2005simulations,buneman1993computer}, ORB5 \cite{jolliet2007global}, XGC \cite{chang2008spontaneous}, OSIRIS \cite{fonseca2002osiris}, IMPACT-T \cite{qiang2006three},  OPAL \cite{adelmann2019opal} and Warp-X \cite{vay2018warp}, to name a few.

    Some of these production codes have already begun their journey towards performance portability as, evidenced in 
    \cite{mniszewski2021enabling,myers2021porting}. It is expected that a lot more will also do so sometime
    soon, in order to benefit from the high performance of existing and future advanced computing architectures. Mini-applications (or mini-apps) which are performance portable
    can greatly help in this respect \cite{heroux2009improving}. These are light-weight proxy codes which contain 
    performance critical components of the application codes. They can be used for implementing new algorithms, optimizing 
    implementations, and provide reliable performance expectations for the real application of interest on different computing architectures
    \cite{heroux2009improving}. As such, mini-apps serve as an interface between applied mathematics, high performance computing  and production codes. 

    Our objective in this work is to study the performance portability and scaling of the components of PIC schemes through a set of mini-apps (Alpine\footnote{ALPINE: A set of performance portable pLasma physics Particle-in-cell mINi-apps for Exascale}) with applications in 
    plasma physics and particle accelerator modeling, targeting exascale architectures. In the present article, we consider physical situations for which the electrostatic assumption is justified, and for which collisions can be neglected. In future work, we plan to 
    extend the study by enriching the collection of mini-apps with electromagnetic examples, which may or may not include collisions and electron pair production. 
    The mini-apps are built using the performance portable library Independent Parallel Particle Layer (IPPL) which is 
    described in detail in Section \secref{ippl}. The contributions of the current work are summarized below: 
\begin{itemize}
    \item Alpine provides a test bed for implementing new algorithms and/or novel implementations of existing algorithms related to PIC schemes in the context of exascale architectures in a portable way.
    \item The performance study in this work provides important insights on how different components of PIC schemes function on different architectures. 
    
    \item This work serves as guidance for the plasma PIC community to identify the major causes of performance limitations and better prepare for exascale architectures.

    \item So far, portable exascale PIC studies have mostly been conducted for PIC schemes designed for electromagnetic plasma models \cite{myers2021porting,bird2021vpic}. To the best of our knowledge, this is the first study which considers the performance
    of a PIC scheme for an electrostatic plasma model in the portable exascale context.
\end{itemize}

    This paper is organized as follows.\ Section \secref{ippl} describes the IPPL library. The electrostatic PIC 
    scheme implemented in the mini-apps is described in Section \secref{pic}.\ Section \secref{miniapps} describes the mini-apps along with 
    their physical parameters and verification studies.\ The strong and weak scaling results, as well as the performance 
    analysis of different components on four different pre-exascale architectures, are presented in Section \secref{scaling}. Finally, we summarize our results in Section \secref{conclusions}, and propose directions for future work.

\section{IPPL}
\seclab{ippl}
The Independent Parallel Particle Layer (IPPL) \CPP{} library was developed about 20 years ago and was inspired by and partially based on POOMA \cite{doi:10.1063/1.168723}. The general framework is designed to enable the rapid development of Lagrangian, Eulerian, and hybrid Eulerian-Lagrangian solvers such as solvers based on the PIC method. IPPL uses the Message Passing Interface (MPI)  paradigm to distribute fields and particles across multiple processes. It also makes use of expression templates \cite{veldhuizen1995expression} to speed up the computation of mathematical operations on field and particle data.

Besides revising IPPL to the latest \CPP17 standard, we replaced the core data structures of IPPL with Kokkos \cite{CarterEdwards20143202,trott2021kokkos} data structures to enable performance portability across various hardware architectures. Although Kokkos allows more flexibility, for simplicity as well as to avoid costly host-device copies IPPL currently allocates all particles and fields on the target device which is either GPU or CPU, but not both.

\begin{figure}[htp]
    \centering
    \includegraphics{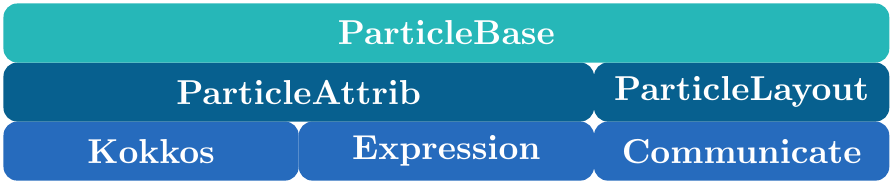}
    \caption{General layout of the particle container class ({\tt ParticleBase}) in IPPL. The underlying container of each particle attribute ({\tt ParticleAttrib}) is a Kokkos View. The memory layout of the particle container is therefore a \textit{struct of arrays}. The communication among processes is managed by its layout ({\tt ParticleLayout}).}
    \label{fig:ippl_particles}
\end{figure}

Figure \ref{fig:ippl_particles} summarizes how IPPL treats particles. It uses the paradigm of \textit{struct of arrays} where each particle attribute ({\tt ParticleAttrib}) is in principle a Kokkos View enhanced with expression templates. Note that the underlying data type of a particle attribute is not restricted to scalar types; hence vector attributes (e.g. position and velocity) themselves are \textit{arrays of struct}. The basic collection of particles consisting of a position ({\tt R}) and unique identity number ({\tt ID}) attributes is represented with {\tt ParticleBase}. Its parallel distribution among processes is managed by the class {\tt ParticleLayout}. Applications must derive their own particle container class from {\tt ParticleBase}.
An example is shown in \Cref{lst:particle_container}. Supplementary particle attributes are easily added with {\tt addAttribute}. This makes sure that the new attributes are included in MPI send and receive operations. In order to send particles from one process to another, all particle attributes are serialized into one buffer. 

These buffers are allocated with an over-allocation factor which is a parameter in IPPL and  can be specified by the users according to their application needs to avoid performance degradation, especially on GPU systems. A high value of this factor is usually favored, as it reduces the number of future allocations after the initial allocation, which are costly on GPUs. However, we may not have enough memory on the GPUs to accommodate such over-allocation, in which case we need to reduce it. It should also be noted that the allocated buffers are never freed until the simulation ends, which may also pose memory constraints during the simulation. With an over-allocation factor of 1.0, the space allocated for each buffer is exactly the requested size. 

\begin{figure}[htp]
    \centering
    \includegraphics[width=1.0\textwidth]{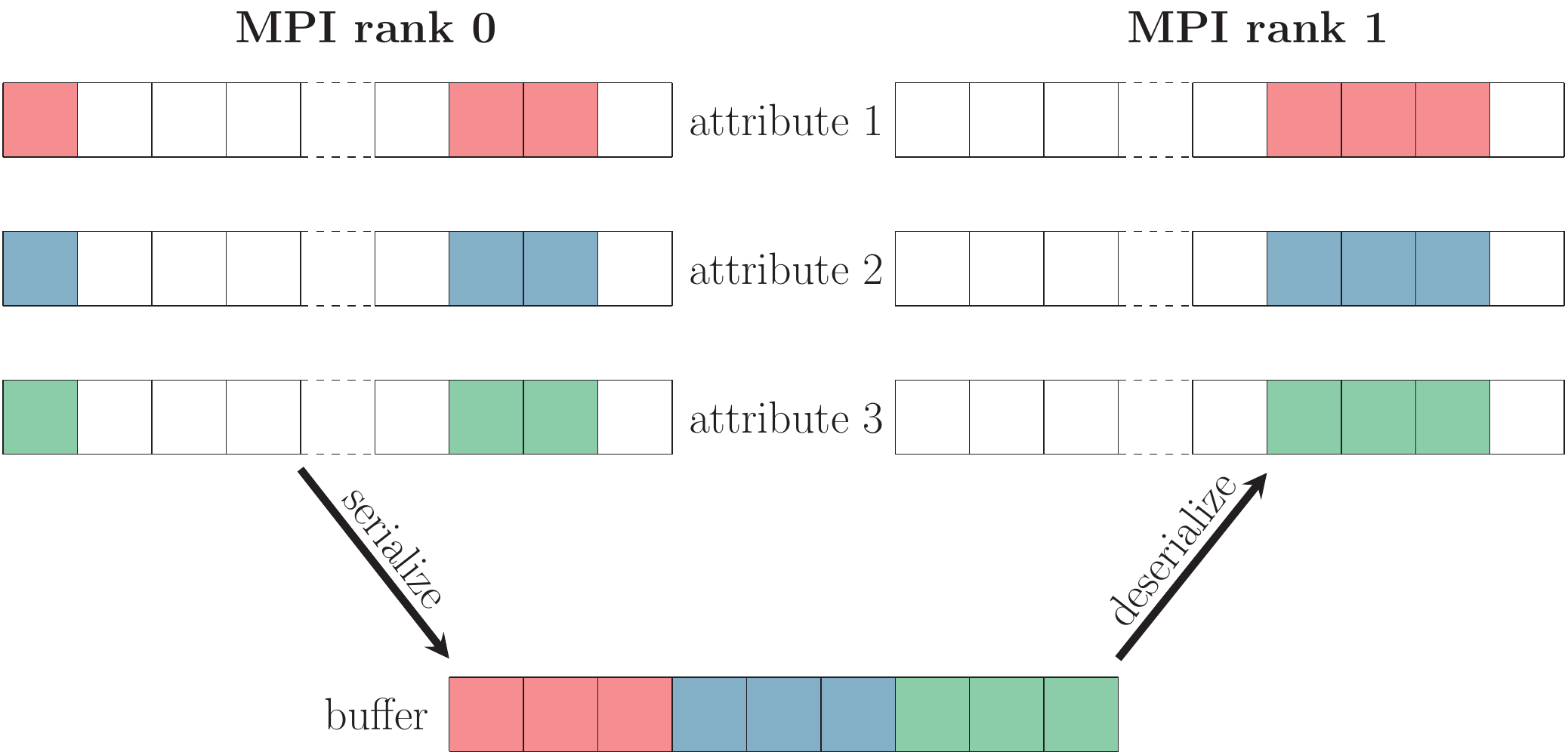}
    \caption{Sending and receiving particles. MPI rank 0 packs all attributes (in this case 3) of all the particles to be sent to rank 1 (again 3 in this case) into a single pre-allocated buffer that is sent to MPI rank 1, which unpacks and appends the received particle data.}
    \label{fig:particle_comm}
\end{figure}

Once the data is sent, the receiver then unpacks and appends the particles to its attribute
containers (see Figure \ref{fig:particle_comm}). The local particle container of each MPI process is also over-allocated to further reduce the amount of dynamic allocations.

To ensure an even workload across all MPI processes, the particles are redistributed regularly using the orthogonal recursive bisection algorithm \cite{quinlan1997mlb}. For this purpose, the particle densities are interpolated onto the grid, which is then divided into subregions, each with approximately the same number of particles. The particles are then passed to the MPI process that occupies the subregion into which they fall.

\begin{lstlisting}[language=C++,
                   basicstyle=\scriptsize\ttfamily,
                   showspaces=false,
                   showstringspaces=false,
                   xleftmargin=\parindent,
                   keywordstyle=\bfseries\color{green!40!black},
                   commentstyle=\bfseries\sffamily\color{red},
                   frame=tb
                   captionpos=b,
                   caption={Example of a particle container class with additional particle attributes (mass and velocity). The design of IPPL allows the user to easily add new attributes. Mathematical expressions compile to a single Kokkos kernel without storing any temporaries, thanks to the use of expression templates.},
                   label={lst:particle_container}
]
using namespace ippl;
template<class PLayout>
struct Bunch : public ParticleBase<PLayout> {
    Bunch(PLayout& playout) : ParticleBase<PLayout>(playout) {
        // add application attributes
        this->addAttribute(mass);
        this->addAttribute(velocity);
    }
    ParticleAttrib<double> mass;
    ParticleAttrib<Vector<double>> velocity;
};

// compiles to single Kokkos kernel
bunch->R = bunch->R + dt * bunch->velocity;
\end{lstlisting}

\begin{figure}[htp]
    \centering
    \includegraphics[width=1.0\textwidth]{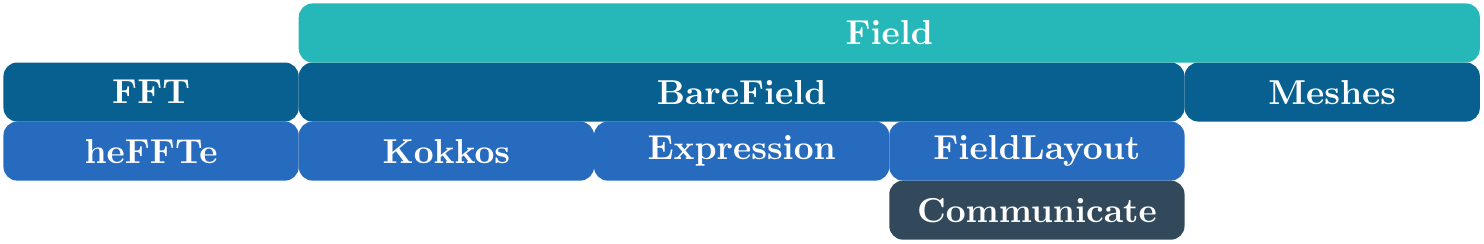}
    \caption{Design layout of fields in IPPL. The underlying data structure is a Kokkos View. IPPL provides an interface to the Fourier transform routines of heFFTe.}
    \label{fig:ippl_fields}
\end{figure}

The design of fields follows a similar structure to that of the particle container class and is shown in Figure \ref{fig:ippl_fields}. The basic structure is again a Kokkos View. We however distinguish between two types of fields: {\tt BareField} and {\tt Field}. The former represents a field without spatial reference. The communication of field data is managed by the field layout ({\tt FieldLayout}). IPPL uses halo (guard or ghost) cells to exchange field data among MPI processes and it follows the same pack-send-receive-unpack set of operations
we described regarding particle communication. The number of halo cells can be set for each field independently. 

An interface to heFFTe \cite{ayala2020heffte} enables us to compute fast Fourier transforms (FFT) in a portable manner. Currently, we have to perform a copy between IPPL fields and the data structures being passed to heFFTe for computing FFTs. This is 
because of the presence of 
halo cells in IPPL fields, which have to be stripped out before passing the field data to heFFTe. Because of this copy operation, we currently have a memory overhead which we will try to remove in future IPPL versions. 

\section{Particle-in-cell method}
\seclab{pic}
\subsection{Particle-in-cell for the Vlasov-Poisson system}
    In this work, we consider the non-relativistic Vlasov-Poisson system with a fixed magnetic field, and introduce the PIC method in that setting. The electrons are immersed in a uniform, immobile, neutralizing background ion population, and the electron dynamics is given by
    \begin{equation}
        \eqnlab{Vlasov}
        \frac{\partial f}{\partial t} + \vb \cdot \nabla_{{\bf x}} f + \frac{q_e}{m_e}\LRp{\Eb + \vb \times \B_{ext}} \cdot \nabla_{\vb} f = 0,
    \end{equation}
where $\Eb = \Eb_{sc} + \Eb_{ext},$ and the self-consistent field due to space charge is given by 
\begin{equation}
    \eqnlab{potential}
    \Eb_{sc} = -\nabla \phi, \quad -\Delta \phi = \rho/\varepsilon_0 = \LRp{\rho_e - \rho_i}/\varepsilon_0.
\end{equation}
In equation \eqnref{Vlasov}, $f({\bf x},\vb,t)$ is the electron phase-space distribution and $q_e$, $m_e$ and $\varepsilon_0$ are the electron charge, mass, and permittivity of free space respectively. The total
electron charge in the system is given by $Q_e=q_e\int\int f \mathrm{d}{\bf x}\mathrm{d}\vb$, the electron charge density by $\rho_e({\bf x}) = q_e\int f \mathrm{d}\vb$ and the
constant ion density by $\rho_i = Q_e/\int \mathrm{d}{\bf x}$. Throughout this paper we use bold letters for vectors and non-bold letters for scalars. The numerical methods and algorithms we discuss in this article can be easily applied to situations involving non-uniform external magnetic fields with finite curvature. On the other hand, physical systems for which relativistic or electromagnetic effects play a significant role will in general require different approaches, and the corresponding codes will perform differently on large scale high performance computing architectures.

The particle-in-cell method discretizes the phase space distribution $f({\bf x},\vb,t)$ in a Lagrangian way by means of macro-particles (hereafter referred to as ``particles" for simplicity). At time $t=0$, the
distribution $f$ is sampled, which leads to the creation of the computational particles. Subsequently, a typical computational cycle in PIC consists of the following steps:

\begin{enumerate}
    \item \label{step1_pic} Assign a shape function - e.g. cloud-in-cell \cite{birdsall2004plasma} - to each particle $p$ and deposit the electron charge onto an underlying mesh. This is known as ``scatter" in PIC.
    \item \label{step2_pic} Use a grid-based Poisson solver to compute $\phi$ by solving $-\Delta \phi = \rho/\varepsilon_0$ and differentiate $\phi$ to 
        get the electric field $\Eb=-\nabla \phi$ on the mesh with appropriate boundary conditions.
    \item Interpolate $\Eb$ from the grid points to particle locations ${\bf x}_p$ using the same interpolation function as in the scatter operation. This is 
        typically known as ``gather" in PIC.
    \item By means of a time integrator, advance the particle positions and velocities using
        \begin{align}
        \eqnlab{momentum_eqn}
            \frac{d\vb_p}{dt} &= \frac{q_e}{m_e}\LRp{\Eb + \vb \times \B_{ext}}|_{{\bf x}={\bf x}_p}, \\
            \frac{d{\bf x}_p}{dt} &= \vb_p.
        \end{align}
\end{enumerate}

\subsection{Numerical implementation}

In our implementation we use the non-dimensional form of the Vlasov-Poisson system and we follow the same non-dimensionalization as in \cite{rodriguez2020implementation} (see Table I). In this case, the equations for 
the PIC scheme look very similar to the dimensional form except that in equation \eqnref{momentum_eqn} the factor $q_e/m_e$ becomes $-1$ and $\varepsilon_0$ in equation \eqnref{potential} is taken as $1$. To avoid introducing new notations, 
in the rest of the article, we denote the non-dimensional quantities with the same notations that we introduced before for
their dimensional counterparts. Now, we succinctly describe the numerical algorithms that are used for the four PIC steps enumerated in the previous section in all our mini-apps. 

        First, we do a random sampling of the particles based on the distribution function $f$ using inverse transform sampling \cite{devroye2006nonuniform}. It is important to create the particles locally in a load balanced manner\footnote{In this paper, when we mention load balanced or imbalanced configurations it is only with respect to particles unless we explicitly mention the fields.} by each MPI rank or GPU. Otherwise, the 
    very high communication required after the particle creation can have a cost which is prohibitive, even if it is a one time cost. It can also lead to 
    a crash in the simulation due to memory requirements during communication or in the creation stage itself, especially on GPUs. Moreover, for simulations on GPUs, we create the particles directly
    on the device using the Kokkos random number generator rather than first creating them on the host (CPU) and then copying them to the device. This
    host-to-device copy would also have very significant cost for the large numbers of particles that we are targeting in this work, and be prohibitively costly even if it is a one time cost. However, it should be noted that the random number generation on GPUs is not deterministic even with a 
    fixed seed and this can present some difficulties with respect to reproducibility. Nevertheless, the results are still reproducible
    in a statistical sense.
        
        We use a cell-centered grid and cloud-in-cell shape function for interpolation from particles to grid and vice versa. We use 
        periodic boundary conditions in all directions and since our grid spacings in the $x$, $y$ and $z$ directions are constants (although potentially different in each direction), 
        we use an FFT-based spectral solver for the computation of the potential and electric field from the charge density.\ The FFTs
        in the field solver are taken using the heFFTe library. For the time integration, we use the synchronized form of the 
        Boris scheme as given in equations 4(a) - 4(c) of \cite{tretiak2019arbitrary}, which gives both the velocity and 
        position of the particles at integer time steps. Having explained the numerical algorithms used in our mini-apps we move on to
        describe the physical parameters for each of the examples considered. 

\section{The Mini-Apps}
\seclab{miniapps}
    In this section, we briefly describe the electrostatic plasma physics problems that we consider for the scaling and performance study. 
        
\subsection{Landau damping}
    The first problem we consider is 
    Landau damping in the weak and strong damping regimes. It is a classical problem which has been studied extensively in the literature \cite{ricketson2016sparse,ho2018physics,myers20174th,kormann2015semi,chen2011energy}. The availability of analytical results makes it an excellent candidate for our verification and performance study. Similar to \cite{ricketson2016sparse}, we consider the following initial distribution
   \begin{equation}
    \eqnlab{landau_f}
        f(t=0) = \frac{1}{\LRp{2\pi}^{3/2}} e^{-|\vb|^2/2} \LRp{1+\alpha\cos(k x)} \LRp{1+\alpha\cos(k y)} \LRp{1+\alpha\cos(k z)},
    \end{equation}
    in the domain $\LRs{0,L}^3$, where $L = 2\pi/k$ is the length in each dimension. We choose the 
    following parameters for our weak Landau damping tests: $k = 0.5$, $\alpha=0.05$. The total electron charge based 
    on our initial distribution is $Q_e = -L^3$. For the case of
    strong Landau damping the difference is that we use a stronger perturbation parameter $\alpha=0.5$. The other parameters are the same as those for the weak damping case. The presence of a stronger perturbation parameter in the strong damping case necessitates particle load balancing as will be shown in Figures \figref{nonlinear_gpus}, \figref{nonlinear_cpus}. 
\begin{figure}[]
    \includegraphics[width=\textwidth]{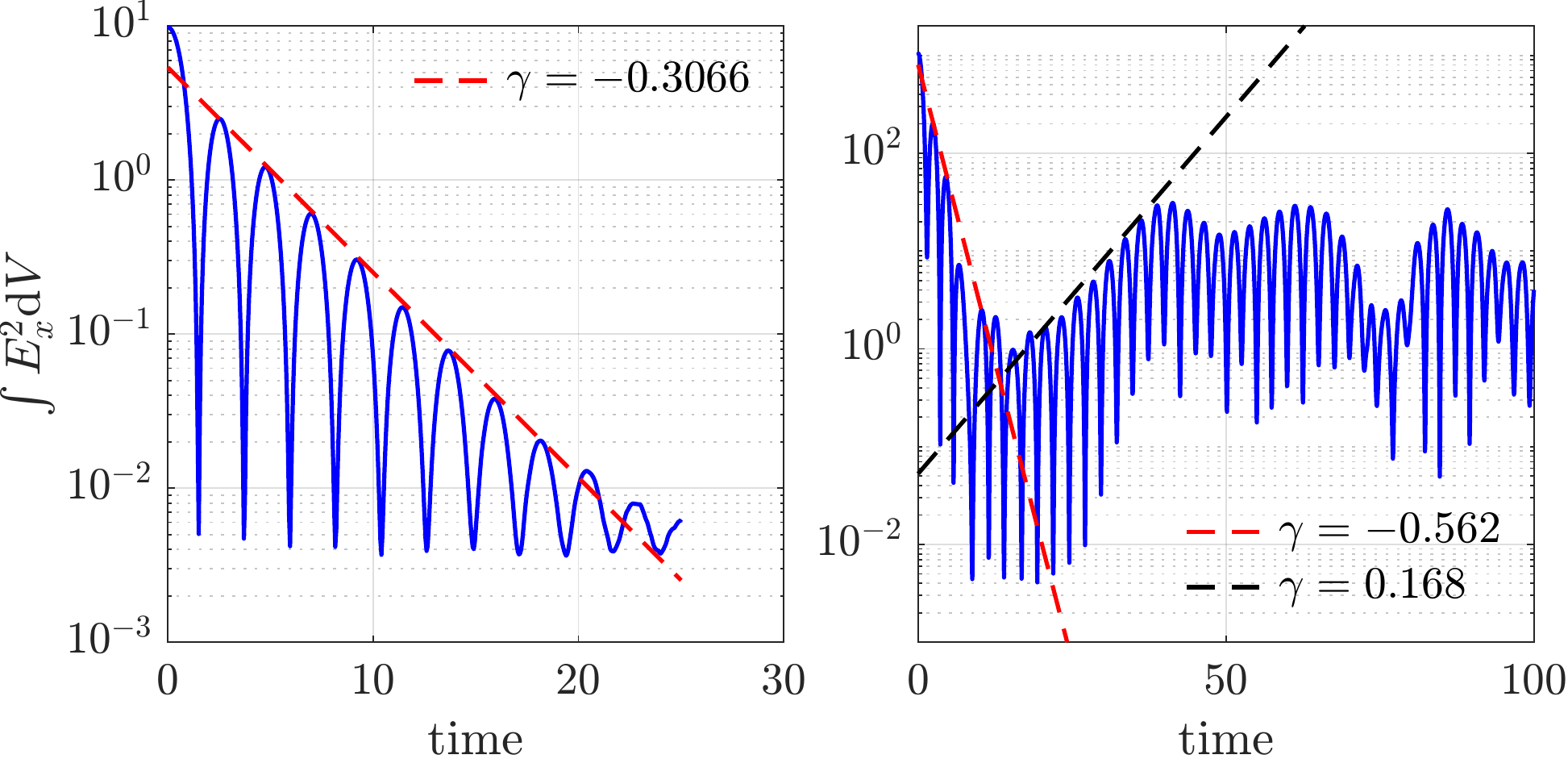}
    \caption{Landau damping of the electric field energy in the $x$-direction ($\int_{V} E_x^2 dV$, where $V$ is the total simulation volume) as a function of time for weak (left) and strong (right) damping cases. The rates match well with the analytical values shown by the dashed lines as well as the results in \cite{ho2018physics}. The number of mesh points, number of particles and the time step in these simulations are $32^3$; $83,886,080$ and $0.05$ respectively.}
    \figlab{Landau_damping_rate}
\end{figure}
As a verification, we show in Figure \figref{Landau_damping_rate} the damping of the electric field energy in the $x$-direction 
for the weak and strong damping cases. Our results agree well with the analytical rates as well as
the results in \cite{ho2018physics}.

\subsection{Bump-on-tail/Two-stream instability} 
    The second mini-app we consider is the two-stream or bump-on-tail instability problem. Similar to Landau damping, this is another
    classical benchmark problem studied in the literature \cite{ho2018physics,myers20174th,kormann2015semi,chen2011energy}, which also has analytical estimates for the growth rates, which can be calculated from the dispersion relation derived from the linearized equations. 

    We consider the following initial distribution of electrons

    \begin{equation}
    \eqnlab{twostream_f}
        f(t=0) = \frac{1}{\sigma^3\LRp{2\pi}^{3/2}}\LRc{\LRp{1-\epsilon}e^{-\frac{|\vb-\vb_{b1}|^2}{2\sigma^2}} + \epsilon e^{-\frac{|\vb-\vb_{b2}|^2}{2\sigma^2}}}\LRp{1+\alpha\cos(kz)},
    \end{equation}
    in the domain $\LRs{0,L}^3$, where $L = 2\pi/k$ is the length in each dimension. Depending on the choice of parameters, 
    we get two flavors of this example. With $\epsilon=0.5$, $\sigma=0.1$, $k=0.5$, 
    $\alpha=0.01$, $\vb_{b1}=\LRc{0,0,-\pi/2}$ and $\vb_{b2}=\LRc{0,0,\pi/2}$ we get the two-stream instability problem 
    studied in \cite{ho2018physics}. On the other hand, choosing $\epsilon=0.1$, $\sigma=1/\sqrt{2}$, $k=0.21$, 
    $\alpha=0.01$, $\vb_{b1}=\LRc{0,0,0}$ and $\vb_{b2}=\LRc{0,0,4}$ we get the bump-on-tail instability problem 
    similar to \cite{sarkar2015bump}. The total charge $Q_e$ is chosen in the same 
    way as in the Landau damping example.

\begin{figure}[]
    \subfigure[Initial velocity distribution in $z-$direction for two-stream (left) and bump-on-tail (right) instabilities.]{\includegraphics[width=\textwidth]{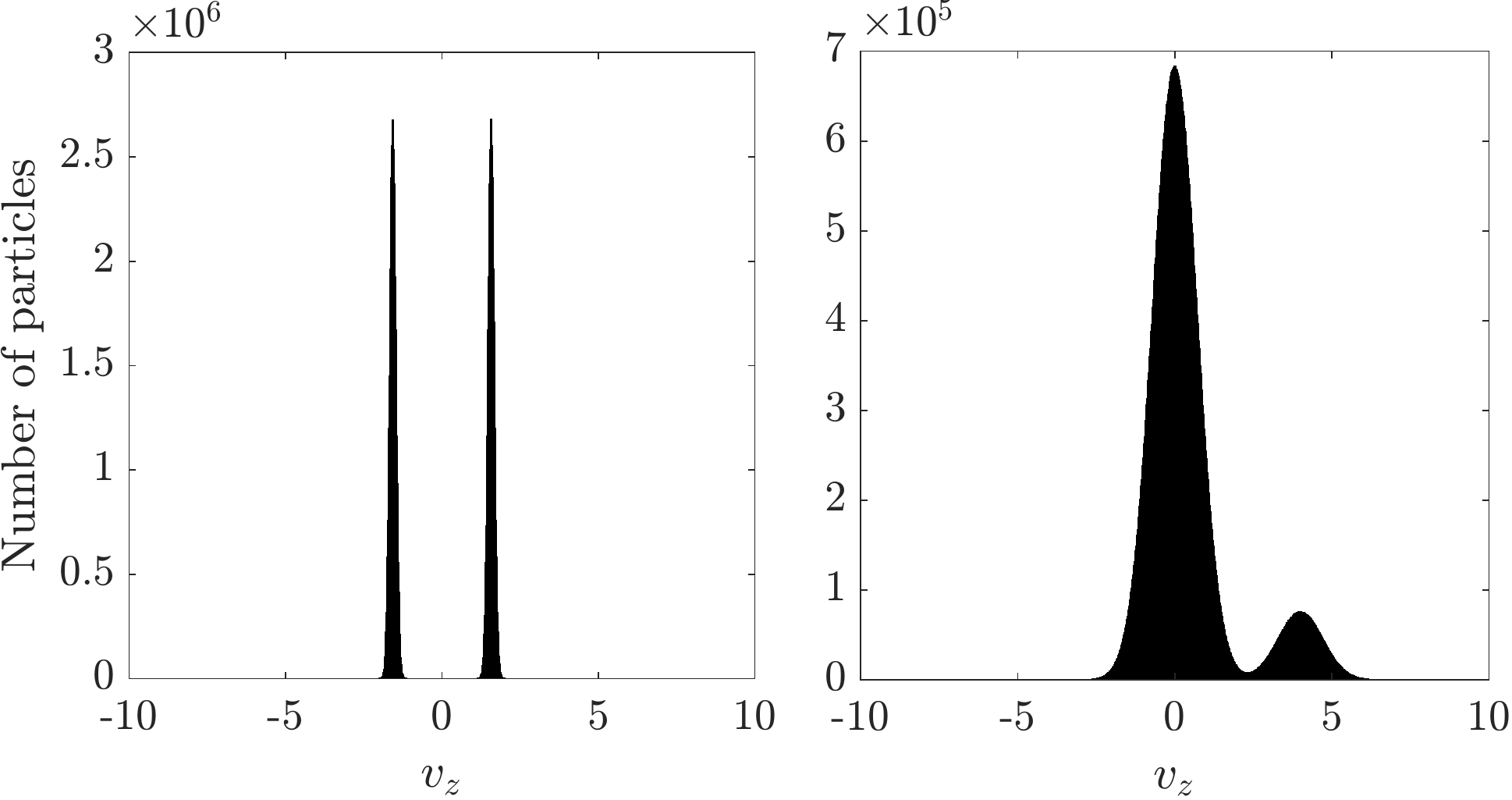}}
    \subfigure[Electric field energy in $z$-direction for two-stream (left) and bump-on-tail (right) instabilities.]{\includegraphics[width=\textwidth]{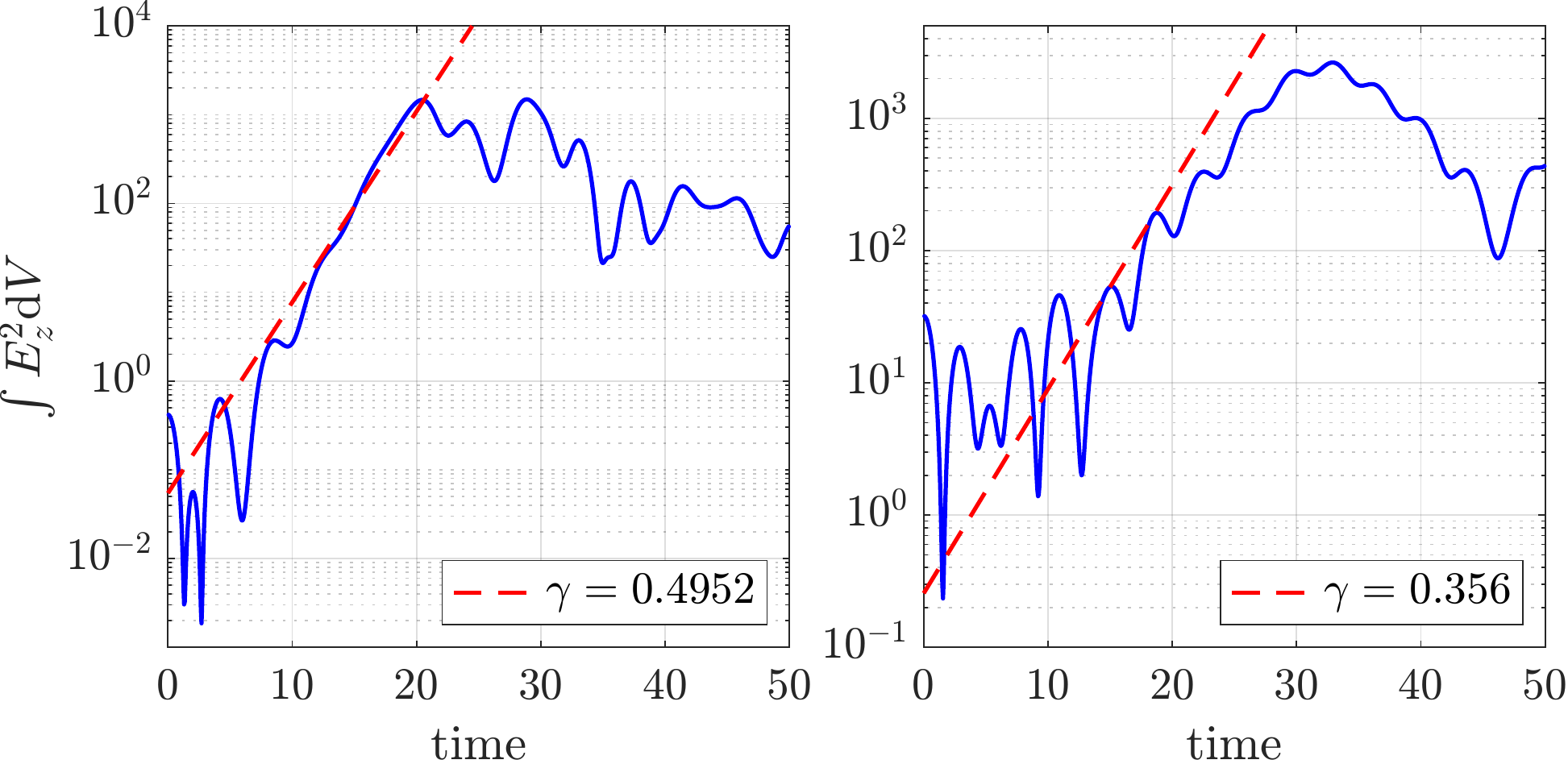}}
    \caption{Initial velocity distribution in the $z$-direction (top row) and electric field energy ($\int_{V} E_z^2 dV$, where $V$ is the total simulation volume) in the $z$-direction as a function of time (bottom row) for the two-stream and bump-on-tail instability test cases. The growth rates agree 
    well with the analytical values shown by the dashed lines as well as the results in \cite{ho2018physics}. The number of mesh points, number of particles and the time step in these simulations are $32^3$; $83,886,080$ and $0.05$ respectively.}
    \figlab{TSI_BTI}
\end{figure}

    In the top row of Figure \figref{TSI_BTI}, we show the 
    velocity distribution in the $z$-direction for these two 
    sets of parameters. In both cases, the electric field energy grows as a function of time as shown in the bottom row of Figure \figref{TSI_BTI}. Similar 
    to Landau damping, we see very good agreement with the 
    analytical rates as well as the results in 
    \cite{ho2018physics}. We would like to mention that even though
    we simulate the bump-on-tail or two-stream instability in 3D-3V in our 
    mini-app, the essential physics for the initial distribution selected occurs predominantly in the 
    $z$-direction, at least for early times.

\subsection{Electron dynamics in a charge neutral Penning trap}
    Our next mini-app corresponds to the dynamics of electrons in a Penning trap with a neutralizing static ion background, as we already considered in \cite{muralikrishnan2021sparse}.
    This problem involves bunching of electrons in configuration space, and therefore presents challenges in terms of 
    field and particle load balancing. The initial conditions for this example, as well as the electron dynamics they lead to, are 
    very similar to that of cyclotrons \cite{adam1995space,yang2010beam,muralikrishnan2021sparse}. Since the particle 
    accelerator library OPAL \cite{adelmann2019opal} will include the portable version of IPPL in the near future, this example is of interest from the point of view of cyclotron simulations.

    Regarding the parameters for this problem, we follow the same setup as in 
    \cite{muralikrishnan2021sparse}. The domain is $\LRs{0,L}^3$, where $L=20$. The external magnetic field is given by $\B_{ext}=\LRc{0,0,5}$ and the quadrapole external electric field by 
    \[
        \Eb_{ext} = \LRc{-\frac{15}{L}\LRp{x-\frac{L}{2}},-\frac{15}{L}\LRp{y-\frac{L}{2}},\frac{30}{L}\LRp{z-\frac{L}{2}}}.
    \]

    For the initial conditions, we sample the phase space using a Gaussian
    distribution in all the variables. The mean and standard deviation for
    all the velocity components are $0$ and $1$, respectively. While the mean
    for all the configuration space variables is $L/2$, the standard 
    deviations are $0.15L$, $0.05L$ and $0.2L$ for $x$, $y$ and $z$, respectively. The total electron charge is 
    $Q_e=-1562.5$. 

\subsection{Uniform plasma test}
    The uniform plasma test is an idealized, non-physical test case. It randomly moves 
all the particles in the range $[-h,h]$ during each time step and carries out the other steps
as in a typical PIC cycle. This mini-app can be used as a gold standard for the performance of different components with which the other
mini-apps can be gauged and understood. 

\section{Scaling results}
\seclab{scaling}
\subsection{Setup}
   In this section, we present representative strong and weak scaling results as well 
   as the performance of different 
   components for the mini-apps described 
   in Section \secref{miniapps}. We do not present the scaling results for the two-stream and bump-on-tail instabilities 
   or the uniform plasma test, as they are very similar to those of the weak Landau damping study; the timings differ by at most $\pm10$ percent for the most part. 

   For the strong scaling, we consider the mesh and particle setups shown in Table \ref{tab:strong_scaling_setup}.
        \begin{table}[h!b!t!]
        \centering
        \begin{tabular}{|r|c|c|c|}
        \hline
            \!\!\! Case \!\!\!\! & \!\!\! Grid ($N_c$) \!\!\!\! & \!\!\! Total number of Particles ($N_p$) \!\!\!\! & \!\!\! Particles per cell ($P_c$) \!\!\!\!\\
        \hline
            A & $512^3$ & $1,073,741,824$ & 8 \\
            B & $1024^3$ & $8,589,934,592$ & 8 \\
            C & $256^3$ & $1,073,741,824$ & 64 \\
            D & $512^3$ & $8,589,934,592$ & 64 \\
        \hline
        \end{tabular}
        \caption{\label{tab:strong_scaling_setup}Cases considered for the strong scaling study with different number of grid points and particles. The total number of particles for the two sets of cases (A,B) and (C,D) are the same, however, the latter has 8 times more particles per cell than the former due to reduced grid resolutions.}
        \end{table}
        For the scaling studies up to Section 
\secref{penning_trap}, the simulations are done
on the CPU and GPU partitions of the Piz Daint
supercomputer at the Swiss National Supercomputing Centre (CSCS), Switzerland. The architectures of Piz Daint's GPU and CPU partitions are as follows:
\begin{itemize}
            \item GPU nodes: Each Cray XC50 compute node has 1 NVIDIA Tesla P100 GPU with 16 GB memory and 12 Intel Xeon E5-2690 v3 cores at 2.6 GHz with 64 GB RAM.
            \item CPU nodes: Each Cray XC40 compute node has $36$ Intel Xeon E5-2695 v4 cores at 2.1 GHz with either 64 or 128 GB RAM.
            \item The interconnect configuration is Aries and the network topology is Dragonfly.
\end{itemize}
    
    For the GPU simulations, we use 1 MPI rank per GPU and no OpenMP thread-based parallelism, whereas for the CPU simulations we use 1 MPI rank and 32 OpenMP threads per node. Basically, we use MPI for communication 
    between GPUs both within a node and between nodes, whereas for CPU-based simulations, we use MPI only 
    for communication between nodes 
    while OpenMP-based shared 
    memory parallelism is used within a node. 
    This setup helps to minimize the particle and field communication costs.

    We conduct the strong scaling study for cases A and C with the following sets of GPU and CPU nodes: number of GPU nodes $\in\LRc{16,32,64,128,256}$, number of CPU nodes $\in\LRc{4,8,16,32,64}$. The minimum number of GPU or CPU nodes is based on 
    the memory requirement for the given number of particles and grid points. For the maximum number of nodes, we stop after increasing the number of MPI ranks or GPUs five times, each time by a factor of 2, as is done in \cite{myers2021porting}. The reason is that the runtimes are
    mostly communication dominated at this stage and the problems stop scaling beyond that. For cases B and D, the number of grid points and particles are both 8 times larger and hence the number of CPU and GPU resources are increased 
    accordingly, i.e. number of GPU nodes $\in\LRc{128,256,512,1024,2048}$, number of CPU nodes $\in\LRc{32,64,128,256,512}$.

     In terms of the time step, for the Landau damping mini-app we choose it to be proportional to the mesh 
     size $h$ as $\Delta t = 0.5h_{min}$. We also make sure that it is below the
    stable time step requirement of $\Delta t \leq 2 \omega_{pe}^{-1}$, where $\omega_{pe}$ is the electron plasma frequency.
    Since the Penning trap simulations involve a lot more particle communication than the 
    Landau damping problem, we choose the time step based on the finest mesh used in our scaling
    studies, i.e. $\Delta t = 0.5(L/2048) \approx 0.005$, just to have the same dynamics, and therefore similar particle communication, for 
    different grid resolutions in a weak scaling study.

    For the mini-apps we choose an over-allocation factor of 2.0 for the weak and strong Landau damping tests,
    whereas for the Penning trap simulations we choose a value of 1.0 due to its high memory requirement per rank. These values are chosen based on 
    numerical experiments.

    For the ease of performance analysis, in Table \ref{tab:parallel_comm_kernels} we split the significant kernels in our code 
    into computation kernels, from which we can expect parallel efficiency, and communication kernels, which are required because of 
    domain decomposition. However, this separation is not perfect since the FFTs required for the field solve 
    are computed using heFFTe and this includes communication in addition to the purely local operations. This is because we use heFFTe as a black box for
    Fourier transforms and hence do not use IPPL timers inside its source code. The total time which is included in the 
    computation kernels column is the time taken for the entire simulation to finish including the communication kernels. However, we exclude the time spent writing output data to files and an initial warm up call to the field solve. The first call to the field solve takes significantly more time than the subsequent ones owing to the initializations performed by heFFTe. In the communication kernels, the ``Fill halo cells" and ``Accumulate halo cells" operations are required during the 
    gather and scatter stages of the PIC cycle, and the ``Particle update" sends the particles to the appropriate ranks once they 
    leave the local subdomain of the current rank. The components not included in Table \ref{tab:parallel_comm_kernels} account for less than
    $10$ percent of the total time in most of our simulations. We therefore do not consider them in the performance study.

\begin{table}[h!b!t!]
\centering
\begin{tabular}{|l||l|}
\hline
    \!\!\! Computation kernels \!\!\!\! & \!\!\! Communication kernels \!\!\!\! \\
\hline
    Gather ({\tt gather}) & Particle update ({\tt updateParticle}) \\
    Scatter ({\tt scatter}) & Fill halo cells ({\tt fillHalo})\\
    Push position ({\tt pushPosition}) & Accumulate halo cells ({\tt accumulateHalo}) \\
    Push velocity ({\tt pushVelocity}) & Particle load balance ({\tt loadBalance}) \\
    Particle BCs ({\tt particleBC}) & \\
    FFT-based field solve ({\tt solve})& \\
    Total time ({\tt total})& \\
\hline
\end{tabular}
    \caption{\label{tab:parallel_comm_kernels}List of computation and communication kernels used for the performance study. Inside the parentheses are the labels which are used to represent the components in the figures.}
\end{table}

We run the simulations for 20 time steps and report in figures the wall time per simulation time step for each of the kernels in Table \ref{tab:parallel_comm_kernels}. As such, our 
performance figures are not indicative of production runs, as one typically needs to run for thousands of time steps in real plasma simulations. Furthermore, depending on the long time dynamics of the problem under consideration the performance can be significantly different. Our objective
here is to assess the performance of different components in the mini-apps across different architectures without spending too many node hours or having to wait for long periods in the job submission queues.  

 The versions of the compilers, MPI, Kokkos, heFFTe and IPPL used for the simulations on each of the computing architectures considered in this work is mentioned in Appendix A. 
For the parameters in heFFTe we chose pencil decomposition, pipelined 
point-to-point communication with no-reordering. We refer the readers to \cite{ayala2020heffte} for descriptions of these parameters. 
These options were chosen based on our heFFTe benchmarking study with all 
the possible parameter combinations, selecting the best one in terms of scalability and time to solution. In terms of domain decomposition for our mini-apps, we use parallelization in all three directions
for the fields, which gives each processor a brick of field data along with one layer of halo cells. 
\subsection{Weak Landau damping}
\seclab{linear_landau}
\subsubsection{Strong scaling: 8 particles per cell}
    
\begin{figure}[]
    \subfigure[Computation kernels for case A (left) and case B (right)]{\includegraphics[width=\textwidth]{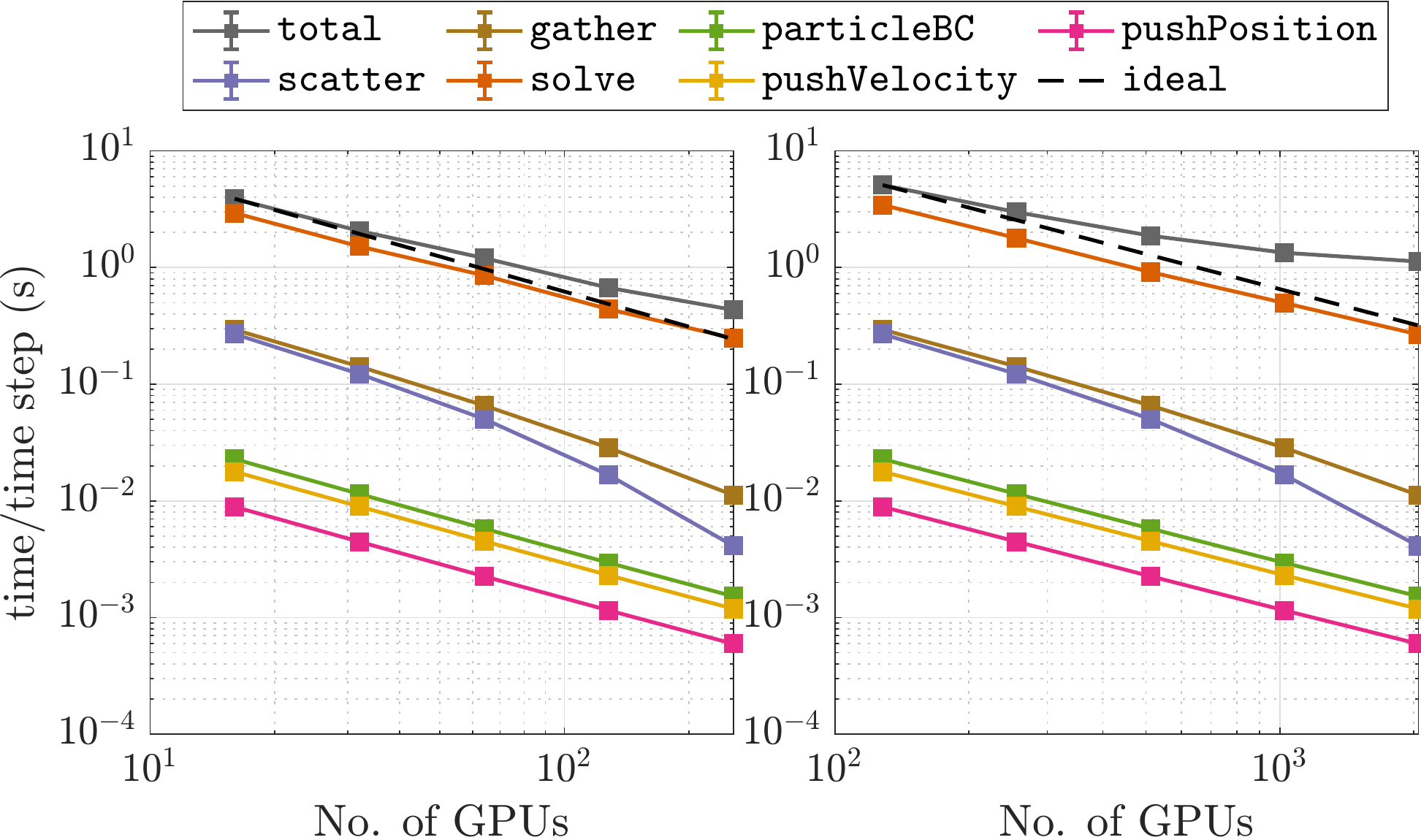}}
    \subfigure[Communication kernels for case A (left) and case B (right)]{\includegraphics[width=\textwidth]{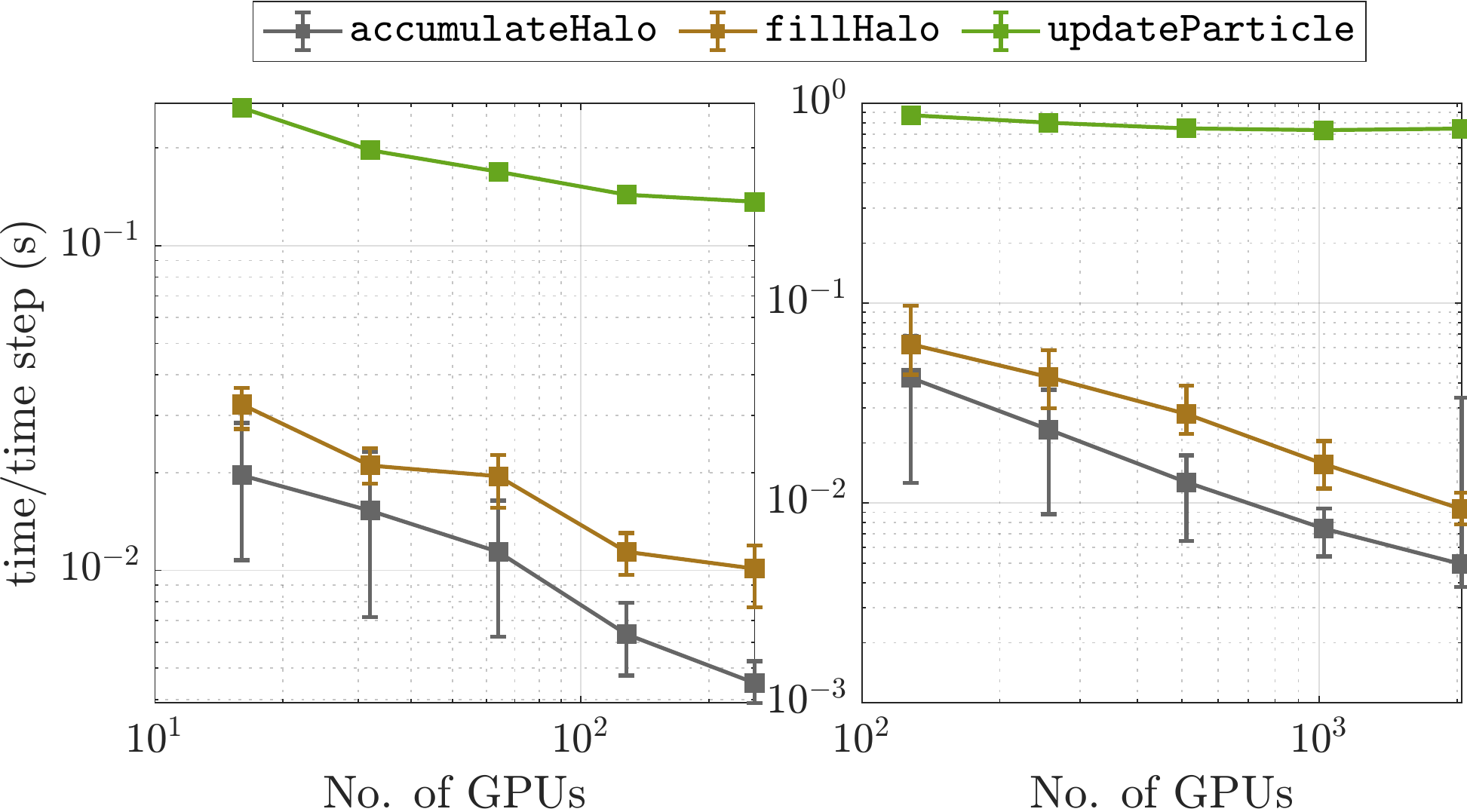}
    \figlab{comm_case_linear_gpu}}
    \caption{Weak Landau damping. Strong scaling on GPUs: scaling of computation and communication kernels for cases A and B.}
    \figlab{linear_gpu1}
\end{figure}

    In this section we present the scaling results for the weak Landau damping problem. We first consider the strong scaling study.
    In Figure \figref{linear_gpu1}, we show the timings of computation and communication kernels listed in Table \ref{tab:parallel_comm_kernels}
    for cases A and B on GPUs. In all the figures we show the average, minimum and maximum wall times across the MPI ranks or GPUs by means
    of square markers and errorbars. It is clear from the figure that the computation kernels scale almost ideally, except for the {\tt solve}. Since the FFT-based field solver also includes communication between GPUs, the scaling deteriorates when we increase the number of GPUs.
    We also observe that the field solve is the dominant computation kernel. It takes an order of magnitude more time than the next dominant kernel which is the {\tt gather}. The runtimes of the field solve are also two orders of magnitude higher than those of the pure particle kernels. 

    One way to improve the field solver performance is by using a 2D or 1D domain decomposition which gives pencils or slabs for the fields rather than bricks as we use in this study. This will reduce the number of transposes required in the 
    FFT computation, which in turn leads to a reduction in runtime of the field solve. However, it can have adverse effects on the field and particle communication costs in the rest of the code. A 
    thorough study of the different domain decompositions, in order to determine the decomposition leading to the best overall performance is beyond the scope of this article. We intend to investigate this question in our future work.

    Regarding the communication kernels, the {\tt updateParticle} takes the most time, as seen in Figure \figref{comm_case_linear_gpu}. Let us therefore examine its components in more detail: 
\begin{enumerate}
    \item The first one is the search for all the local particles across all the MPI ranks to figure 
    out the mapping between particles and ranks after the push. Since each MPI rank knows the global information about the domains owned by the other ranks, this search happens locally without any need for communication. Then, each processor communicates to all the
    other processors how many particles they receive from it. These two operations do not scale in our current implementation, as they involve two loops: one over the 
    local number of particles and another over the total MPI ranks. The product of these quantities is a constant in a strong scaling setup, 
    and increases linearly in a weak scaling setup. 
    \item The next component involves packing and sending the particles which are leaving the local field domain of
    the current processor, and receiving and unpacking the particles which are coming from the other processors. 
        This part scales in both the strong and weak scaling sense\footnote{It should be noted that the statements about the scaling of the components are valid only when there is not much imbalance in the particle loads across ranks.}. 
\end{enumerate}
    When the second component dominates, we see a decrease in {\tt updateParticle} runtime as 
    illustrated by the first four data points in the left column of Figure \figref{comm_case_linear_gpu} for case A. This is because the weak 
    Landau damping simulations have a very small particle load imbalance. In our experiments, we noticed that it does not require load balancing. However, the first component will eventually start 
    dominating, and we then see a flat regime in the particle update cost. For case B this happens almost from the beginning, as we can observe from the right column of Figure \figref{comm_case_linear_gpu}. This is due to both the higher particle count and the 
    higher GPU count compared to case A.

    From Figure \figref{comm_case_linear_gpu}, the field communication kernels {\tt fillHalo} and \\{\tt accumulateHalo} take much less time compared to the {\tt updateParticle}. They also decrease in a strong scaling setup with increasing number of GPUs, as the number of halo cells to be communicated with the field neighbors decreases.

\begin{figure}[]
    \subfigure[Computation vs communication kernels for case A (left) and case B (right)]{\includegraphics[width=\textwidth]{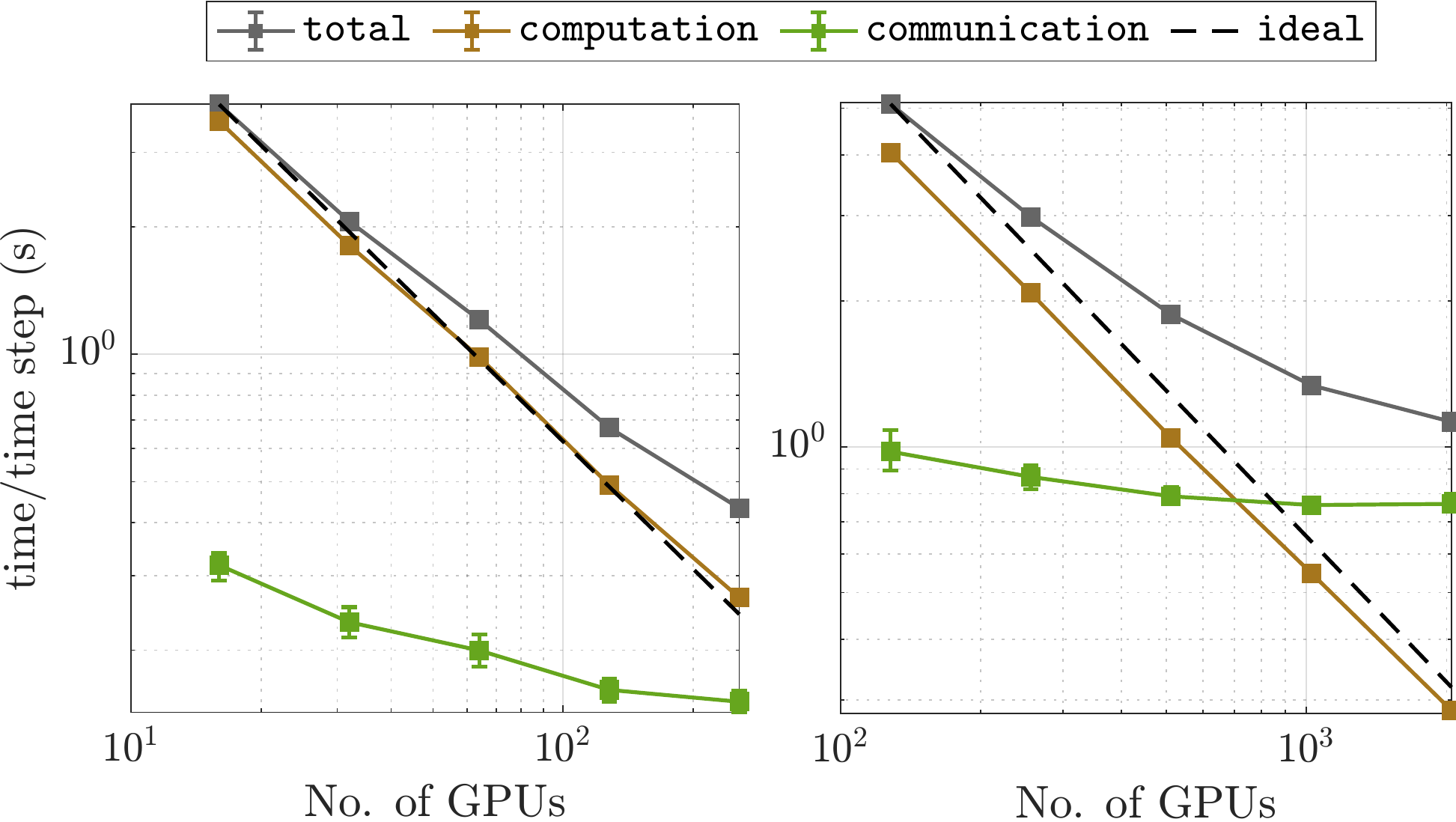}
    \figlab{parallel_comm_case_linear_gpu}}
    \subfigure[Efficiency of computation kernels for case A (left) and case B (right)]{\includegraphics[width=\textwidth]{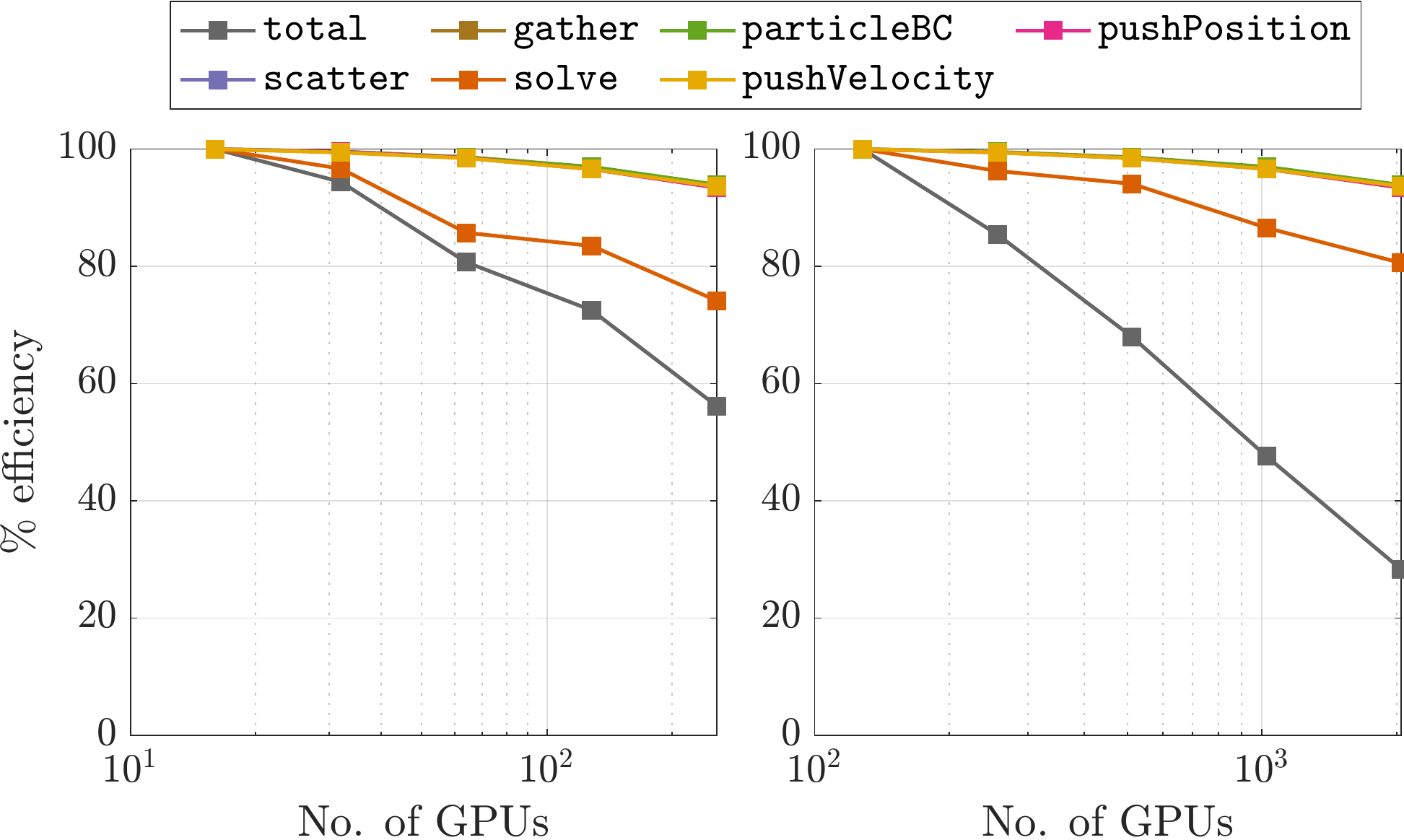}
    \figlab{eff_case_linear_gpu}}
    \caption{Weak Landau damping. Strong scaling on GPUs: cross over point between computation and communication kernels and
    parallel efficiencies as a function of the number of GPUs for cases A and B.}
    \figlab{linear_gpu2}
\end{figure}

    In Figure \figref{parallel_comm_case_linear_gpu} the cumulative time of computation 
    kernels (except total time) and the communication kernels are compared with the total time, for increasing number of GPUs. While in case A the cross over point between 
    communication and computation costs is not seen for our set of parameters, in case B the communication kernels dominate over the computation kernels for the last two 
    data points. The efficiency is shown in Figure \figref{eff_case_linear_gpu} for the computation
    kernels. We see that the total efficiency is comparable to the field solver efficiency until the communication kernels become comparable or start dominating, at which point we observe a significant drop in the total efficiency. All other computation kernels except for the field solver exhibit near ideal efficiency, as visually observed in Figure \figref{linear_gpu1}.

\begin{figure}[]
    \subfigure[Computation kernels for case A (left) and case B (right)]{\includegraphics[width=\textwidth]{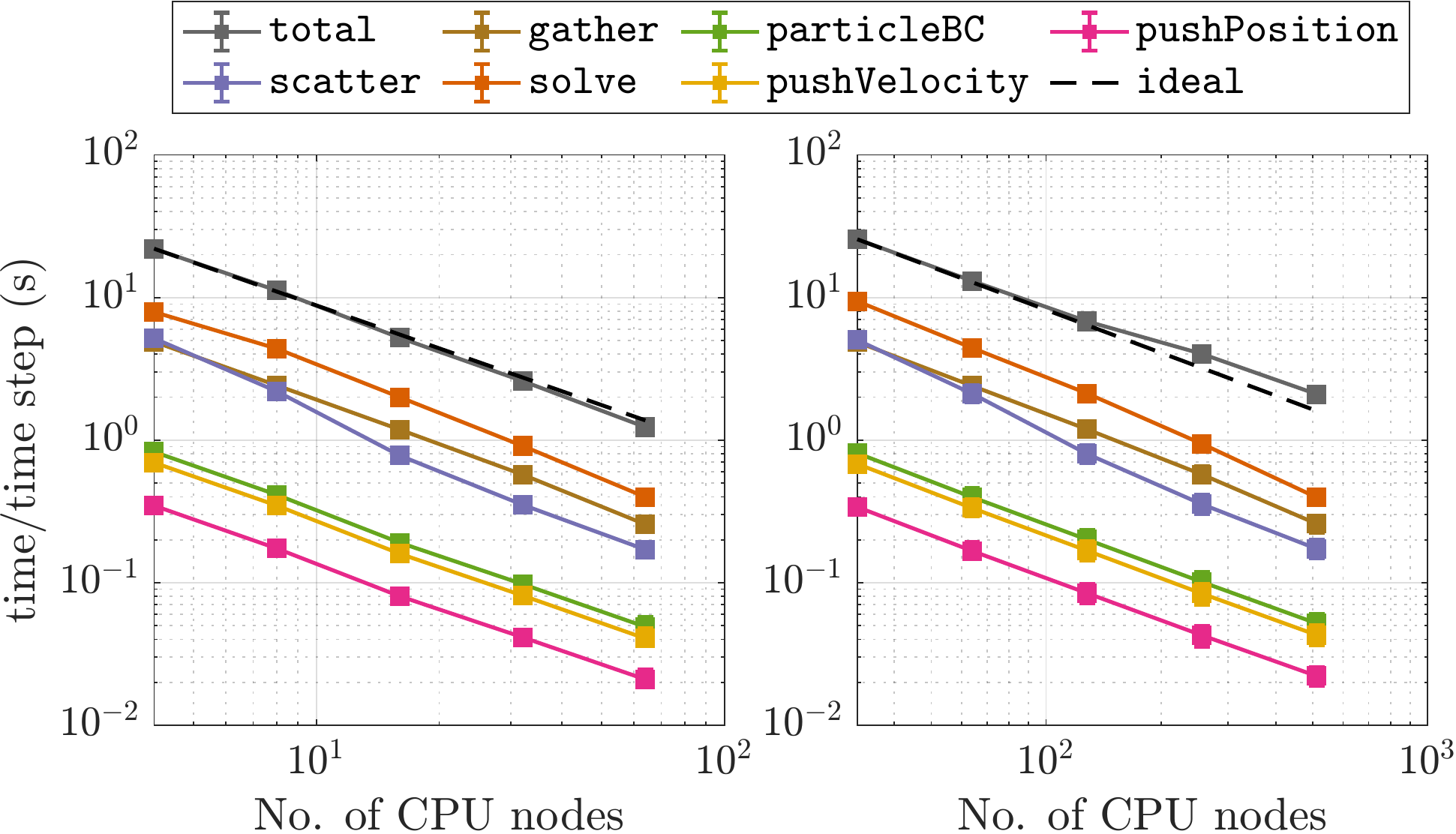}
    \figlab{parallel_case_linear_cpu}}
    \subfigure[Communication kernels for case A (left) and case B (right)]{\includegraphics[width=\textwidth]{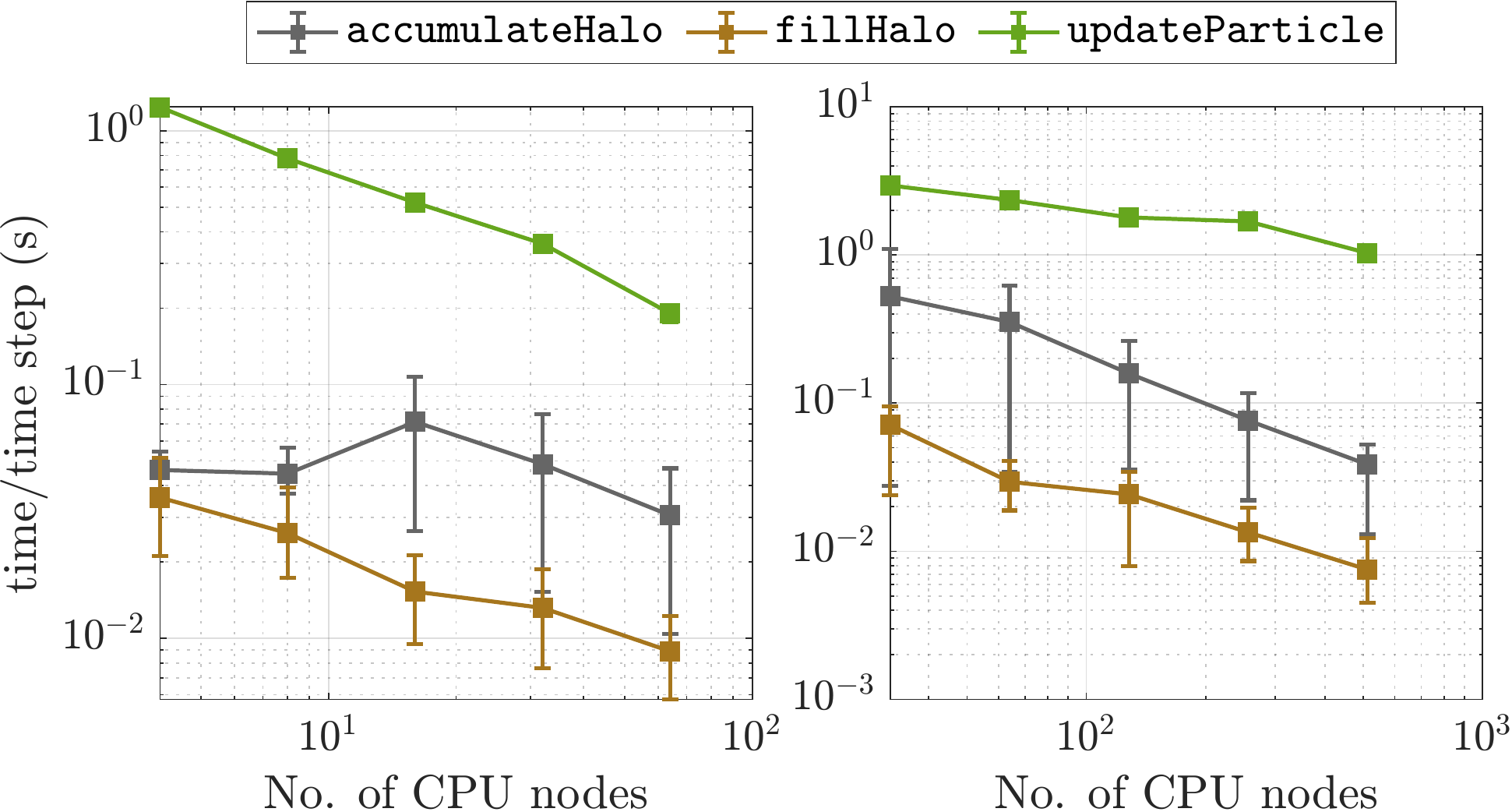}}
    \caption{Weak Landau damping. Strong scaling on CPUs: scaling of computation and communication kernels for cases A and B.}
    \figlab{linear_cpu1}
\end{figure}

\begin{figure}[]
    \subfigure[Computation vs communication kernels for case A (left) and case B (right)]{\includegraphics[width=\textwidth]{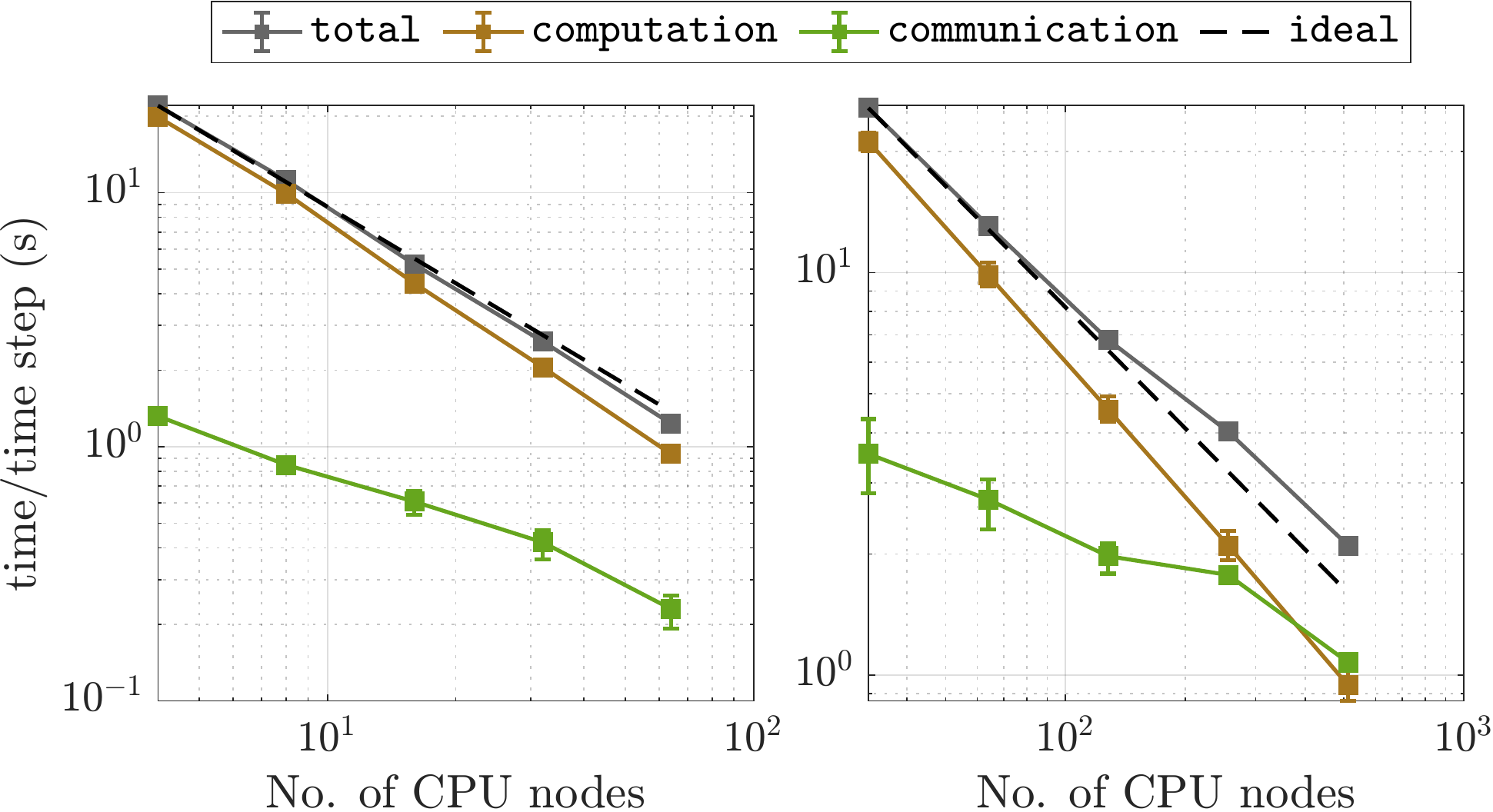}}
    \subfigure[Efficiency of computation kernels for case A (left) and case B (right)]{\includegraphics[width=\textwidth]{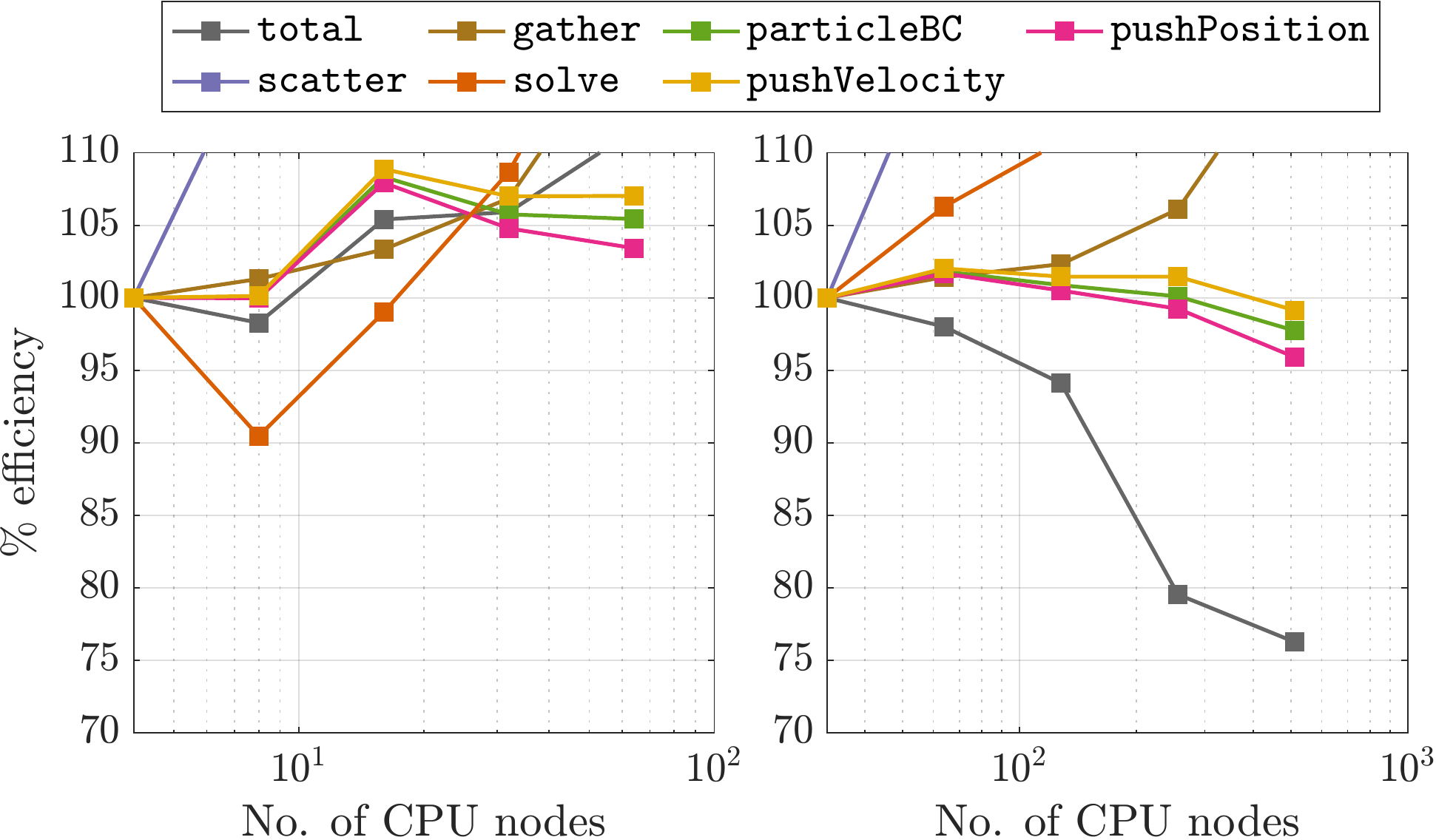}}
    \caption{Weak Landau damping. Strong scaling on CPUs: cross over point between computation and communication kernels and
    parallel efficiencies as a function of the number of CPU nodes for cases A and B.}
    \figlab{linear_cpu2}
\end{figure}

    In Figures \figref{linear_cpu1} and \figref{linear_cpu2} we show the results for the CPU-based strong scaling study. Here, we only highlight the differences with the GPU simulations. We first observe from Figure \figref{parallel_case_linear_cpu} that even though
    field solve is still the dominant computation kernel, the difference between {\tt solve} and the next dominant kernel, {\tt gather}, as well as the pure particle kernels, is much
    less than on the GPUs. 
    Next, we note that the field solve efficiency and the total efficiency are higher for the CPU cases than the GPU ones. This is partly due to the fact
    that we can use fewer MPI ranks in the CPU simulations than the GPU simulations, because the CPU nodes have more memory. This in turn leads to lower communication costs and increased efficiency.  Apart from these differences, the other observations we made for the GPU simulations hold for the CPU cases too.

\begin{figure}[]
    \subfigure[Computation kernels for case C (left) and case D (right)]{\includegraphics[width=\textwidth]{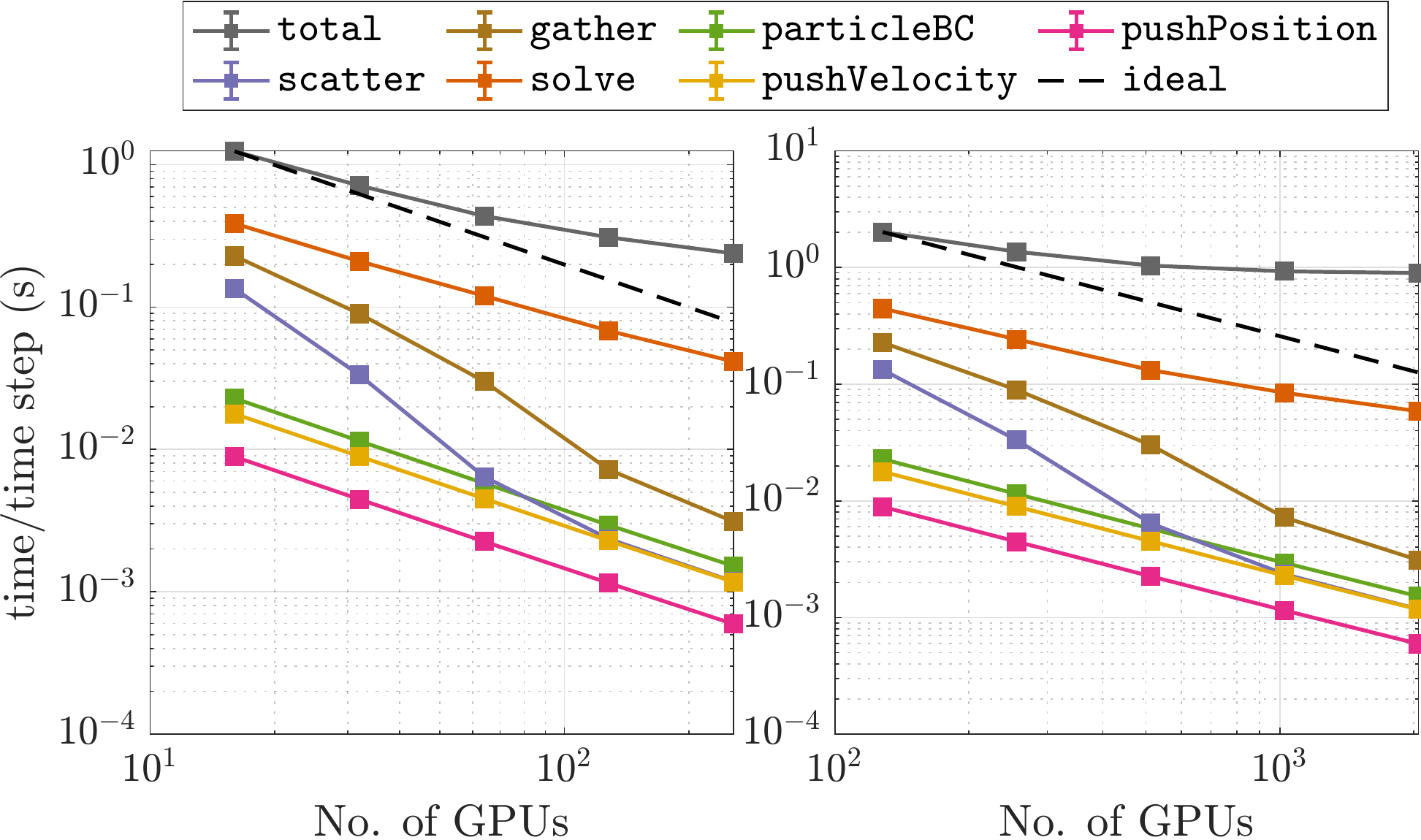}
    \figlab{parallel_caseC_D_linear_gpu}}
    \subfigure[Computation vs communication kernels for case C (left) and case D (right)]{\includegraphics[width=\textwidth]{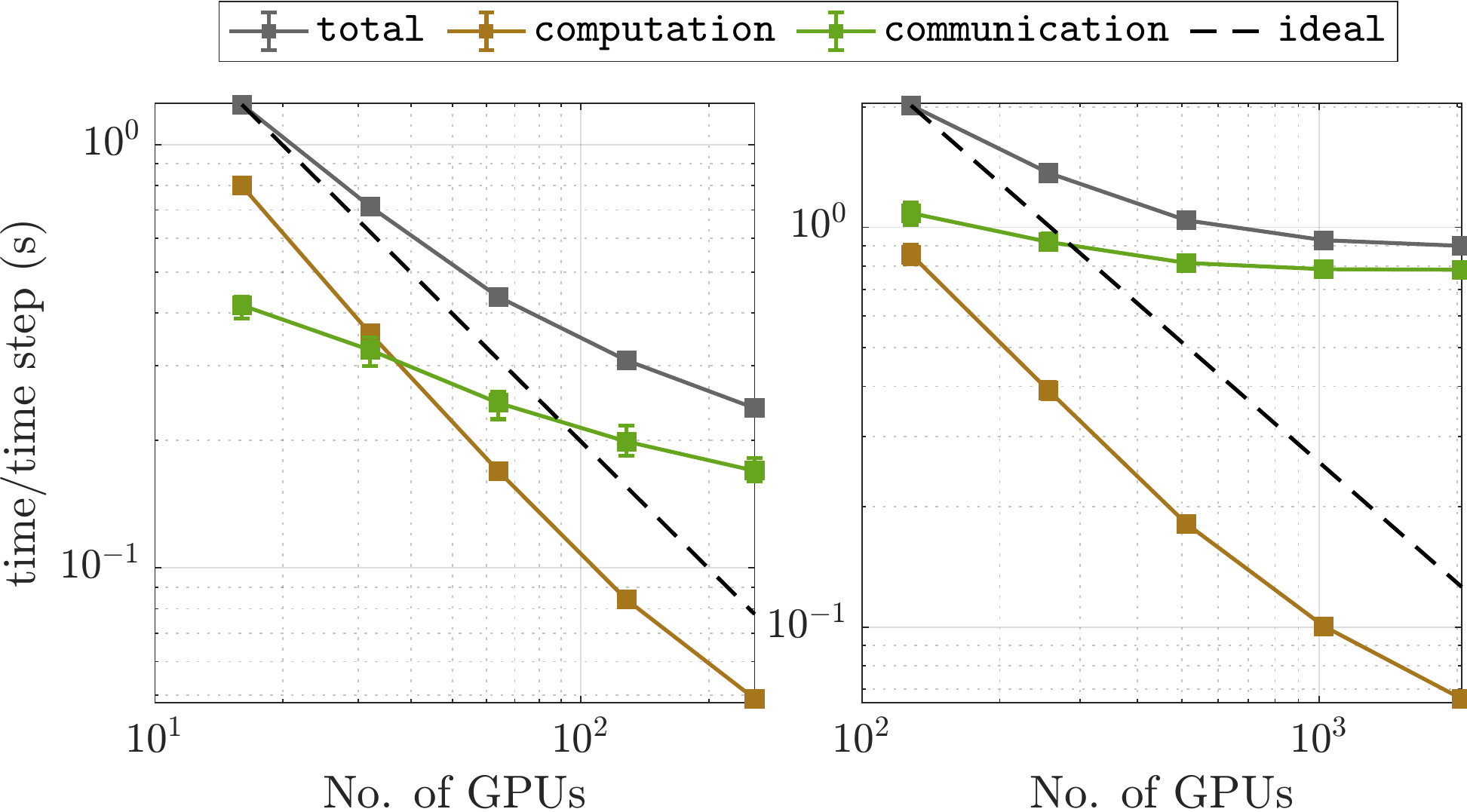}
    \figlab{parallel_comm_caseC_D_linear_gpu}}
    \caption{Weak Landau damping. Strong scaling study on GPUs with higher number of particles per cell ($P_c=64$).}
    \figlab{linear_gpu_case_CD}
\end{figure}

\begin{figure}[]
    \subfigure[Computation kernels for case C (left) and case D (right)]{\includegraphics[width=\textwidth]{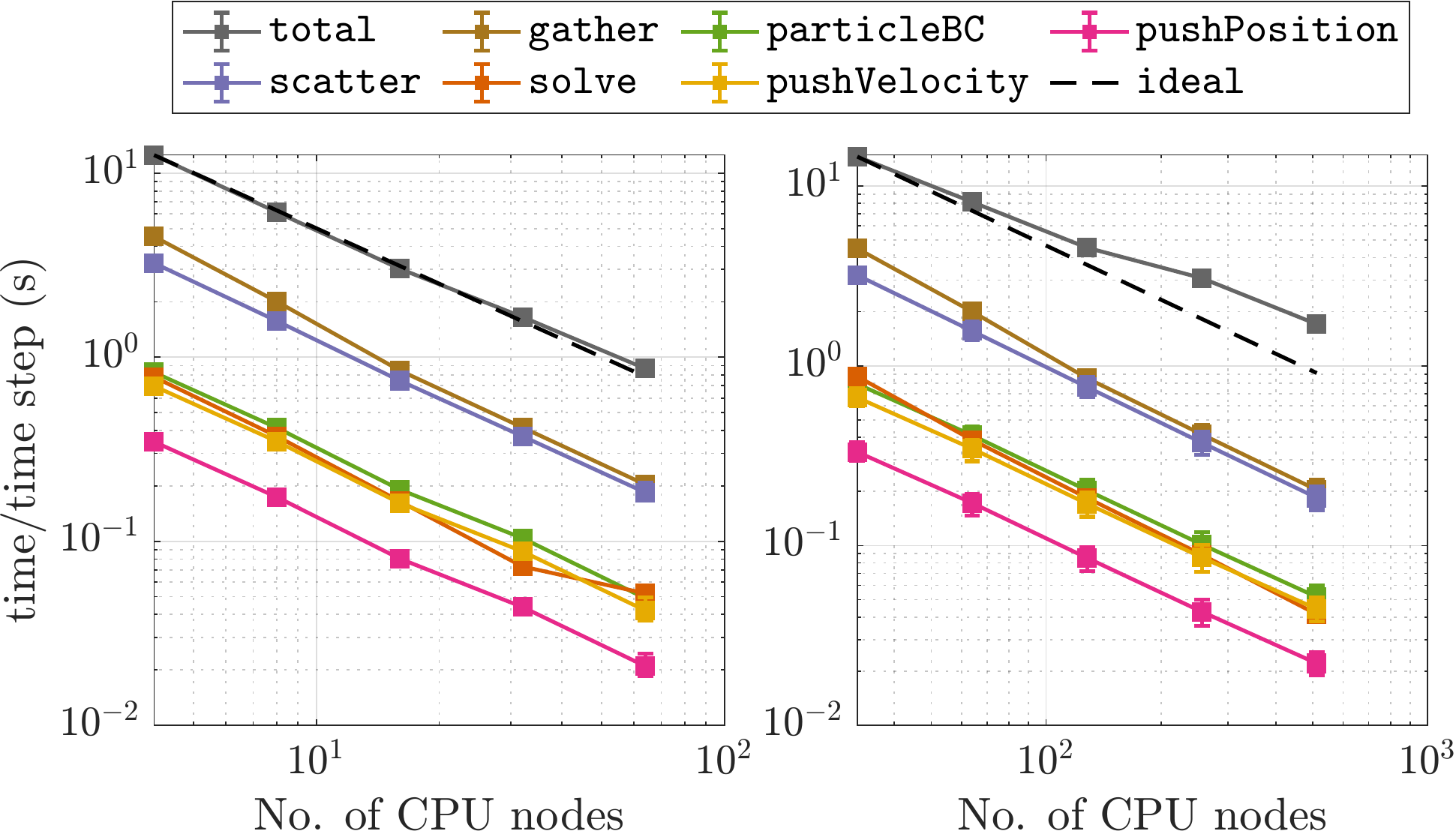}
    \figlab{parallel_caseC_D_linear_cpu}}
    \subfigure[Computation vs communication kernels for case C (left) and case D (right)]{\includegraphics[width=\textwidth]{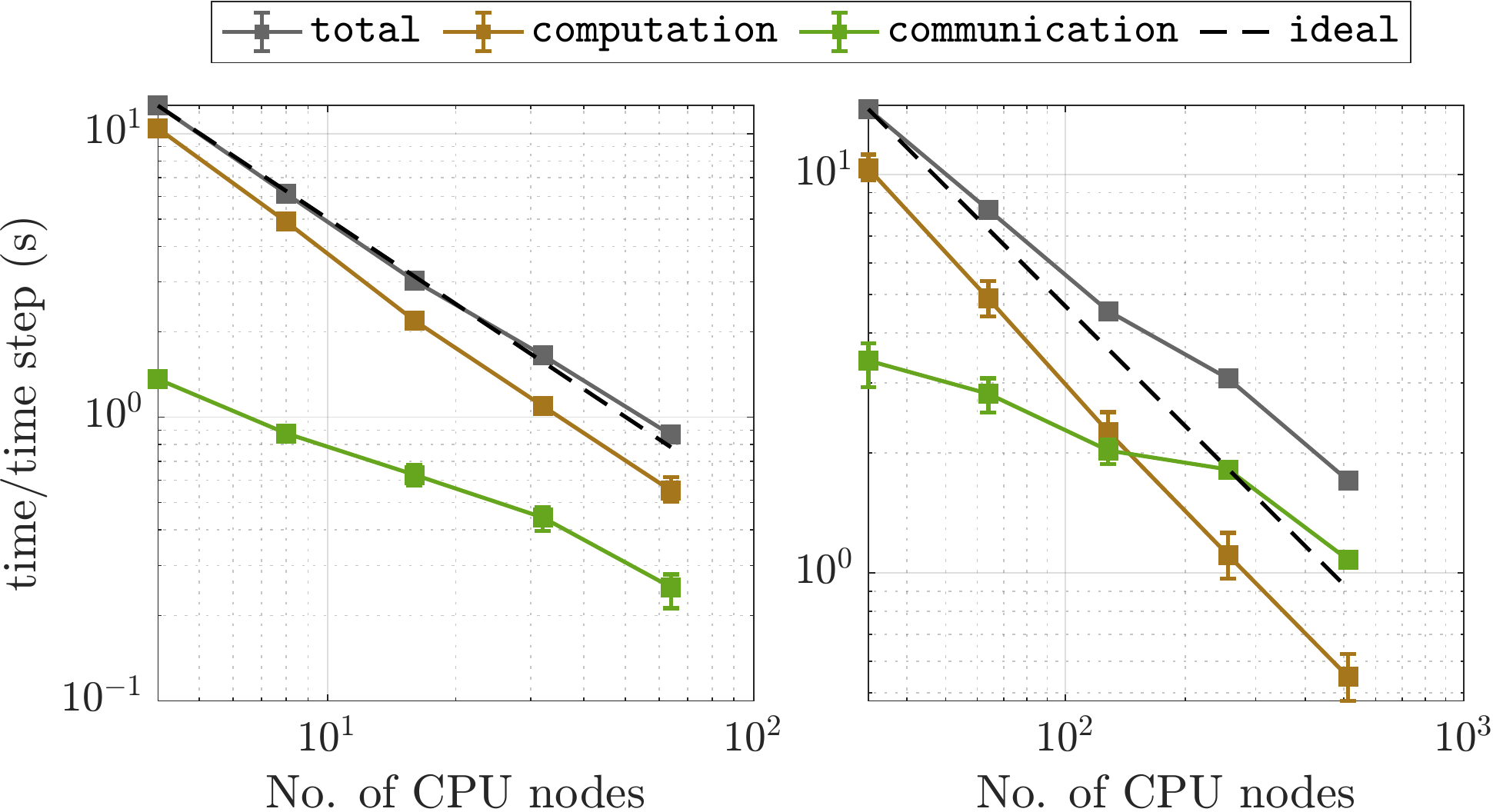}
    \figlab{parallel_comm_caseC_D_linear_cpu}}
    \caption{Weak Landau damping. Strong scaling study on CPUs with higher number of particles per cell ($P_c=64$).}
    \figlab{linear_cpu_case_CD}
\end{figure}

\subsubsection{Strong scaling: 64 particles per cell}
    So far, we have considered the strong scaling study on GPUs and CPUs with 8 particles per cell for two different mesh sizes. However, 
    many of the plasma physics codes typically use a much higher number of particles per cell to reduce the statistical noise in  the simulations.
    Hence, to obtain representative results for this scenario we consider the strong scaling study in the cases C and D with 64 particles per cell and
    reduced grid resolutions of $256^3$ and $512^3$, respectively. This is because we have an upper bound on the total number of
    particles that we can simulate with the considered number of GPUs and CPU nodes due to memory requirements. Hence, to have a setup with a higher 
    number of particles per cell we keep the same total number of particles for cases C and D as in cases A and B but with reduced grid 
    resolutions.

    From Figure \figref{parallel_caseC_D_linear_gpu} we observe that the field solve is still the 
    dominant computation kernel on GPUs, although the runtime difference between the {\tt solve} and the other particle-based kernels is reduced compared
    to cases A and B due to the reduced number of grid points. We also notice that the reduced grid sizes adversely affect the scaling of the field solver,
    due to the smaller amount of local work compared to the communication costs in the FFTs. On the other hand, for the CPU simulations we observe from Figure \figref{parallel_caseC_D_linear_cpu} that the field solve 
    is in fact one of the least dominant computation kernels. It is at a level comparable to the particle pushes. The scaling is also better 
    than for the GPUs since fewer MPI ranks are involved. This is possible because CPU nodes have more memory compared to GPUs. For the GPU simulations, we indeed utilize only the device memory without transferring field or particles to the host, because this operation is very costly. This forces us to involve more GPUs.

    We do not show the figures for communication costs here, because the particle communication costs are very similar to cases A and B, given that the total number of particles is the same in both sets of cases. The field communication costs are smaller than for cases A and B due to the reduced grid sizes and lower number of halo cells that need to be communicated. 

    Finally, from Figure \figref{parallel_comm_caseC_D_linear_gpu} we see that the cross over point between
    computation and communication happens much sooner on GPUs compared to cases A and B, especially for case D which is completely communication dominated. This is because the field solver
    now takes less time due to the smaller numbers of grid points. As a result, the efficiency is significantly reduced compared to the other cases, especially for case D. For CPUs this effect is somewhat less pronounced, with case C looking very similar to case A. For case D the cross 
    over point happens a bit earlier than for case B as seen from Figure \figref{parallel_comm_caseC_D_linear_cpu}. This can be understood as follows. The field solver is not the dominant kernel anymore, but the gather operation, which takes only slightly less time than the field solver in Figure \figref{parallel_case_linear_cpu}, now dominates the computation kernels and has similar runtimes for both sets of cases.

    We can see from Figure \figref{linear_gpu_case_CD} that cases C and D show poor scaling. This is due to the following reason. Because of the small memory capacity of Piz Daint GPUs (16 GB/GPU), we have to use a lot of GPUs to have enough memory for the particles. However, we do not have enough
    work to utilize the computing power of these GPUs, due to the reduced field sizes. Hence, it does not make much sense
    to run the higher particles per cell cases like D on large numbers of GPUs (like 1024 and 2048) as we get very little efficiency out of  
    it. That is why for the other examples in the remainder of this article, we focus only on cases A and B, as we observed very 
    similar results for cases C and D in the other mini-apps. In future work, we will reduce the communication costs corresponding to the particle search, by first searching over the field neighbors, and then only searching for the leftover particles among the remaining ranks. Since in explicit PIC the time step is selected such that most of 
    the particles do not travel more than one mesh cell during a time step, this can help to reduce the particle communication cost.

\begin{figure}[]
    \begin{center}
    \subfigure[Computation kernels on GPUs (left) and CPUs (right)]{\includegraphics[width=0.8\textwidth]{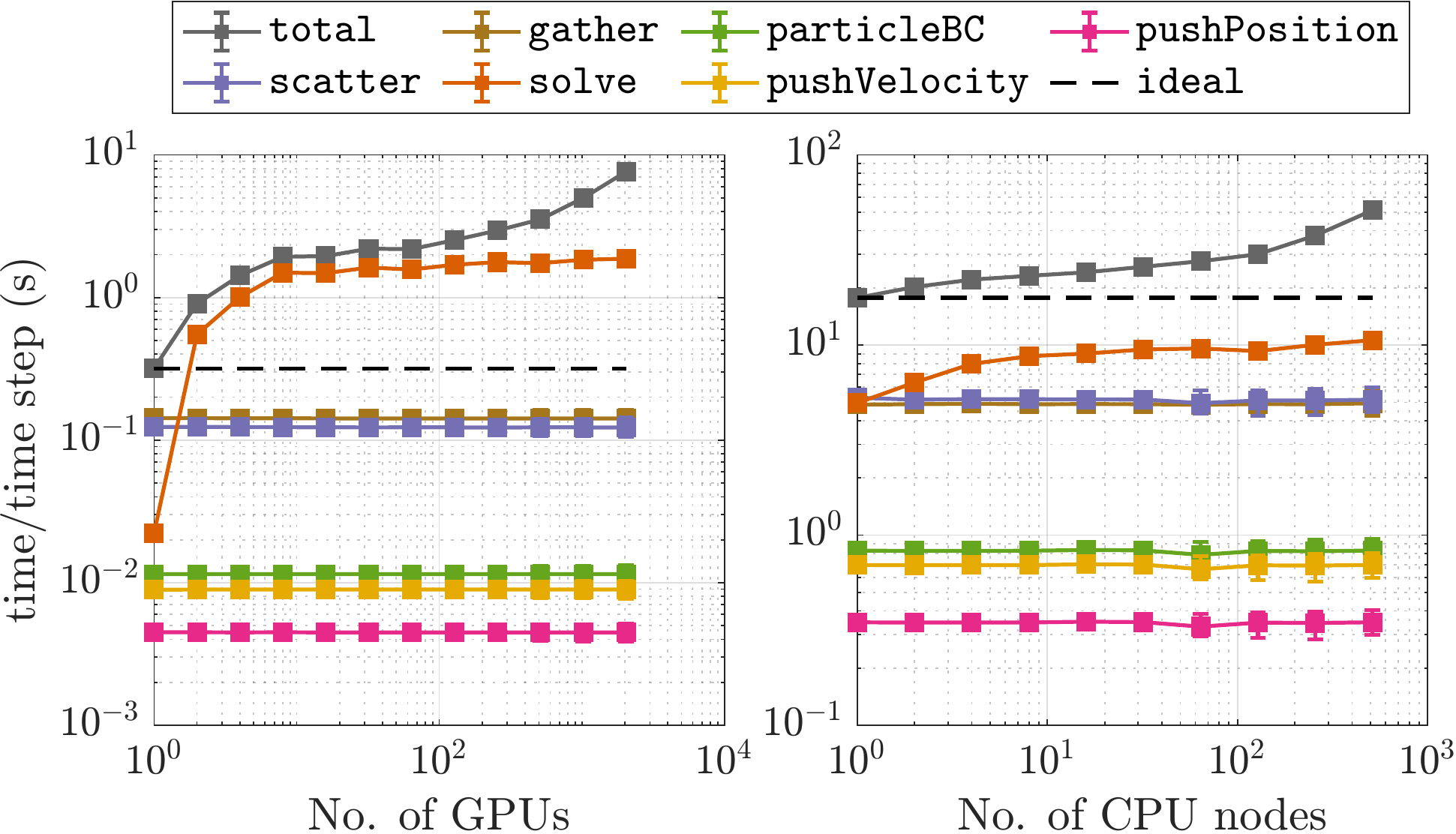}
    \figlab{parallel_weak_linear_cpu_gpu}}
    \subfigure[Communication kernels on GPUs (left) and CPUs (right)]{\includegraphics[width=0.8\textwidth]{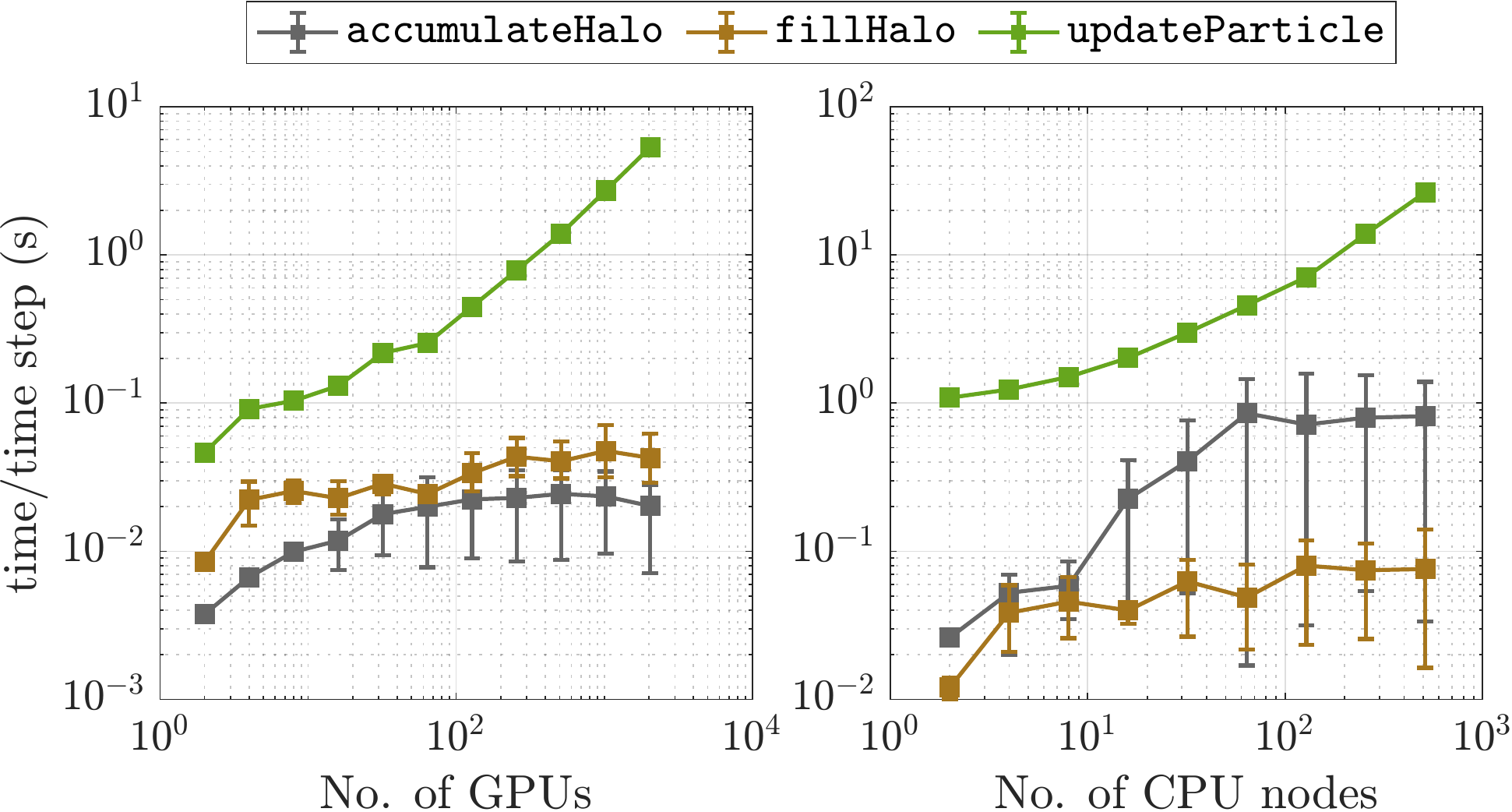}
    \figlab{comm_weak_linear_cpu_gpu}}
    \subfigure[Computation vs communication kernels on GPUs (left) and CPUs (right)]{\includegraphics[width=0.8\textwidth]{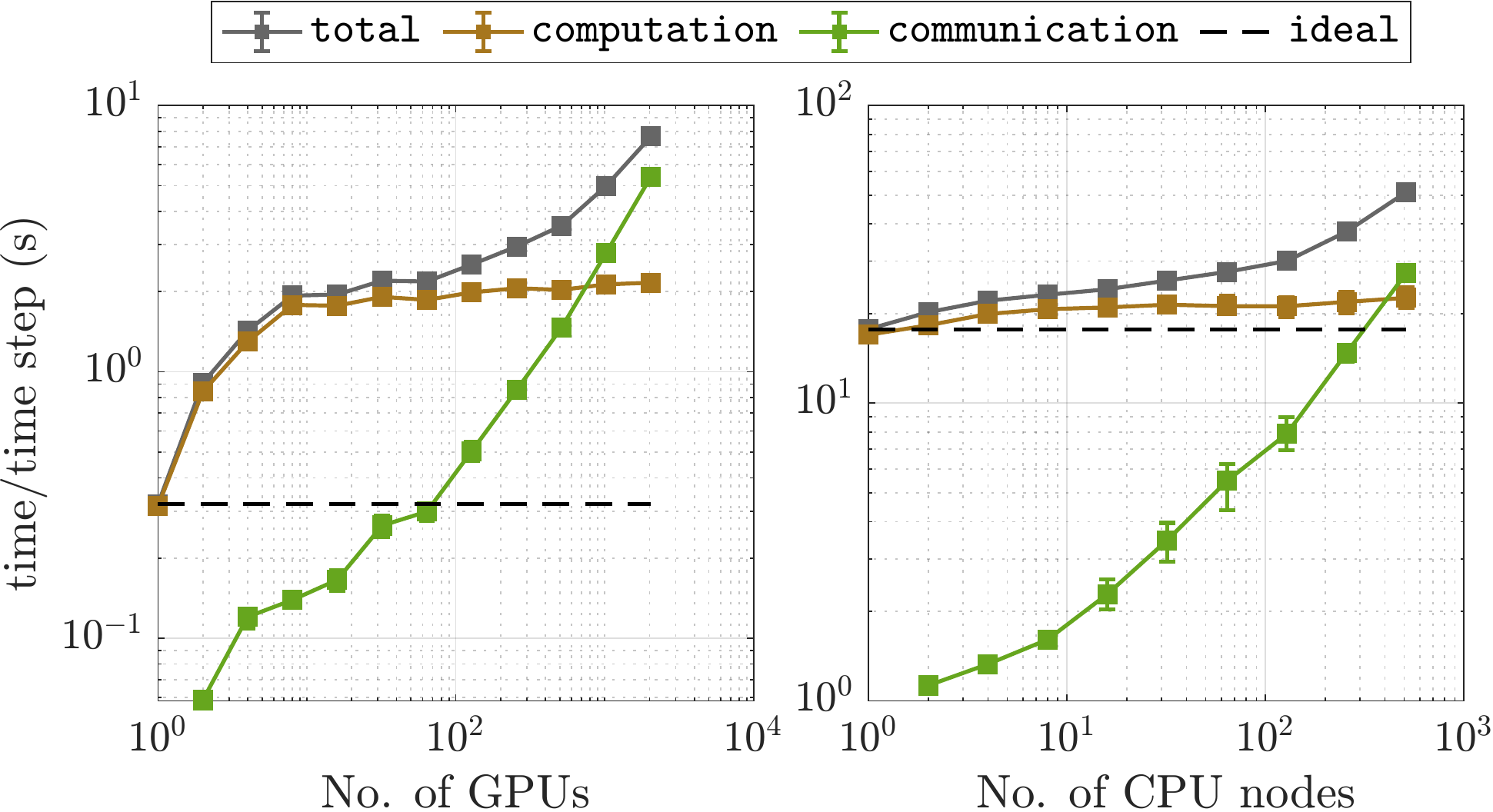}
    \figlab{parallel_comm_weak_linear_cpu_gpu}}
    \end{center}
    \vspace{-5mm}
    \caption{Weak Landau damping. Weak scaling of computation and communication kernels and cross over point between computation and communication on GPUs and CPUs.}
    \figlab{linear_weak_cpu_gpu}
\end{figure}

\begin{figure}[]
    \begin{center}
    \includegraphics[width=0.8\textwidth]{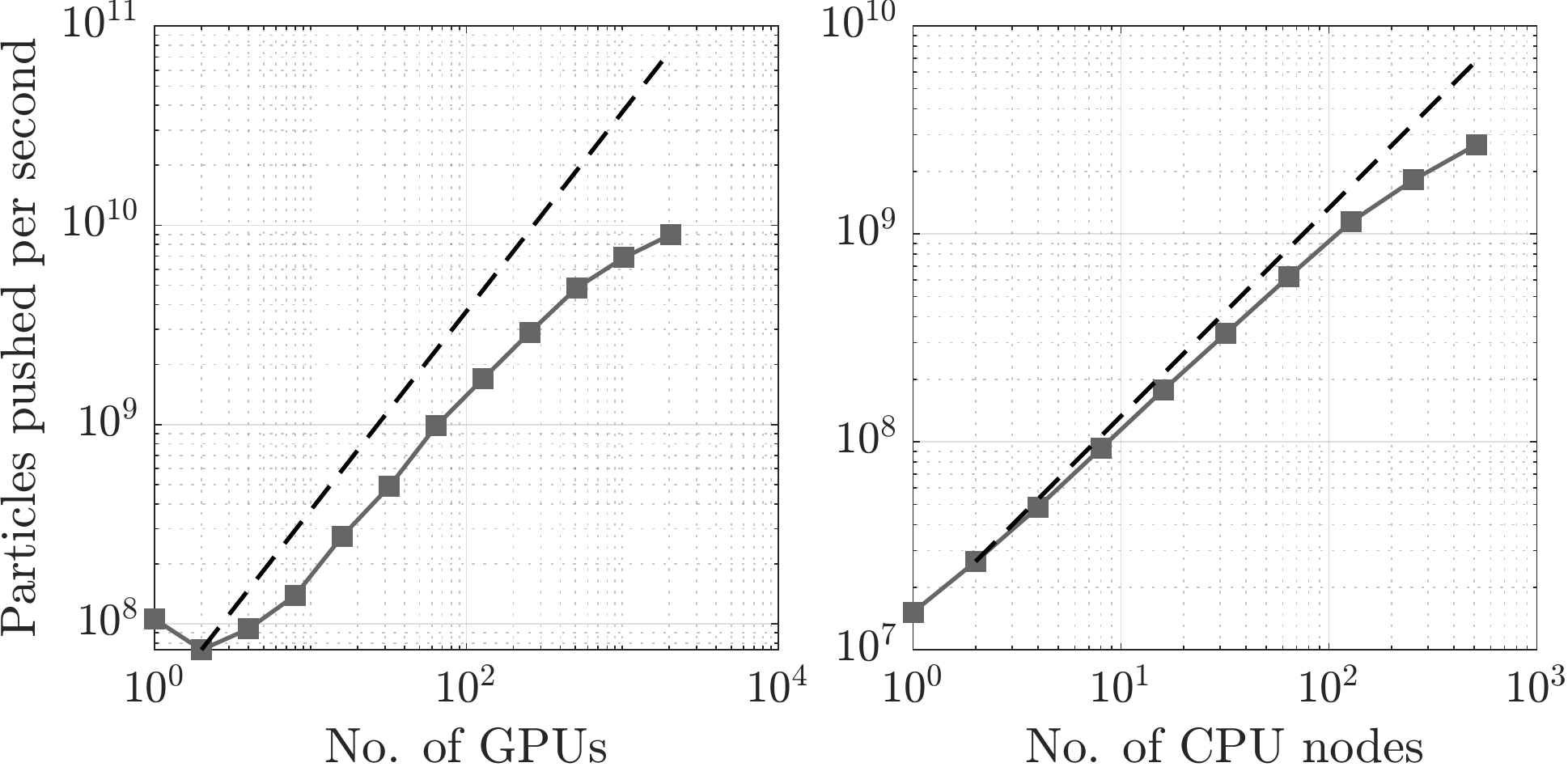}
    \end{center}
    \vspace{-5mm}
    \caption{Weak Landau damping. Weak scaling: Number of particles pushed per second on GPUs (left) and CPUs (right).}
    \figlab{linear_particles_pushed}
\end{figure}

\subsubsection{Weak scaling}
    Having studied the strong scaling so far, we will now turn our attention to the weak scaling. We first describe the weak scaling setup for the weak Landau damping test problem before discussing the results. For 
    GPU simulations, we take the base case for 1 GPU as a $256\times128^2$ grid whereas 
    for CPUs we take a $512\times256^2$
    grid, because of the greater memory availability. For both CPUs and GPUs our simulations have 8 particles 
    per cell. The maximum grid size and number of particles for the GPU simulations 
    are $N_c=2048^3$ and $N_p=68,719,476,736$ on 2048 GPUs, 
    whereas for CPU simulations they are $N_c=4096\times2048^2$ 
    and $N_p=137,438,953,472$ on 512 nodes = 16,384 cores.

    In Figure \figref{linear_weak_cpu_gpu}, all the computation 
    kernels scale ideally, except for the field solve and total simulation, as we had already seen for the strong scaling study. The scaling of the field solver is also close to ideal starting from 8 GPUs or CPU nodes. 
    This can be explained as follows. Starting from 8 MPI ranks, each processor has a brick of field data and the number of transposes performed during the FFT remains the same. 
    On the other hand, for 1, 2 and 4 ranks, we have either no communication or fewer transposes due to slab or pencil decompositions. Hence 
    they take less time than the 8 ranks case. 
    The total time is mostly dictated by the 
    field solve, except for the last few data points, where communication kernels dominate. We also see a
    linear increase asymptotically in the particle communication cost in Figure \figref{comm_weak_linear_cpu_gpu}, due to the particle search and communication as explained 
    in the strong scaling study. In Figure \figref{linear_particles_pushed}, we show the number of particles pushed per second on GPUs and CPUs as measured in this weak scaling
    study. This metric is obtained by dividing the total number of particles by the wall time per simulation time step. We get a maximum of approximately $10^{10}$ particles per second on GPUs and $3\times10^9$ particles per second on
    CPUs.

\subsection{Strong Landau damping}
\begin{figure}[]
    \subfigure[Initial density]{\includegraphics[trim=12.0cm 2cm 12.0cm 4cm,clip=true,width=0.31\textwidth,height=0.28\textwidth]{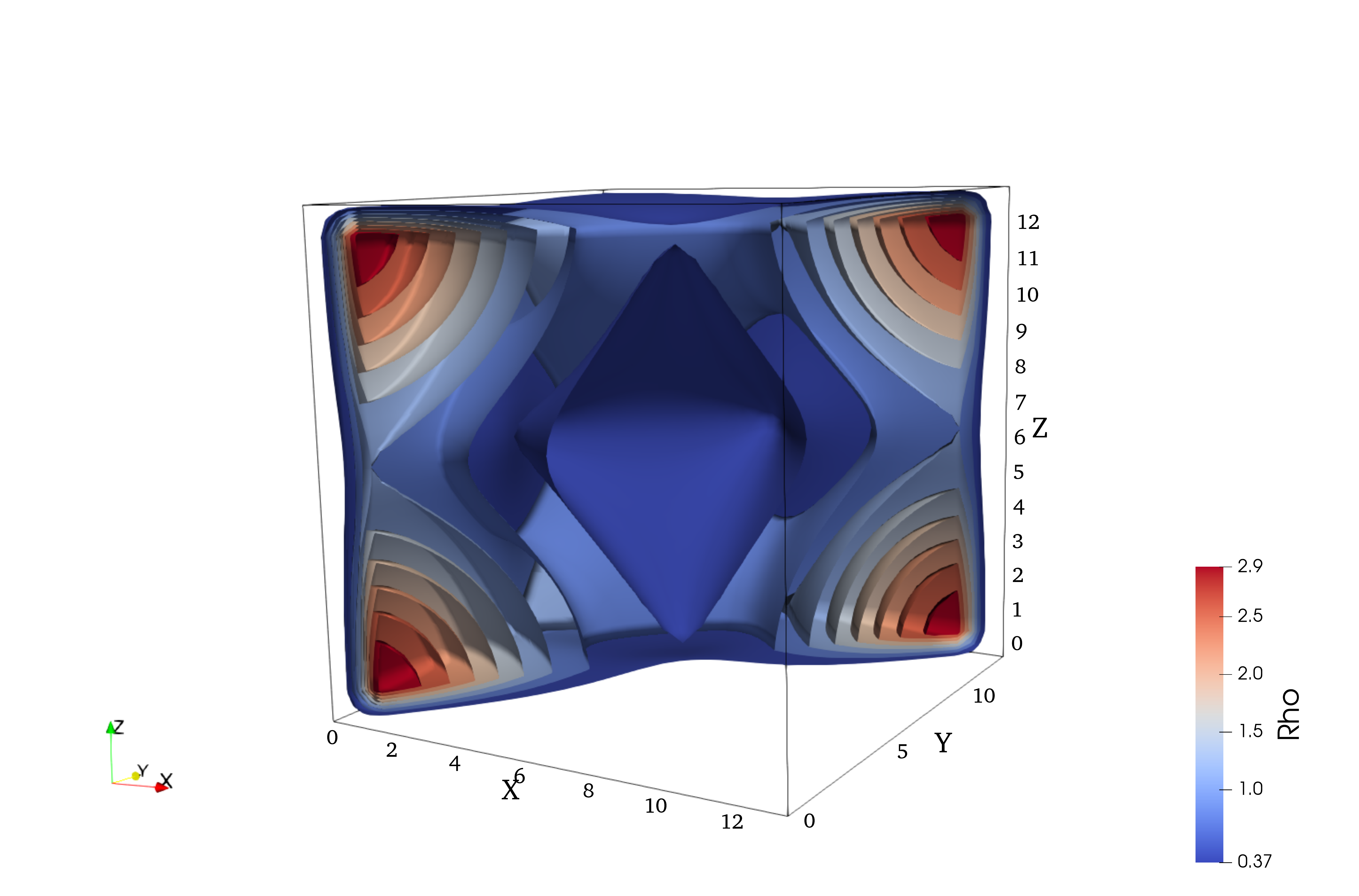}
    \figlab{initial_density_nonlinear}}
    \subfigure[No. of GPUs=128]{\includegraphics[trim=3.5cm 1cm 2cm 2cm,clip=true,width=0.31\textwidth]{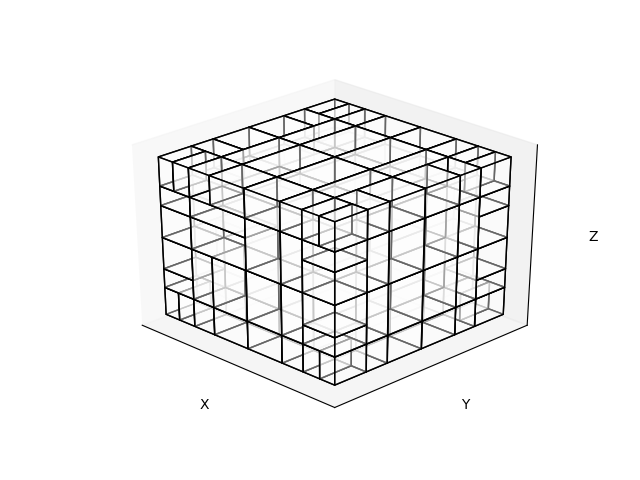}
    \figlab{decomp_128_nonlinear}}
    \subfigure[No. of GPUs=2048]{\includegraphics[trim=3.5cm 1cm 2cm 2cm,clip=true,width=0.31\textwidth]{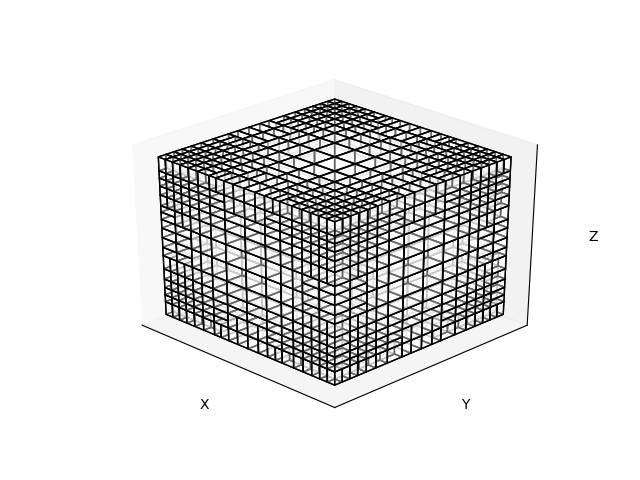}
    \figlab{decomp_2048_nonlinear}}
    \caption{Strong Landau damping: Cross sectional view of the initial particle density profile showing high (red) and low (blue) densities, and particle load balanced domain decompositions for $128$ and $2048$ GPUs.}
    \figlab{domain_decomp_nonlinear}
\end{figure}

\begin{figure}[]
    \subfigure[Case B: Computation vs communication kernels with (left) and without (right) load balancing]{\includegraphics[width=\textwidth]{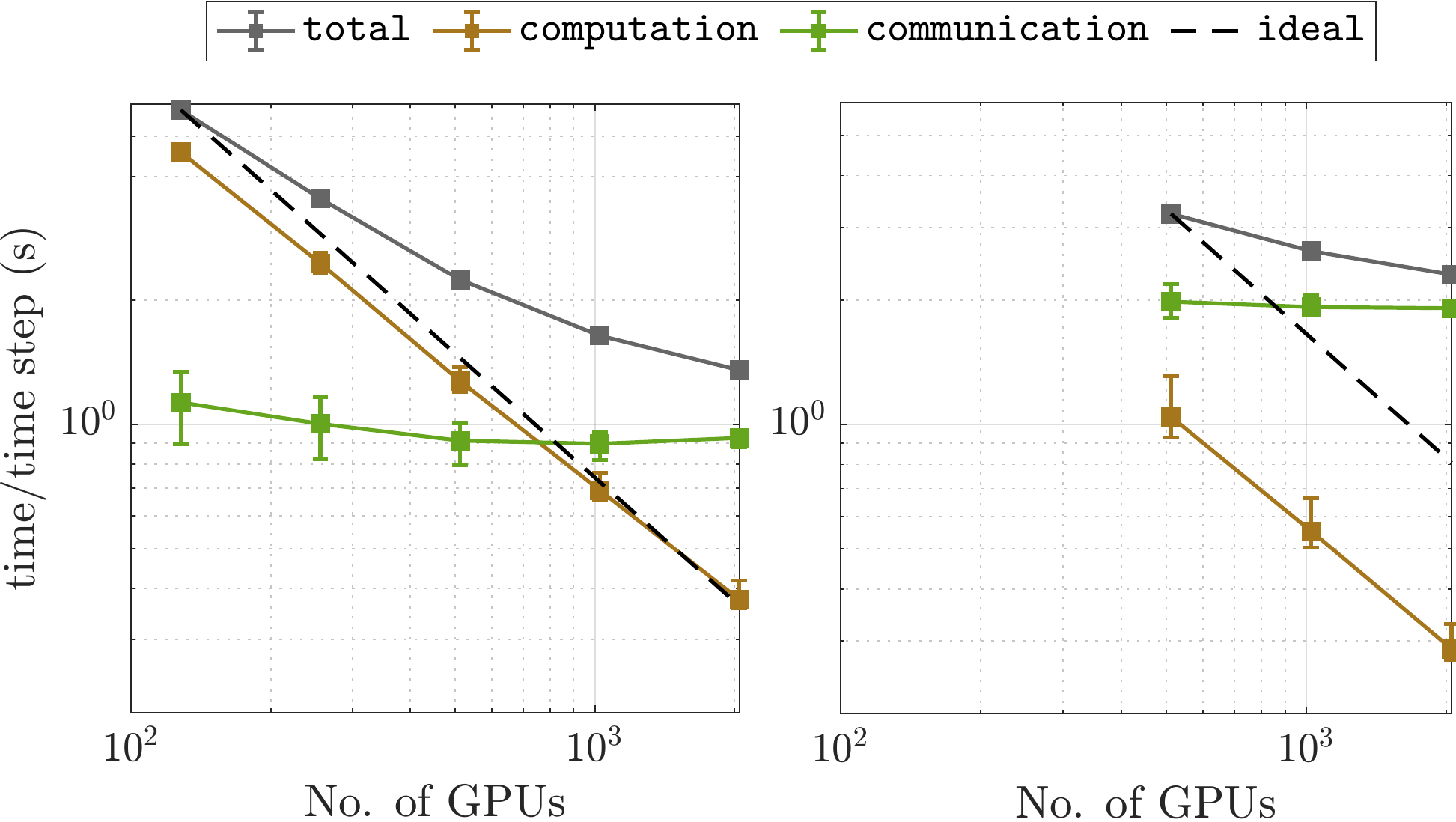}
    \figlab{parallel_comm_nonlin_gpus}}
    \subfigure[Case B: Communication kernels with (left) and without (right) load balancing]{\includegraphics[width=\textwidth]{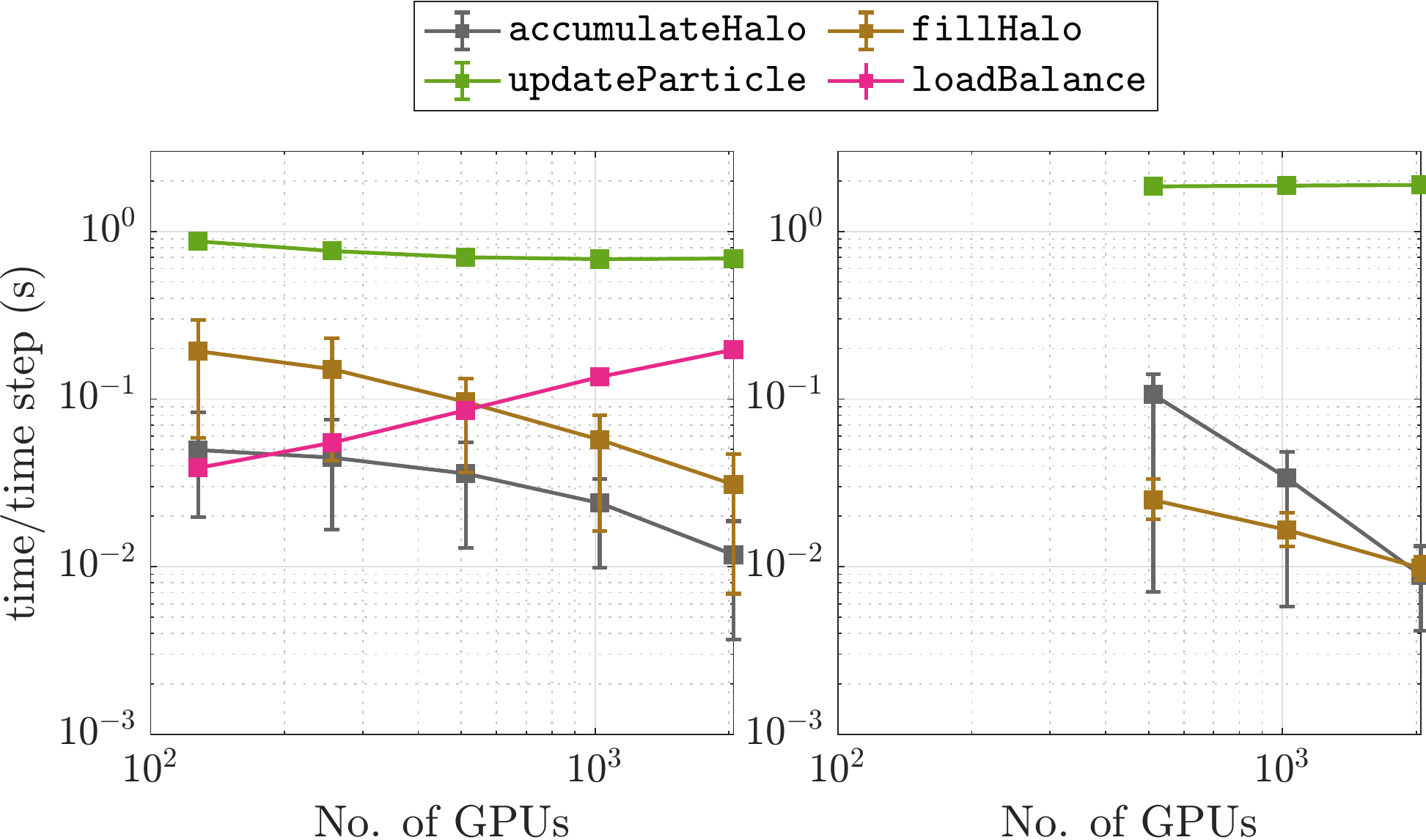}
    \figlab{comm_nonlin_gpus}}
    \caption{Strong Landau damping. Strong scaling on GPUs: comparison of scaling and performance with and without particle load balancing for case B. The right column, corresponding to the case without load balancing, is missing the first two data points because these simulations fail due to high memory imbalance leading to lack of memory.}
    \figlab{nonlinear_gpus}
\end{figure}

\begin{figure}[]
    \subfigure[Case B: Computation vs communication kernels with (left) and without (right) load balancing]{\includegraphics[width=\textwidth]{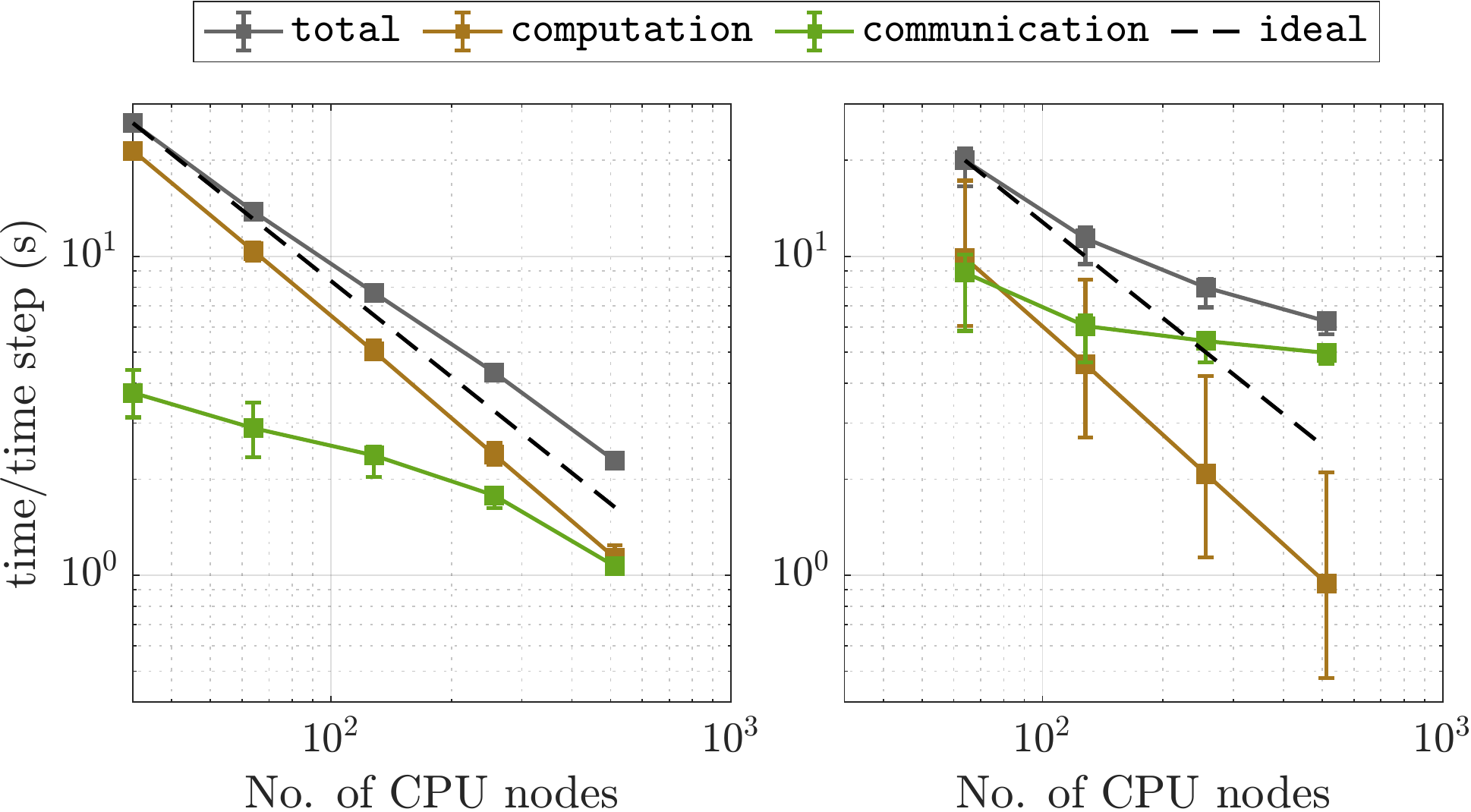}
    \figlab{parallel_comm_nonlin_cpus}}
    \subfigure[Case B: Communication kernels with (left) and without (right) load balancing]{\includegraphics[width=\textwidth]{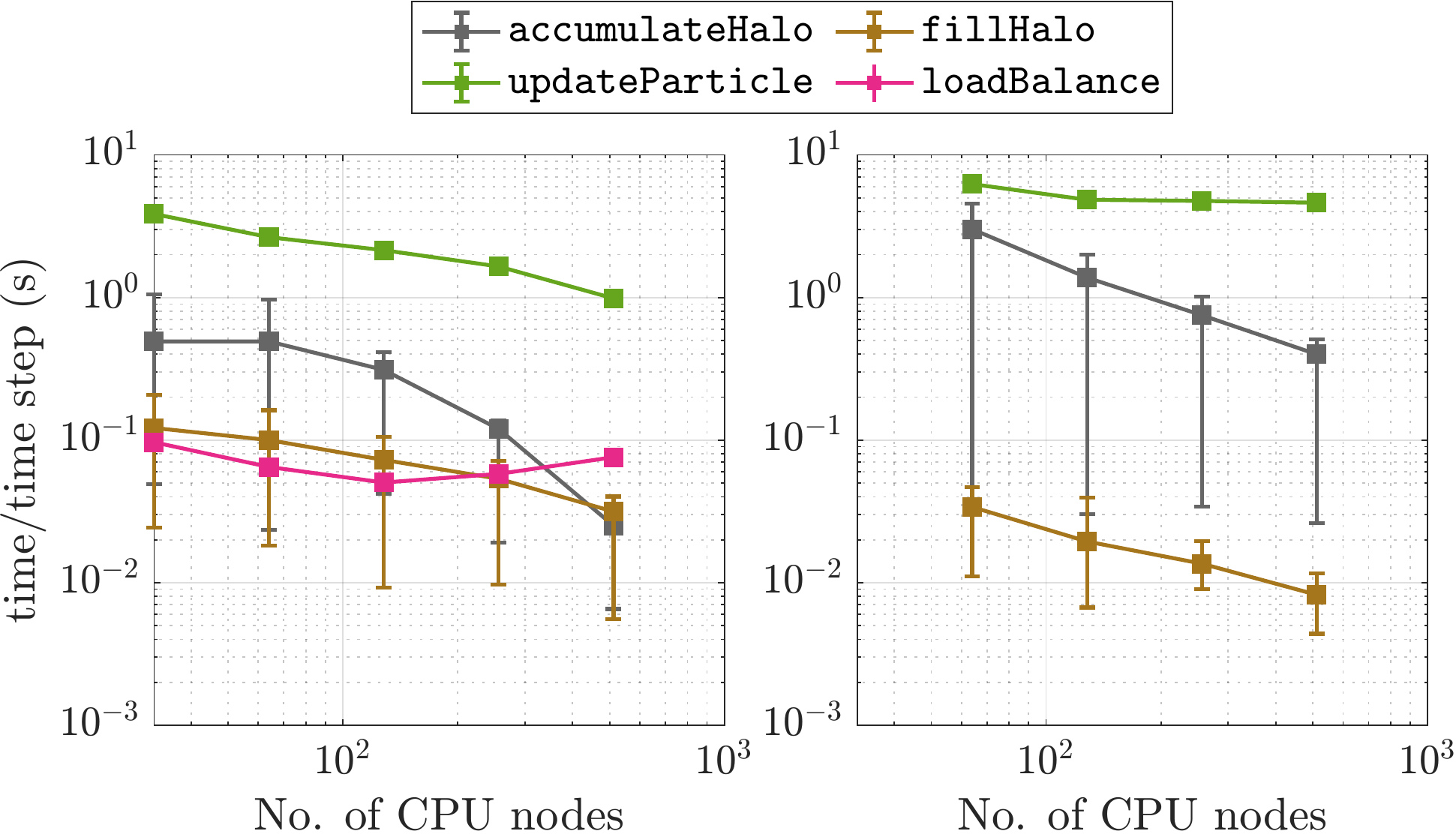}
    \figlab{comm_nonlin_cpus}}
    \caption{Strong Landau damping. Strong scaling on CPUs: comparison of scaling and performance with and without particle load balancing for case B. The right column, corresponding to the case without load balancing is missing the first data point because this simulation fails due to high memory imbalance leading to lack of memory.}
    \figlab{nonlinear_cpus}
\end{figure}

    In this section, we consider the case of strong Landau damping, which corresponds to a larger perturbation parameter $\alpha$ in equation \eqnref{landau_f} as compared to the weak damping case. 
This leads to a higher particle load imbalance among the MPI ranks with equal distribution of field domains. This in turn leads to an
increased total time for the simulation compared to the load balanced case, as the MPI rank or GPU which takes the maximum time for the computation and communication kernels determines the overall simulation time. An even more important problem is the memory requirement, as the GPU or CPU node which has the greatest number of particles may not have enough memory to store them all, while the memory in the other GPUs or CPU nodes is under-utilized. Hence, particle load balancing is of critical importance in this 
example. We create the particles in a balanced way by means of orthogonal recursive bisection using the initial analytical density profile 
as shown in Figure \figref{initial_density_nonlinear}. 
Figures \figref{decomp_128_nonlinear} and \figref{decomp_2048_nonlinear} show the representative initial particle load balanced domain decompositions we obtain for 
$128$ and $2048$ GPUs. After the initial load balancing, we perform the balancing if the imbalance percentage given by 
\[
    \text{imbalance\%}= \LRp{\frac{|n_{loc} - n_{ideal}|}{N_p}} \times 100
\]
in any rank exceeds a given threshold. Here, $n_{loc}$ is the local number of particles in each rank or GPU and $n_{ideal}=N_p/N_{ranks}$ is the ideal 
local number of particles. We set the threshold to 1\% for this problem, as well as the Penning trap example in the next section, based on numerical experiments.

In Figure \figref{nonlinear_gpus}, we compare the results for case B with and without load balancing on GPUs. We do not show results for 
case A, as the scaling and performance of cases with and without load balancing are similar, and 
they are also comparable to the weak Landau damping results shown in Figures \figref{linear_gpu1} and \figref{linear_gpu2}, due to the smaller total number of particles. We can see from Figure \figref{comm_nonlin_gpus} that the particle communication cost is higher in the case without
load balancing. This leads to an earlier cross over point between computation and communication, and therefore loss of scaling as seen from Figure \figref{parallel_comm_nonlin_gpus}. This is because without load balancing some ranks 
contain a lot more particles than the others, and also have to communicate more. The particle communication has a synchronization step, and since it has to wait for the last arriving rank, this leads to an increase in the {\tt updateParticle} cost. Moreover, due to particle imbalance, the ranks which contain a lot of particles take more time in the computation kernels than the others, and this additional time also gets reflected in the synchronization steps, which in our case correspond to the
{\tt updateParticle}. We also notice from the right column of Figure \figref{nonlinear_gpus} that the case without load balancing is missing data points for $128$ and $256$ GPUs as these runs fail due to lack of memory. As explained before, this is due to the need for higher memory in some GPUs,
which have to store a lot of particles. We draw similar conclusions from the 
CPU simulations for case B in Figure \figref{nonlinear_cpus}. 

Thus we can see from this example that our particle load balancing strategy is effective in cases with non-uniform particle distributions,
without which the simulations either fail due to lack of memory, or have poor scaling. We should however note that the load balancing 
strategy comes with a price which we describe now. First, the additional overhead has to be small relative to the costs of the other kernels for it to be beneficial. This is the case in our tests shown in Figures \figref{nonlinear_gpus} and \figref{nonlinear_cpus} as, apart from the initial load balancing, the routine is never invoked 
for the set imbalance threshold of $1\%$. Second, our strategy of particle load balancing creates a load imbalance in the fields, as every
rank now owns a brick of varied size in contrast to the uniform field distribution in the case without particle load balancing. This is visible from Figures \figref{decomp_128_nonlinear} and \figref{decomp_2048_nonlinear}. Although this does not affect the field solver scaling much in the case of 
strong Landau damping, for problems with highly non-uniform particle distributions, such as our Penning trap simulations, it can have a significant impact, as will be explained in Section \secref{penning_trap}.

The weak scaling results for the strong Landau damping simulations with load balancing are very similar to those of the weak Landau damping simulations shown in Figure \figref{linear_weak_cpu_gpu}. We therefore do not show them, to avoid repetition.

\subsection{Electron dynamics in a charge neutral Penning trap}
\seclab{penning_trap}
\begin{figure}[]
     \subfigure[Initial density]{\includegraphics[trim=12cm 0.5cm 10cm 1cm,clip=true,width=0.31\textwidth,height=0.27\textwidth]{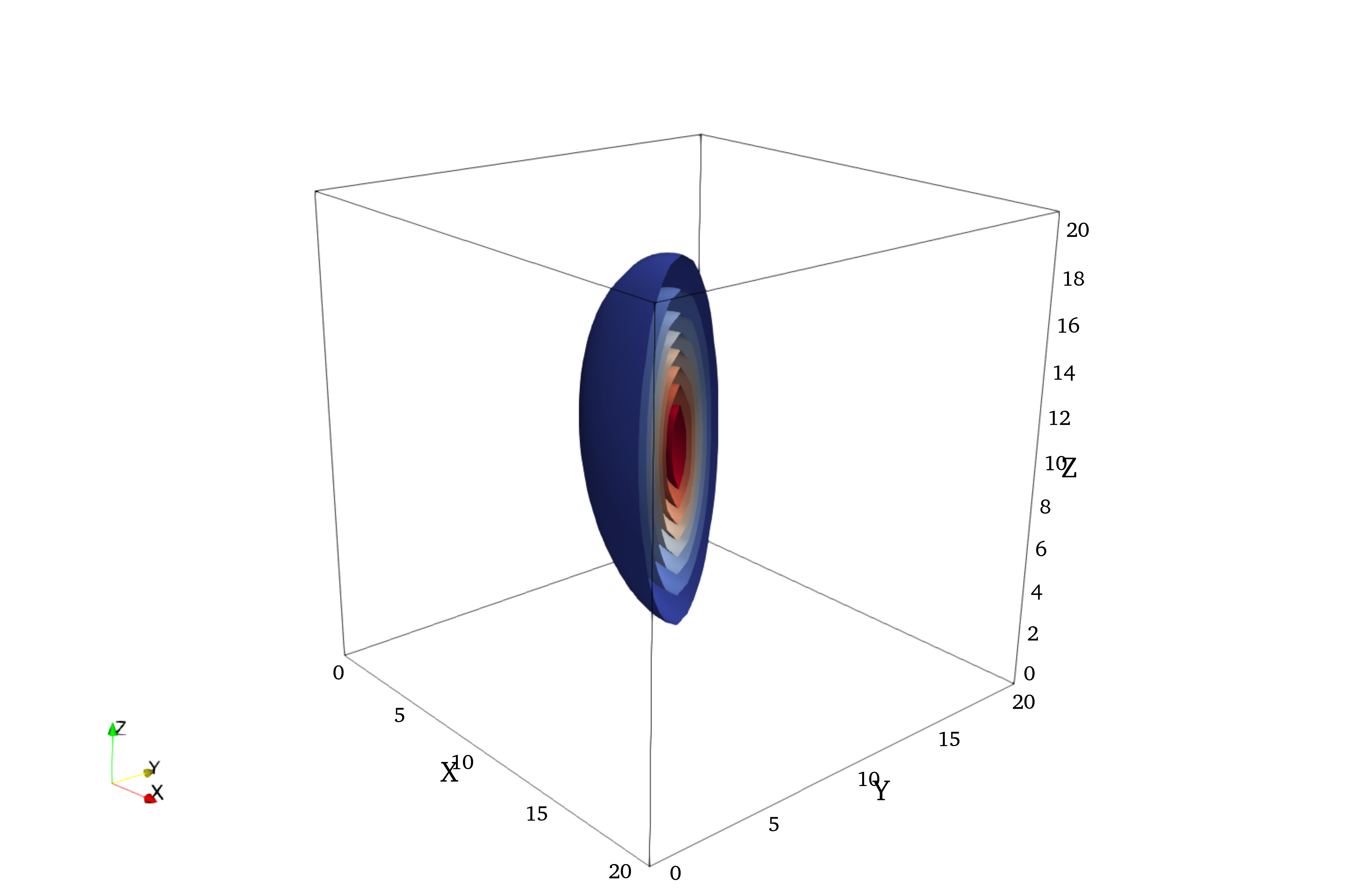}
    \figlab{initial_density_penning}}
    \subfigure[No. of GPUs=256]{\includegraphics[trim=3.5cm 1cm 2cm 2cm,clip=true,width=0.31\textwidth]{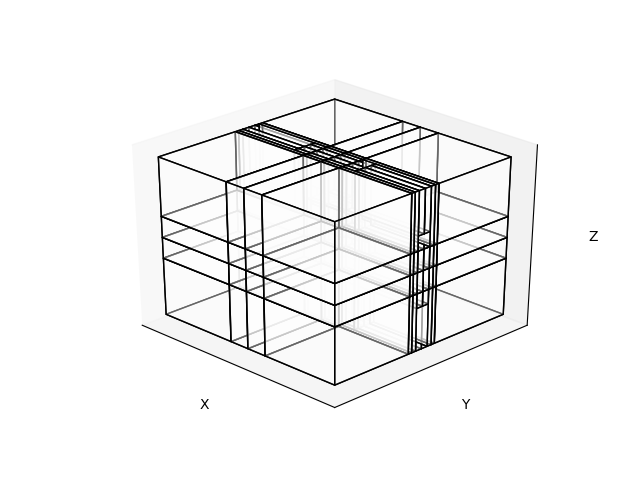}
    \figlab{decomp_256_penning}}
    \subfigure[No. of GPUs=2048]{\includegraphics[trim=3.5cm 1cm 2cm 2cm,clip=true,width=0.31\textwidth]{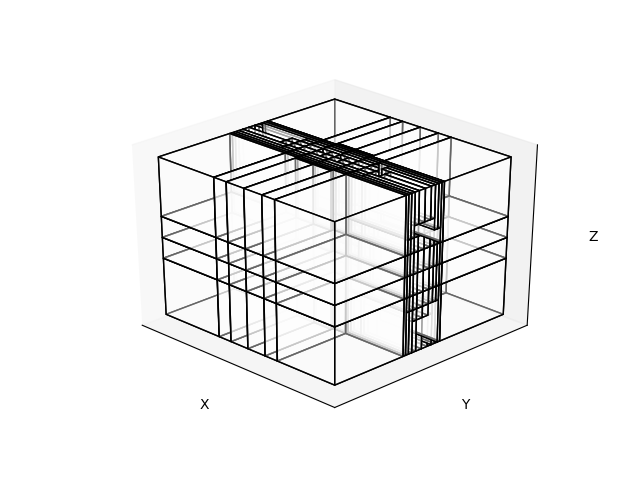}
    \figlab{decomp_2048_penning}}
    \caption{Penning trap: Cross sectional view of the initial particle density profile showing high (red) and low (blue) densities, and particle load balanced domain decompositions for $256$ and $2048$ GPUs.}
    \figlab{domain_decomp_penning}
\end{figure}

    In this section, we present the strong and weak scaling results for the Penning trap mini-app. In Figure \figref{initial_density_penning}, the initial electron density is
    shown, which is a Gaussian bunch. In contrast to the Landau damping examples, this highly non-uniform particle distribution leads to a highly non-uniform field distribution after particle load balancing, as shown in Figures 
    \figref{decomp_256_penning} and \figref{decomp_2048_penning}. In that respect the Penning trap mini-app is a harder test case in terms of 
    field and particle load balancing compared to the Landau damping examples. 

\begin{figure}[]
\vspace{-5mm}
    \begin{center}
    \subfigure[Computation kernels for case A (left) and case B (right)]{\includegraphics[width=0.8\textwidth]{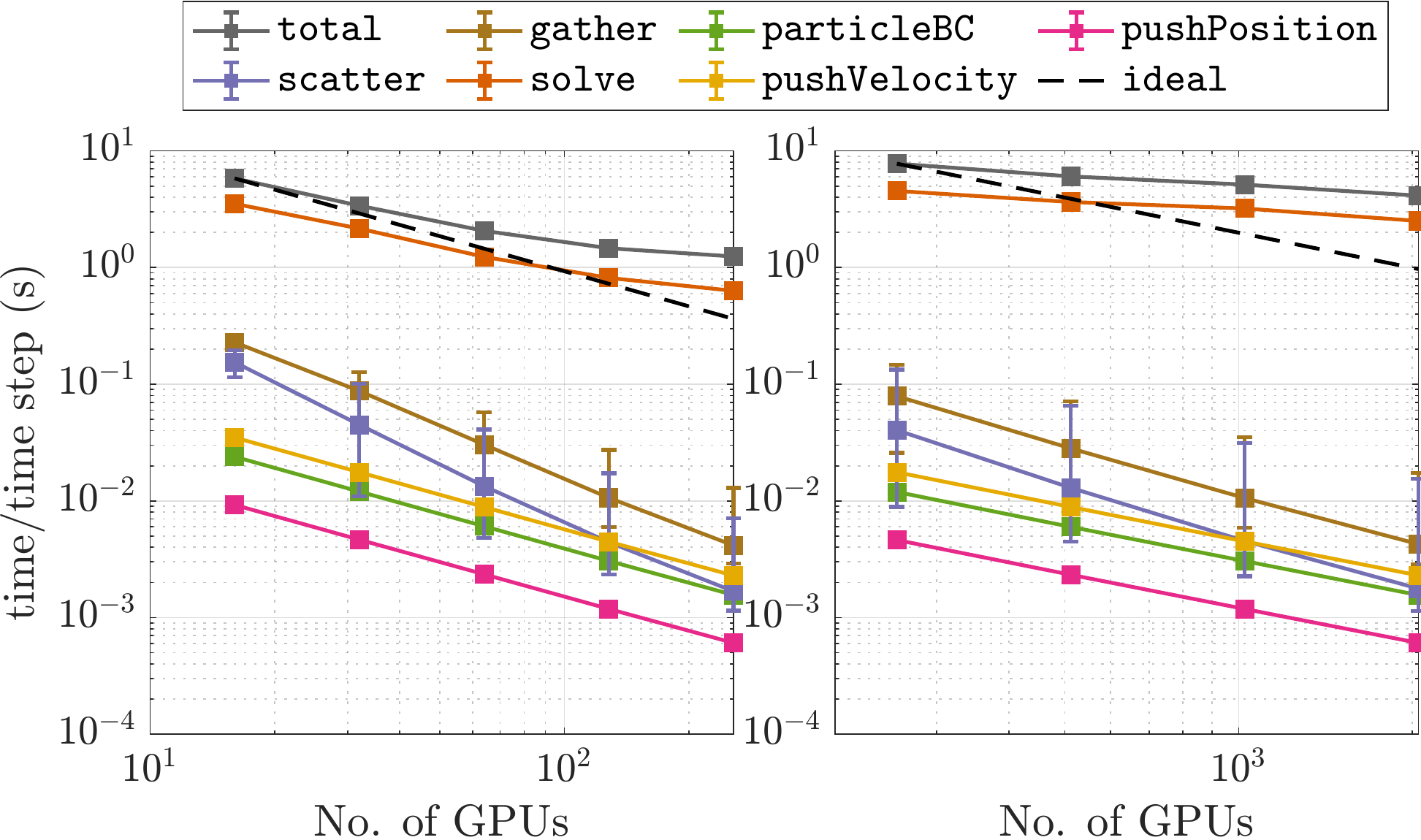}
    \figlab{parallel_penning_gpus}}
    \subfigure[Communication kernels for case A (left) and case B (right)]{\includegraphics[width=0.8\textwidth]{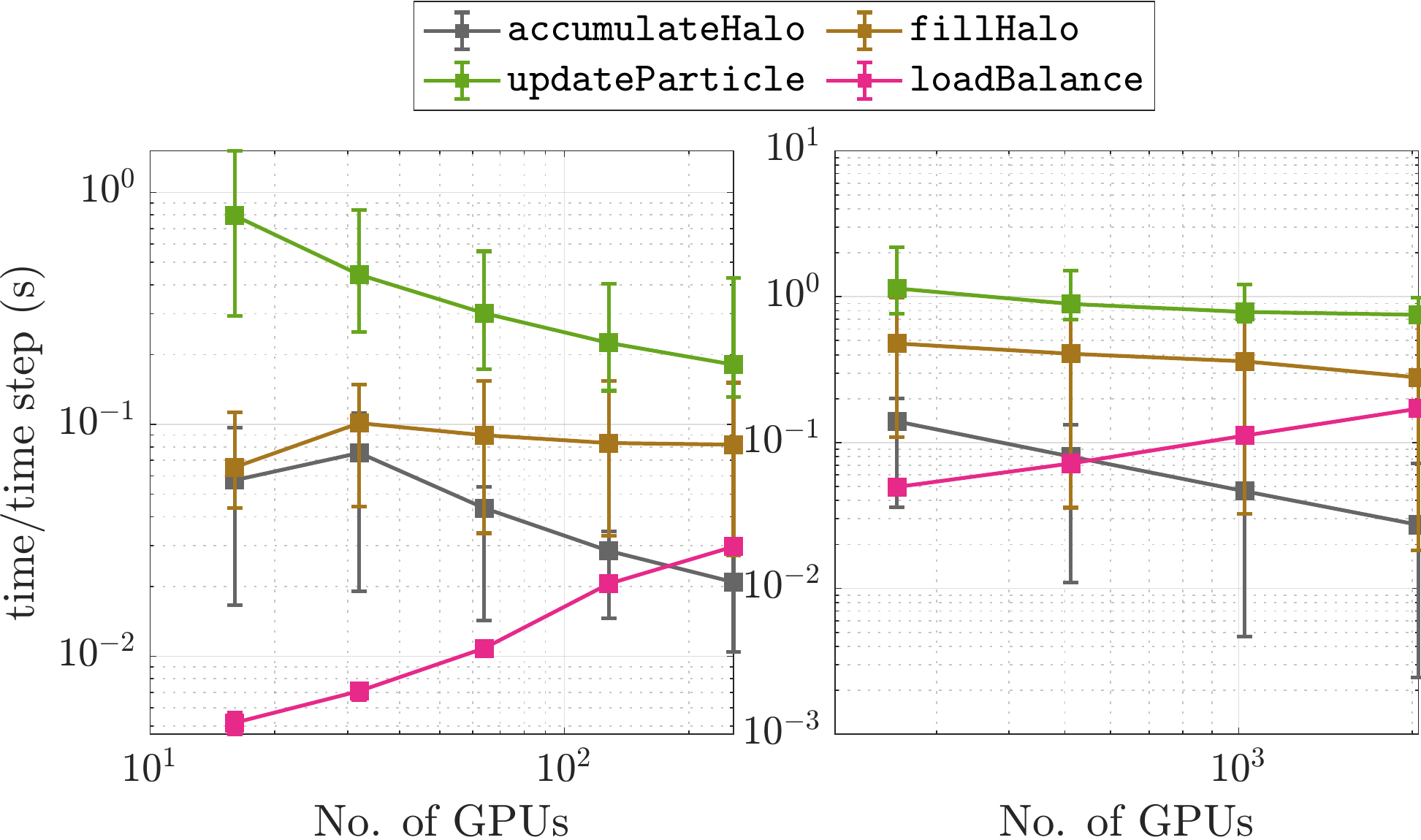}
    \figlab{comm_penning_gpus}}
    \subfigure[Computation vs communication kernels for case A (left) and case B (right)]{\includegraphics[width=0.8\textwidth]{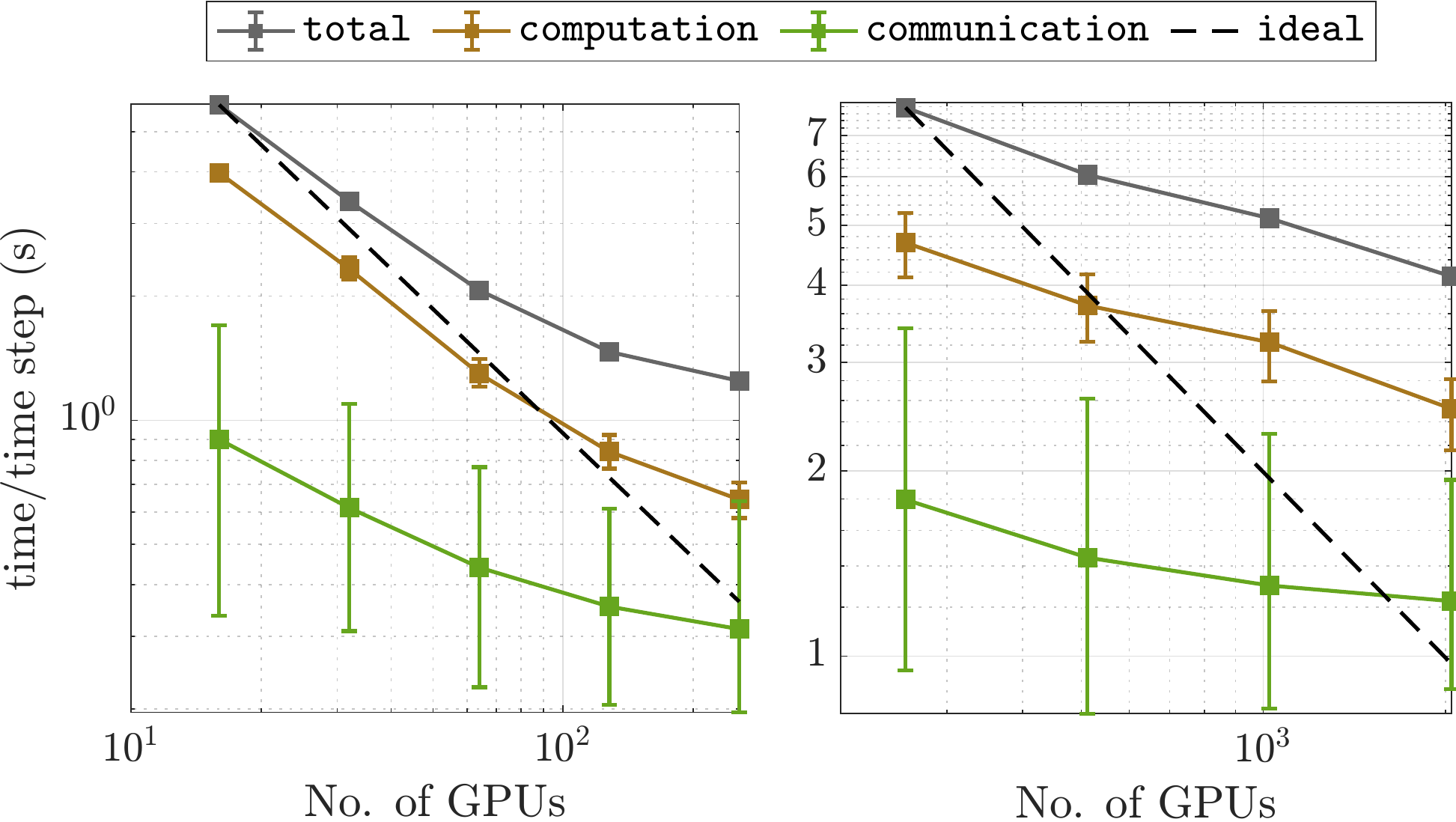}}
    \end{center}
    \vspace{-7mm}
    \caption{Penning trap. Strong scaling on GPUs: scaling of computation kernels, communication kernels, and cross over point between them for cases A and B with load balancing.}
    \figlab{strong_penning_gpus}
\end{figure}

    In Figure \figref{strong_penning_gpus}, we show the strong scaling results for cases A and B with load balancing on GPUs. For this
    example, we are unable to run most of the simulations without load balancing, due to memory requirements. Even in the few cases
    for which we are able to run the simulations, we did not observe scaling. We therefore only present results with load balancing. We observe from Figure \figref{parallel_penning_gpus} that the field solver takes more time and that the scaling is much worse than for the Landau damping cases. This is more significant for case B due to the large level of load imbalance in the fields, as shown in Figures \figref{decomp_256_penning} and \figref{decomp_2048_penning}. We also observe that even with particle load balancing, we are 
    unable to run our simulation for case B with 128 GPUs due to insufficient memory for particle communication operations during time stepping. These 
    observations clearly show that for highly non-uniform particle distributions, as in this Penning trap example, our current load balancing
    strategy has to be improved in order to effectively carry out these simulations with large numbers of particles on 
    large numbers of nodes. We will investigate this in future work. 

    We notice from Figure \figref{comm_penning_gpus} that even with particle load balancing, we have
    significant variation in timings across the GPUs for the {\tt updateParticle}, which we did not observe in the strong Landau damping simulations.
    This again is a symptom of the highly non-uniform electron distribution in the Penning trap simulations. 

\begin{figure}[]
    \begin{center}
    \subfigure[Computation kernels for case A (left) and case B (right)]{\includegraphics[width=0.8\textwidth]{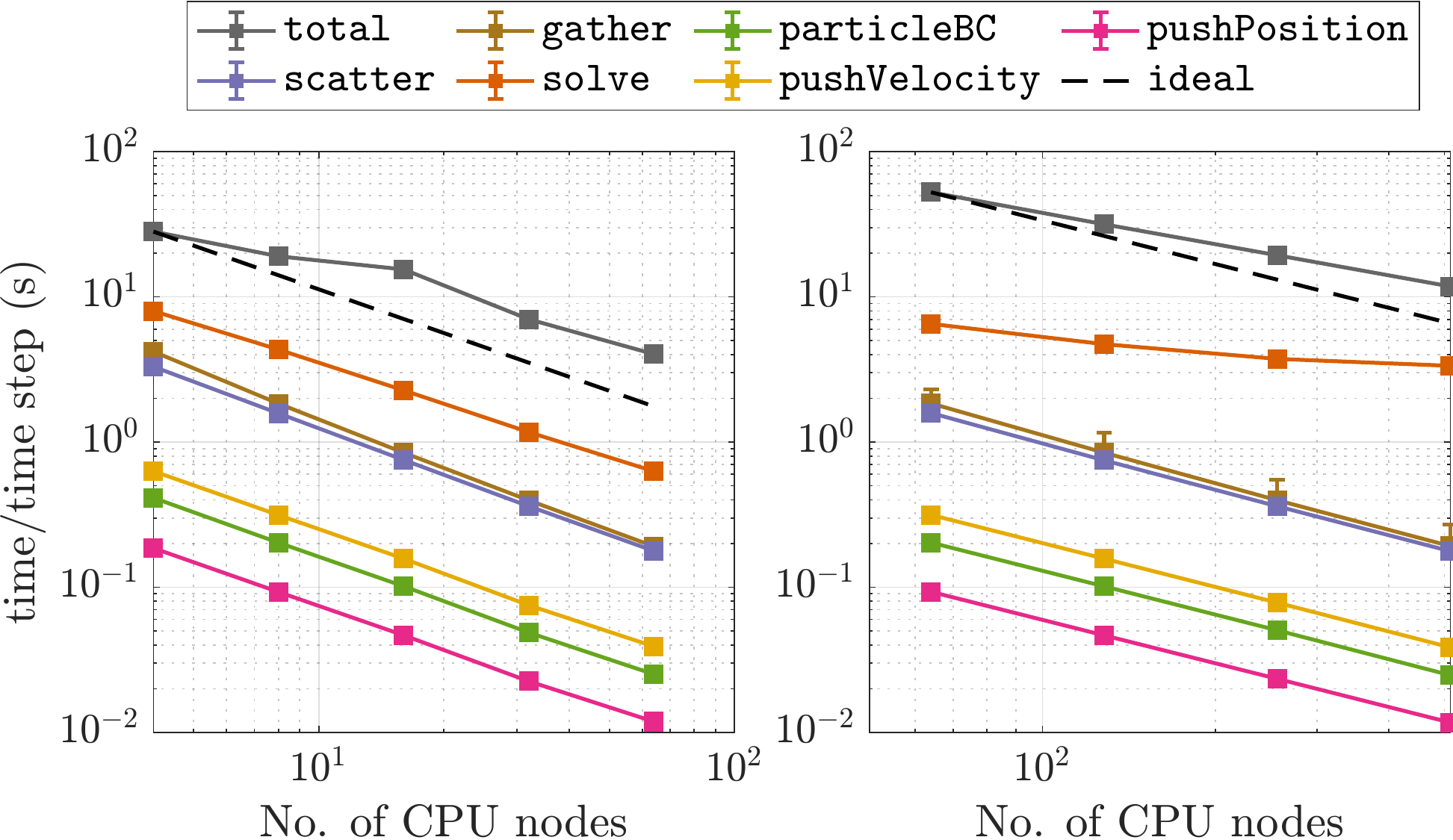}
    \figlab{parallel_penning_cpus}}
    \subfigure[Communication kernels for case A (left) and case B (right)]{\includegraphics[width=0.8\textwidth]{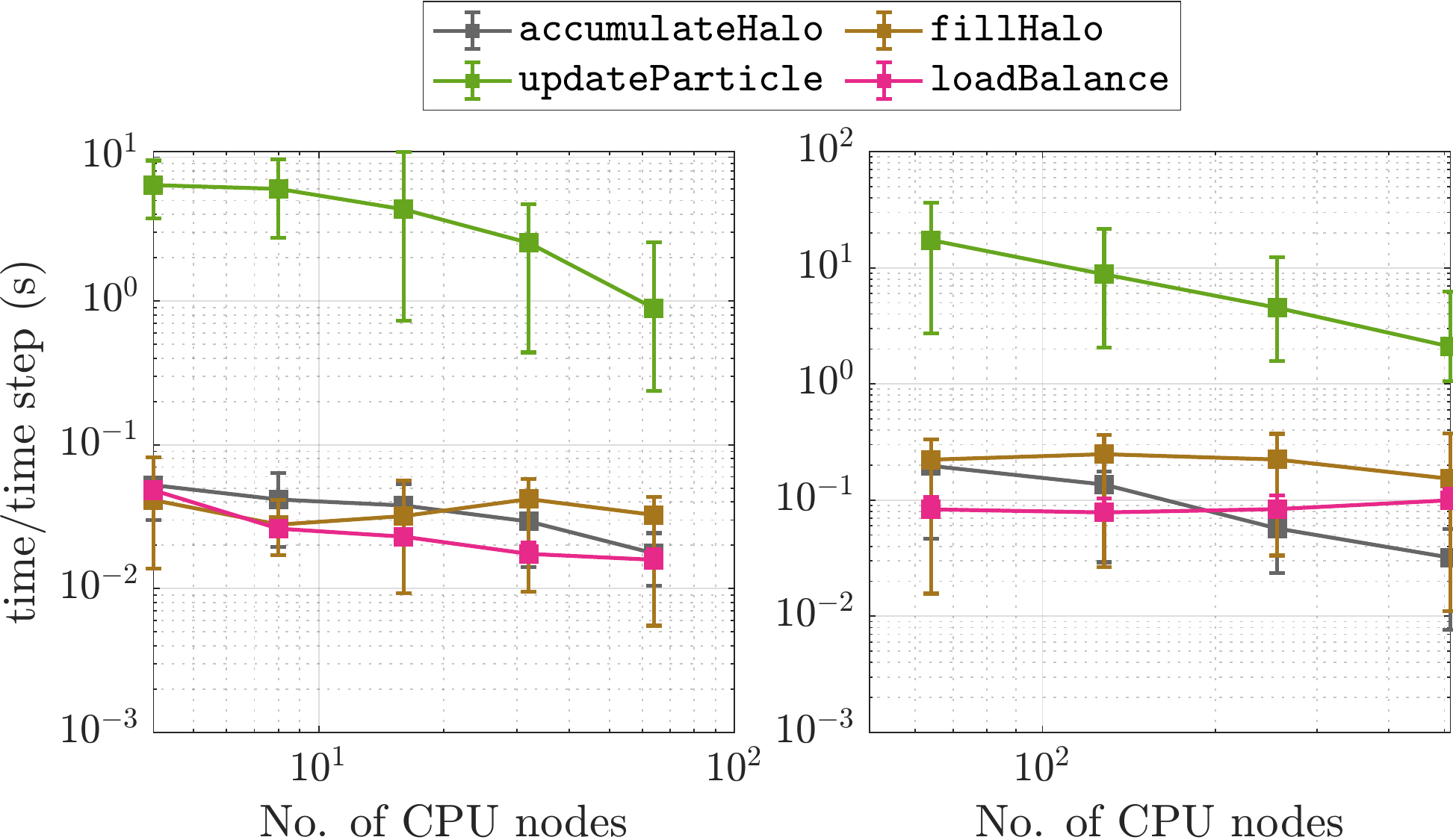}
    \figlab{comm_penning_cpus}}
    \subfigure[Computation vs communication kernels for case A (left) and case B (right)]{\includegraphics[width=0.8\textwidth]{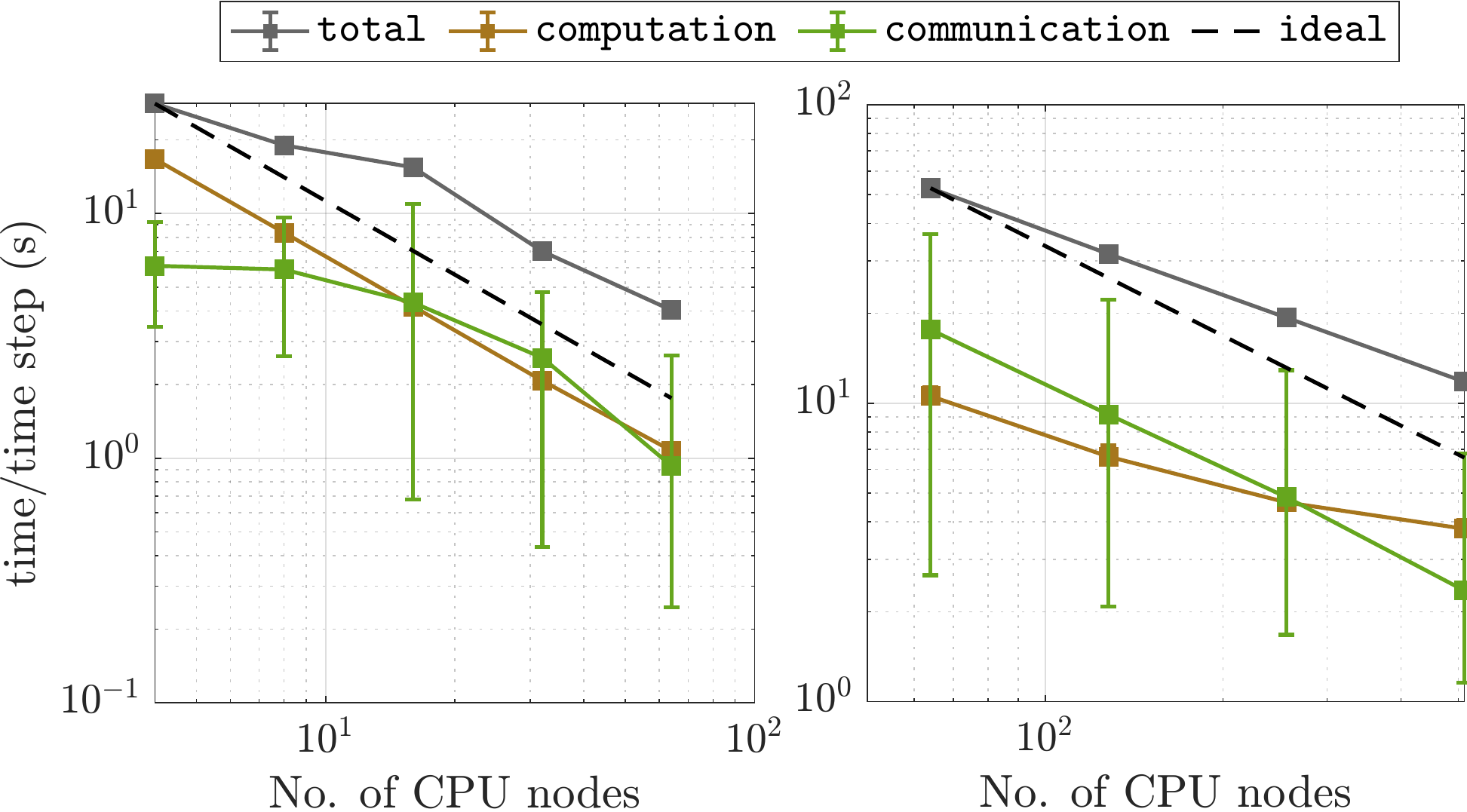}}
    \end{center}
    \vspace{-7mm}
    \caption{Penning trap. Strong scaling on CPUs: scaling of computation kernels, communication kernels, and cross over point between them for case A and case B with load balancing.}
    \figlab{strong_penning_cpus}
\end{figure}

    In Figure \figref{strong_penning_cpus}, we show the strong scaling study on CPUs. Because of the higher memory per node, we are able to run case A without load balancing, whereas for case B we still require load balancing. We observed loss of scaling and increased communication costs for case A without load
    balancing. Since the results are very similar to the results we obtained for the strong Landau damping simulations, we do not show them in this article. 

    For case B, from the right column of Figure \figref{parallel_penning_cpus} we see that the field solver efficiency is lower than in previous examples, as it is for the GPU simulations. Furthermore, even with load balancing, we are unable to run it on 32 nodes, due to the high memory requirements. Comparing the right column of Figure \figref{comm_penning_cpus} and the left column of Figure \figref{comm_nonlin_cpus}, we see that the {\tt updateParticle} 
    cost is higher in the Penning trap simulations than in the strong Landau damping simulations. The decrease of the time per time step as a function of the number of CPU nodes suggests that the particle sending and receiving operations dominate the {\tt updateParticle} in this case, as explained in Section \secref{linear_landau}.

\begin{figure}[]
    \begin{center}
    \subfigure[Computation kernels on GPUs (left) and CPUs (right)]{\includegraphics[width=0.8\textwidth]{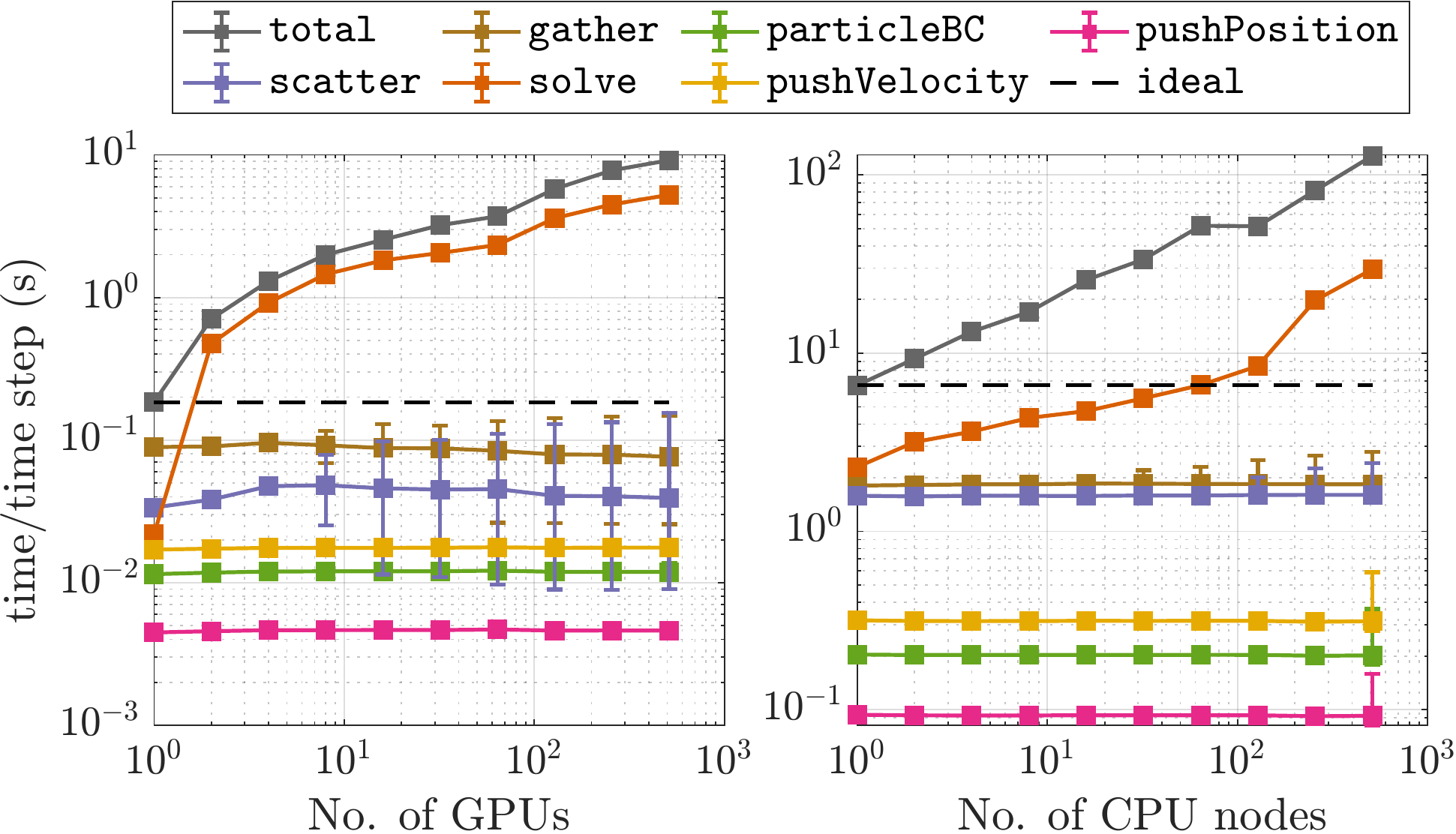}
    \figlab{parallel_weak_penning_cpu_gpu}}
    \subfigure[Communication kernels on GPUs (left) and CPUs (right)]{\includegraphics[width=0.8\textwidth]{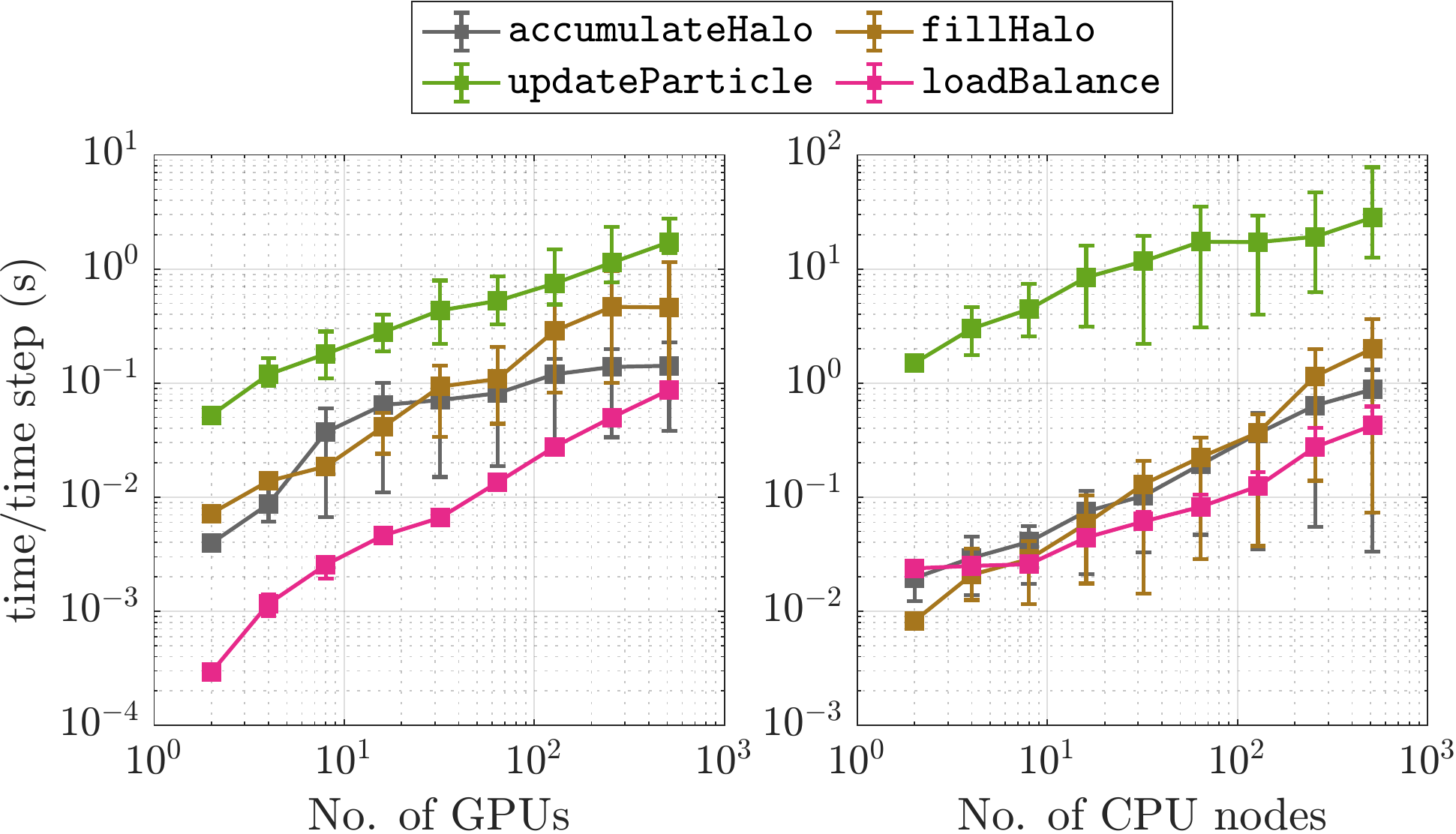}
    \figlab{comm_weak_penning_cpu_gpu}}
    \subfigure[Computation vs communication kernels on GPUs (left) and CPUs (right)]{\includegraphics[width=0.8\textwidth]{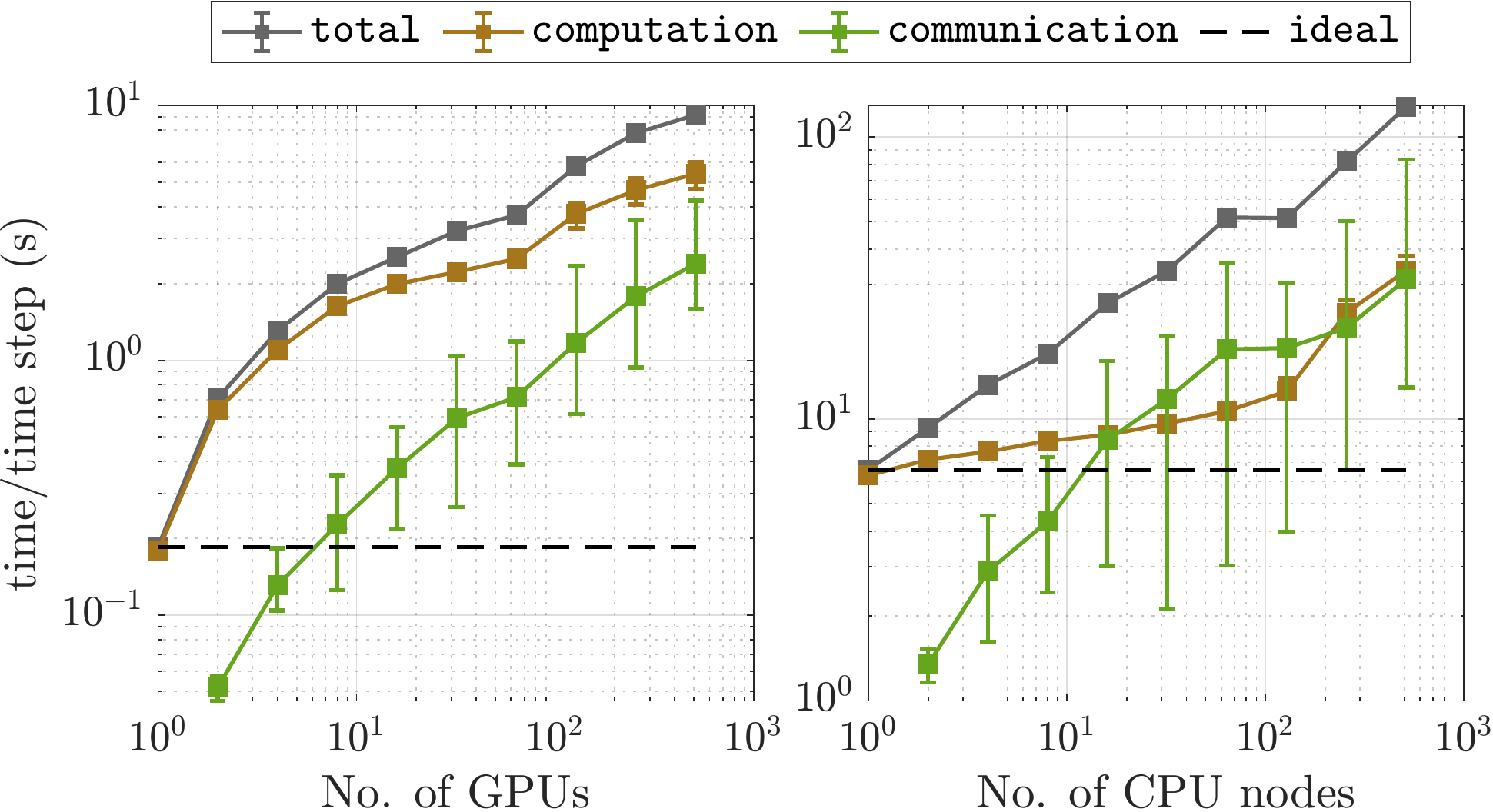}
    \figlab{parallel_comm_weak_penning_cpu_gpu}}
    \end{center}
    \vspace{-7mm}
    \caption{Penning trap. Weak scaling of computation kernels, communication kernels, and cross over point between them for GPUs and CPUs.}
    \figlab{weak_penning}
\end{figure}

    Finally, we consider the weak scaling study for the Penning trap test case. For this, we take the same base case of a $256\times128^2$ grid per GPU as we had for the Landau damping tests, whereas for CPUs we take a $256^3$ grid because of memory requirements imposed by field imbalance as well as field and particle communications. Similar to 
    Landau damping for both CPUs and GPUs we take 8 particles per cell. The runs on 1024 and 2048 GPUs fail due to memory requirements. Hence, the maximum grid size and number of particles for the GPU simulations 
    are $N_c=2048\times1024^2$ and $N_p\approx1.6\times10^{10}$ on 512 GPUs, 
    whereas for CPU simulations they are $N_c=2048^3$ and $N_p\approx6.9\times10^{10}$ on 512 nodes = 16,384 cores.

    In Figure \figref{weak_penning}, we show the weak scaling results on GPUs and CPUs for the Penning trap case. As noticed in the strong scaling study, the field solver scaling is affected due to the imbalance in the fields on both GPUs and CPUs. This leads to higher total time than the Landau damping problem, for which the results were shown in Figure \figref{linear_weak_cpu_gpu}.
    It should also be noted that the increase in time on CPUs is 
    despite having half as many particles and grid points than the Landau damping case. We also see from the right column of Figure 
    \figref{parallel_comm_weak_penning_cpu_gpu} that the communication costs become comparable to the computation costs much earlier than we observed for the Landau damping simulations in Figure \figref{parallel_comm_weak_linear_cpu_gpu}. Thus in cases in which the particle density has strong non-uniformity, such as this Penning trap test case, the field and particle load balancing have to be carefully investigated in order to determine the setup that would lead to the least overall time. 
    This requires improvements in the current load balancing strategy and also partly depends on the number of particles, grid points and 
    MPI ranks or GPUs used in the simulation.  

\subsection{Performance comparison across different architectures}
    Until now we have considered the performance of different mini-apps on Piz Daint CPU and GPU partitions. In this section, we consider the weak Landau damping mini-app and 
    evaluate its performance across different architectures. For this purpose we consider the CPU and GPU architectures in Table \ref{tab:cpu_gpu_partitions}. 

    \begin{table}[h!b!t!]
    \centering
    \begin{tabular}{|l||l|}
    \hline
        \!\!\! GPU architectures \!\!\!\! & \!\!\! CPU architectures \!\!\!\! \\
    \hline
        Piz Daint P100 & Piz Daint (Intel Xeon E5-2695 v4 @ 2.1GHz)\\
        Summit V100  & Cori KNL (Intel Xeon Phi 7250 @ 1.4 GHz)\\
        Perlmutter A100 & \\
    \hline
    \end{tabular}
        \caption{\label{tab:cpu_gpu_partitions}List of CPU and GPU partitions used for the performance comparison across different architectures.}
    \end{table}

\begin{figure}[]
    \subfigure[Total time (left) and efficiency (right) on different GPU architectures]{\includegraphics[width=\textwidth]{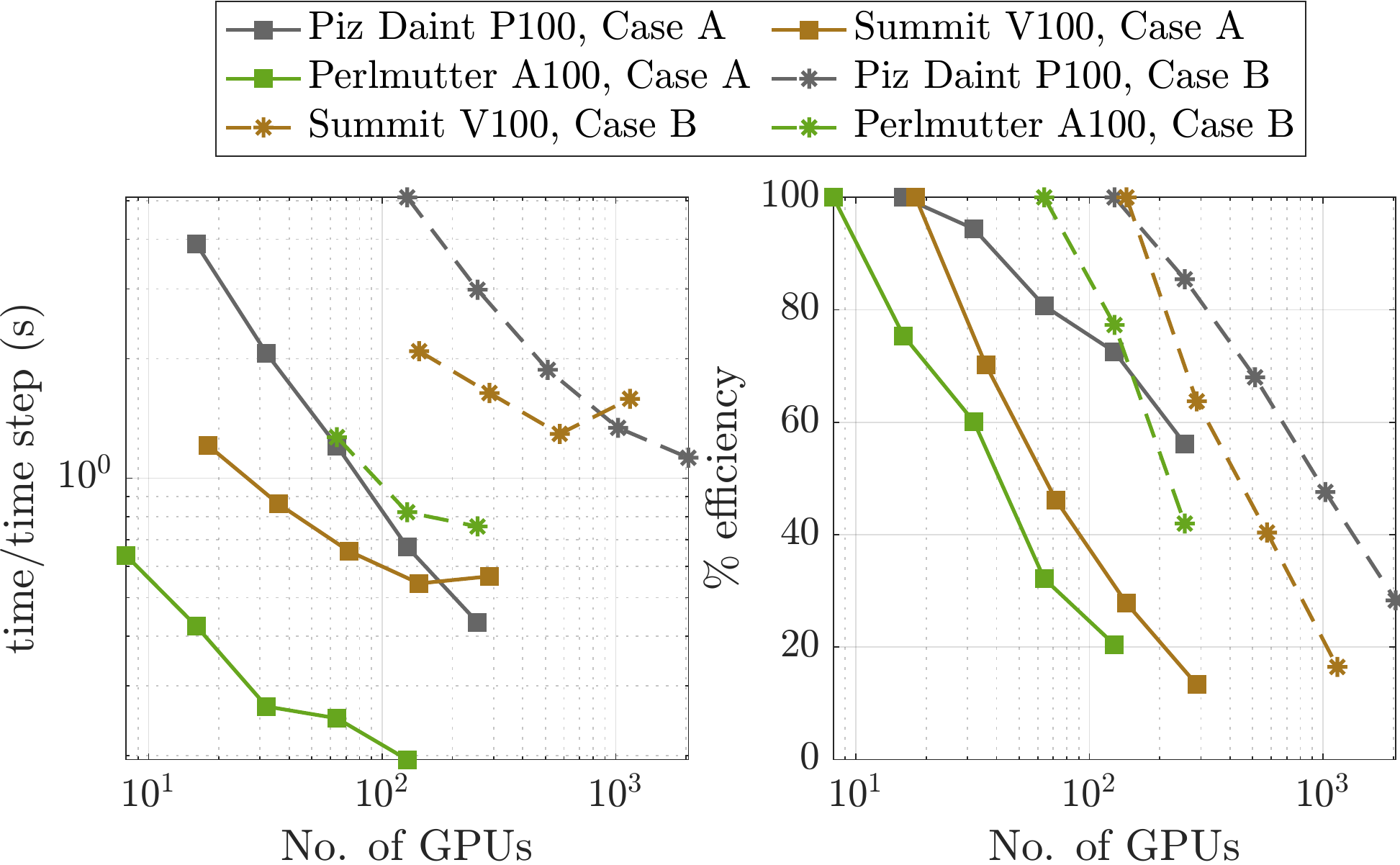}
    \figlab{time_eff_gpu_comb}}
    \subfigure[Total time (left) and efficiency (right) on different CPU architectures]{\includegraphics[width=\textwidth]{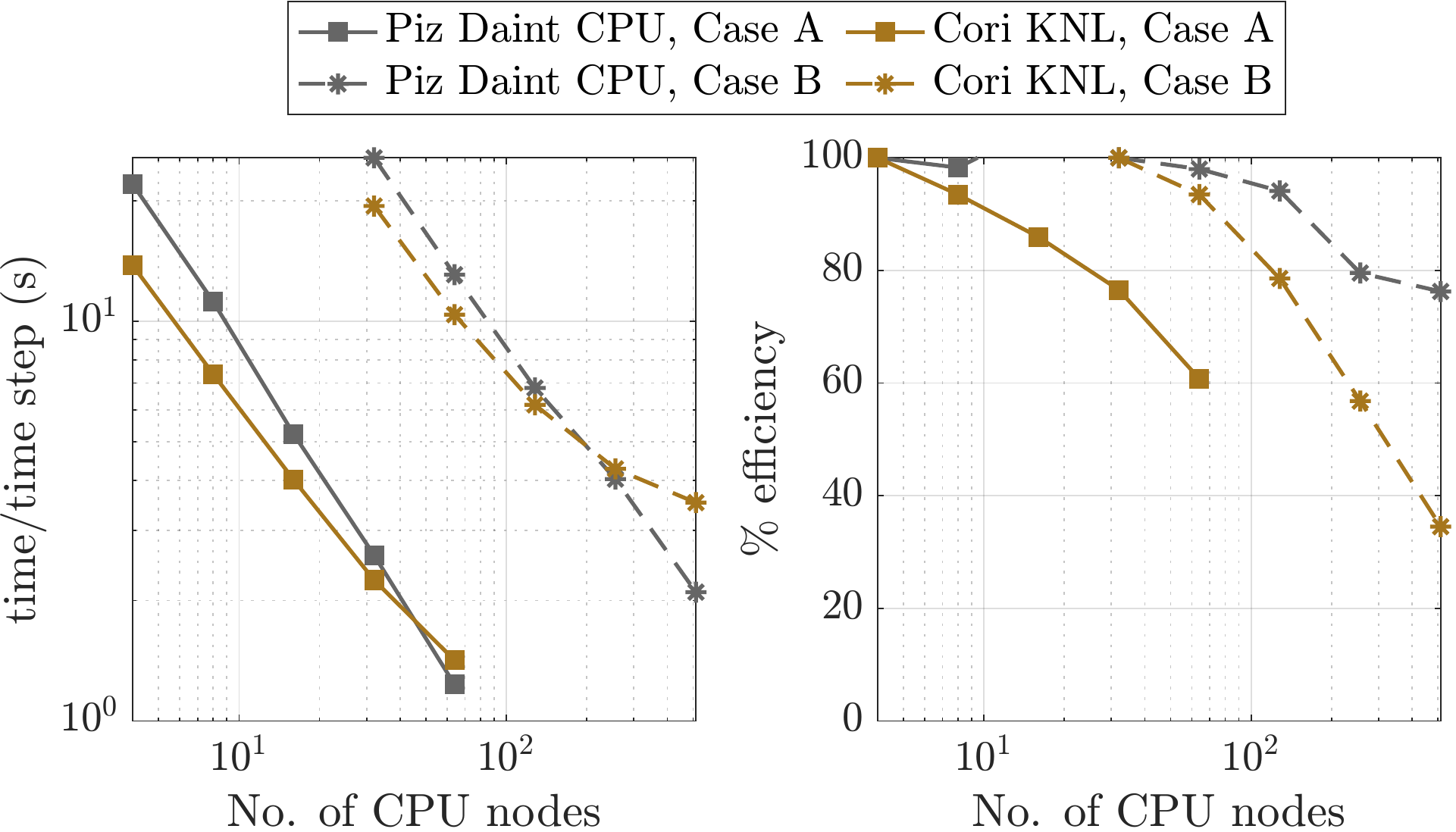}
    \figlab{time_eff_cpu_comb}}
    \caption{Comparison of different GPU architectures (top row) and CPU architectures (bottom row) with respect to strong scaling for the weak Landau damping mini-app.}
\end{figure}
    Each node of the Piz Daint GPU partition consists of 1 Tesla P100 GPU, whereas the Summit and Perlmutter nodes consist of 6 Tesla V100 GPUs and 4 Ampere A100 GPUs, respectively. The memory per GPU is 16 GB 
    for both P100 and V100 GPUs, whereas for A100 it is 40 GB. In the CPU partitions, 
    each node of the Piz Daint CPU partition consists of 36 cores, whereas the Cori KNL nodes have 68 cores each. The interconnect configuration
    is Aries and the network topology is Dragonfly for Piz Daint CPU, Piz Daint GPU and Cori KNL architectures. For Perlmutter, the interconnect configuration is HPE Cray Slingshot with a three-hop Dragonfly topology whereas for Summit, it is Mellanox EDR 100G InfiniBand with a Non-blocking Fat Tree topology.

    We utilize all the GPUs per node in the GPU architectures, whereas for the CPU ones we use 
    32 and 64 cores per node on the Piz Daint and Cori systems, respectively. The multithreading is turned off for the CPU runs. Similar to the previous study on Piz Daint, we use 1 MPI rank per GPU for the 
    GPU architectures, whereas for the CPU ones we use 1 MPI rank per node and use 32 and 64 OpenMP threads on Piz Daint and Cori respectively.

    In Figure \figref{time_eff_gpu_comb} the total time and efficiencies are compared across the three GPU architectures for the strong scaling study corresponding to cases A and B. In the case of Perlmutter, 
    we can start the scaling study with half as many GPUs as the Piz Daint and Summit partitions, thanks to the higher memory configuration of A100 GPUs (40 GB) compared to those of P100 and V100 GPUs (16 GB). We make the following observations. In terms of the absolute wall time per simulation time step, Perlmutter is the fastest, followed by Summit and then Piz Daint. In terms of scaling efficiencies, Piz Daint has the highest efficiency followed
    by Perlmutter and then Summit. We can understand the total time and efficiencies better by looking at the cumulative computation and communication kernels across the three architectures for cases A and B in Figure
    \figref{time_gpu_comp_case_a_b}. The computation kernels of Summit have a speedup of 3-4 compared to Piz Daint until the scaling stops, whereas Perlmutter computation kernels have a speedup of 
    roughly 10 compared to Piz Daint. This is because of the architecture and CUDA compute capability of the GPUs: the latest A100 GPUs are more powerful than the V100, which in turn are more powerful than the P100 GPUs. In terms of communication
    costs, Perlmutter has the least communication cost for both cases A and B. Comparing Summit and Piz Daint, for case A Summit has a higher communication cost than Piz Daint, 
    whereas for case B both are comparable. The higher communication cost of Summit is also reflected in the scaling of the computation kernels, where the field solver loses scaling earlier than Piz Daint and Perlmutter.

    We now consider the CPU architectures, and compare the strong scaling performance on Piz Daint and Cori KNL nodes. In Figure \figref{time_eff_cpu_comb} we show the total time and efficiency 
for these two systems. For small number of nodes, the wall time per simulation time step of Cori nodes are slightly less than for Piz Daint nodes. However, the better scaling and efficiency of Piz Daint eventually leads to a lower runtime than Cori. 
From the computation and communication kernels in Figure \figref{time_cpu_comp_case_a_b} we notice that the computation kernels perform slightly better on Cori than on Piz Daint, but the communication time of Cori is 
higher, which eventually leads to a lower efficiency than Piz Daint. However, overall the performance per number of nodes of the two systems are very similar.  

\begin{figure}[]
    \subfigure[Computation vs communication kernels for case A (left) and case B (right)]{\includegraphics[width=\textwidth]{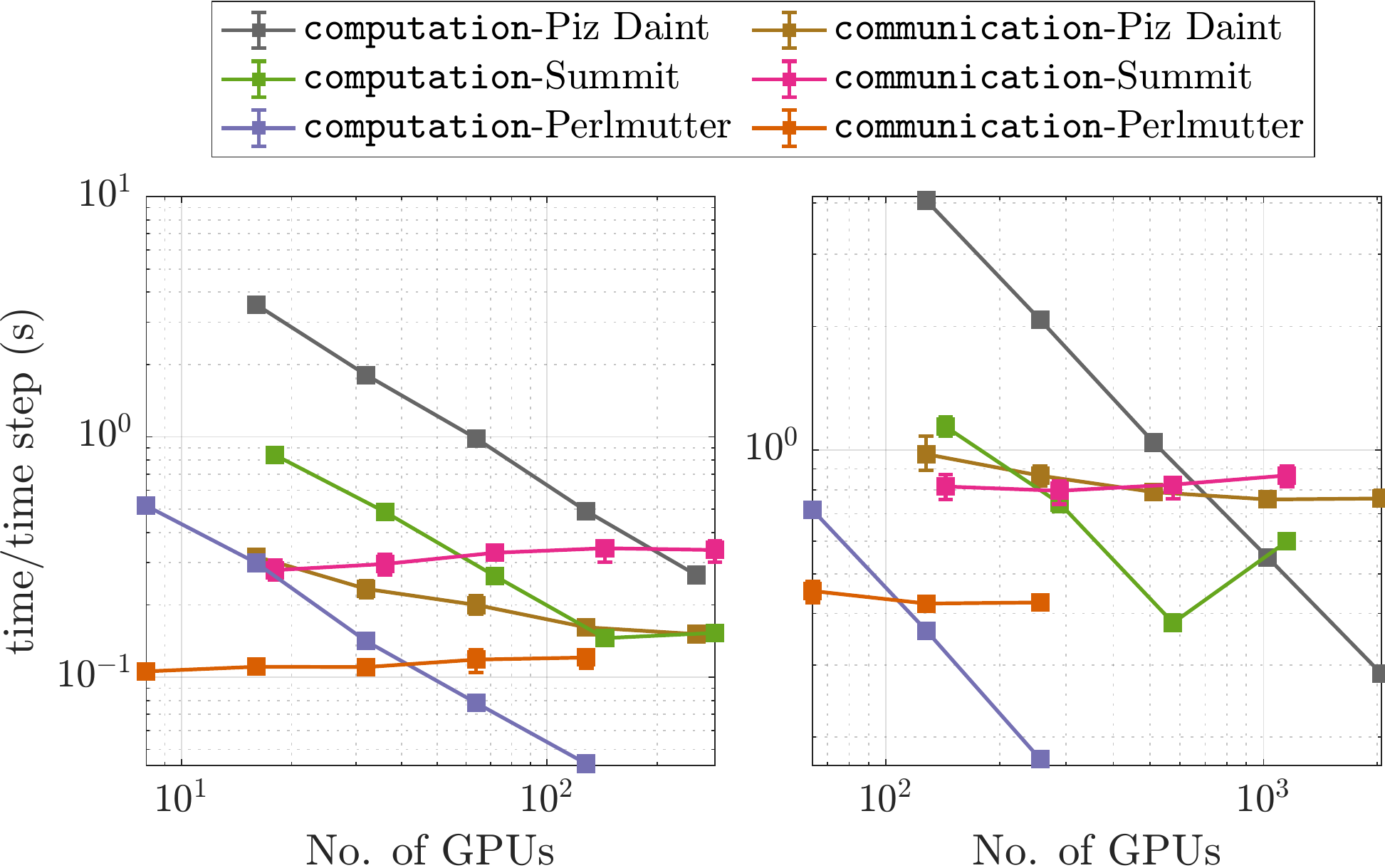}
    \figlab{time_gpu_comp_case_a_b}}
    \subfigure[Computation vs communication kernels for case A (left) and case B (right)]{\includegraphics[width=\textwidth]{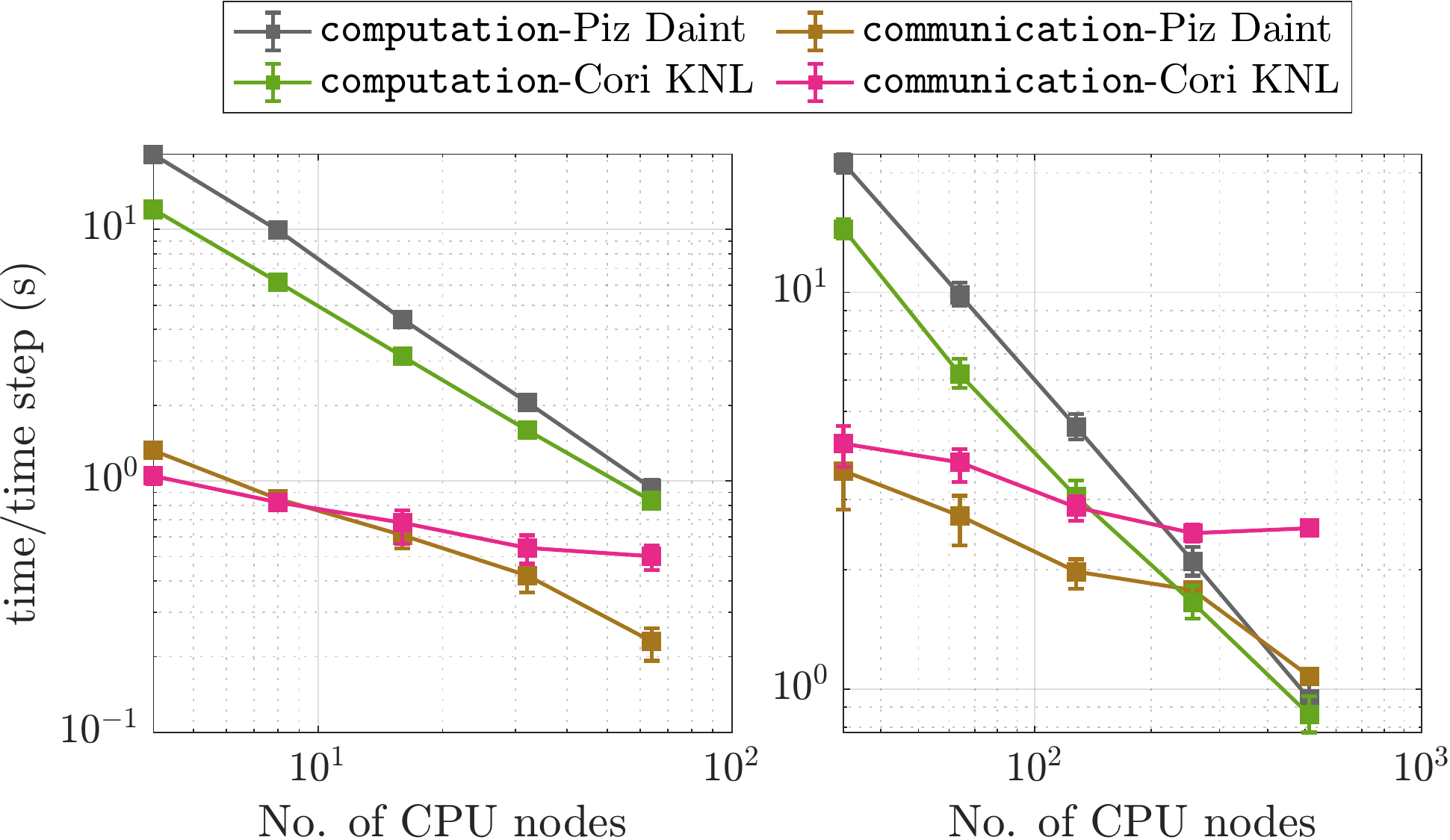}
    \figlab{time_cpu_comp_case_a_b}}
    \caption{Timings of computation and communication kernels for different GPU architectures (top row) and CPU architectures (bottom row) with respect to strong scaling for the weak Landau damping mini-app.}
\end{figure}
\section{Conclusion}
\seclab{conclusions}
    In this work, we performed a scaling and performance portability study of the particle-in-cell scheme for plasma physics applications by means of a set of mini-apps, namely ``Alpine", targeting exascale architectures. The mini-apps include weak and strong Landau damping, 
    the dynamics of an electron bunch in a quasi-neutral Penning trap, and the two-stream and bump-on-tail instabilities, which are commonly used as benchmarks for electrostatic PIC studies. Our scaling and performance analysis shows that
    the weak Landau damping simulations perform the best among the mini-apps in terms of scalability and time to solution. This is because the particle distribution is relatively uniform in this case, and the particle communication 
    timings therefore remain small compared to the timings of computation kernels. We obtained a maximum particle push rate of 
    approximately $10^{10}$ particles per second on GPUs and $3\times10^9$ particles per second on CPUs in the weak scaling study. 
    The scaling and performance of strong Landau damping simulations with particle load balancing are very similar to those of the weak Landau damping simulations, whereas without particle
    load balancing the particle communication costs as well as memory imbalance are much higher, which leads to poor scaling. The Penning trap mini-app corresponds to the toughest test case, due to the highly non-uniform particle distribution. In this case, particle load balancing leads to a significant imbalance in the fields, which then affects the scaling of the field solve. Our current load balancing strategy needs to be improved to handle such cases.
    
    A performance comparison across different GPU and CPU architectures for the weak Landau damping mini-app shows that Perlmutter with the latest NVIDIA A100 GPUs performs the best in terms of wall time per simulation time step, with
    almost an order of magnitude speedup compared to the Piz Daint P100 GPUs, and approximately a three times speedup as compared to Summit with V100 GPUs. In the comparison of CPU architectures, the wall time per simulation time step of Piz Daint and Cori KNL nodes
    are very similar, with Piz Daint showing better scaling than Cori. 

    In terms of future work, we will optimize and improve the current load balancing strategy and particle communication. We will continue our benchmarking
    studies with Perlmutter and other upcoming architectures, to test for higher numbers of particles, grid points and GPUs and CPU cores. Finally, we would also like to extend our current study by adding more
    mini-apps to the Alpine collection, with test problems requiring the inclusion of collisions and the implementation of an electromagnetic PIC scheme. 

\section*{Availability}
Alpine and IPPL are open source projects. The sources can be downloaded from \href{https://gitlab.psi.ch/OPAL/Libraries/ippl}{https://gitlab.psi.ch/OPAL/Libraries/ippl}.

\section*{Acknowledgments}
The authors would like to thank the Kokkos team for helping us with all the queries during the development of IPPL and Alpine. We would also like to thank Marc Caubet Serrabou from PSI for his help with all the installations during the development of IPPL. This project has received funding from the European Union's Horizon 2020 research and innovation program under the Marie Sk{\l}odowska-Curie grant agreement No. 701647 and from the United States National Science Foundation under Grant No. PHY-1820852. This research used resources of the National Energy Research Scientific Computing Center (NERSC), a U.S. Department of Energy Office of Science User Facility located at Lawrence Berkeley National Laboratory, operated under Contract No. DE-AC02-05CH11231 using NERSC award ASCR-ERCAPM888. We acknowledge access to Piz Daint at the Swiss National Supercomputing Centre, Switzerland under the PSI's share with the project IDs psi07 and psigpu. Finally, this research also used resources of the Oak Ridge Leadership Computing Facility at the Oak Ridge National Laboratory, which is supported by the Office of Science of the U.S. Department of Energy under Contract No. DE-AC05-00OR22725.

\section*{Appendix A. Compilers and libraries used for the benchmarks on different computing architectures}
We used Kokkos version 3.5.0 and heFFTe version 2.2.0 for all our simulations. With respect to IPPL, we used the tag {\tt Scaling\_study\_for\_Alpine\_paper} for all the scaling studies. It can be obtained from the IPPL gitlab repository \href{https://gitlab.psi.ch/OPAL/Libraries/ippl}{https://gitlab.psi.ch/OPAL/Libraries/ippl}.

For each of the architectures, the compiler type, version and the MPI used are specified in Table \ref{tab:compilers} below.
\begin{table}[h!b!t!]
        \centering
        \begin{tabular}{|r|c|c|}
        \hline
            \!\!\! Architecture \!\!\!\! & \!\!\! Compiler \!\!\!\! & \!\!\! MPI \!\!\!\!\\
        \hline
            Piz Daint GPU & {\tt gcc/9.3.0}  & {\tt OpenMPI/4.1.2} with {\tt CUDA/11.2} \\
            Piz Daint CPU & {\tt gcc/11.2.0}  & {\tt cray-mpich/7.7.18} \\
            Cori KNL & {\tt intel/19.1.1.217}  & {\tt cray-mpich/7.7.18} \\
            Summit GPU & {\tt gcc/9.1.0}  & {\tt spectrum-mpi/10.4.0} with {\tt CUDA/11.0} \\
            Perlmutter GPU & {\tt gcc/11.2.0} & {\tt OpenMPI/4.1.2} with {\tt CUDA/11.4} \\
        \hline
        \end{tabular}
        \caption{\label{tab:compilers}Compiler type, version and MPI used for each of the architectures used for the scaling study in Section \secref{scaling}.}
        \end{table}

For the Piz Daint CPUs, we also conducted the benchmarking study with the following Intel compiler. 
\begin{itemize}
    \item {\tt intel/2021.3.0}
    \item {\tt cray-mpich/7.7.18}
\end{itemize}
The results were very comparable to those obtained with gcc. However, we sometimes observed an inconsistent ``Bus error", the reason of which is
yet unknown and under investigation.

\section*{References}

\bibliographystyle{elsarticle-num}

\bibliography{references}

\begin{thebibliography}{10}
\expandafter\ifx\csname url\endcsname\relax
  \def\url#1{\texttt{#1}}\fi
\expandafter\ifx\csname urlprefix\endcsname\relax\def\urlprefix{URL }\fi
\expandafter\ifx\csname href\endcsname\relax
  \def\href#1#2{#2} \def\path#1{#1}\fi

\bibitem{hockney1988computer}
R.~W. Hockney, J.~W. Eastwood, Computer simulation using particles, CRC Press,
  1988.

\bibitem{birdsall2004plasma}
C.~K. Birdsall, A.~B. Langdon, Plasma physics via computer simulation, CRC
  press, 2004.

\bibitem{dawson1983particle}
J.~M. Dawson, Particle simulation of plasmas, Reviews of modern physics 55~(2)
  (1983) 403.

\bibitem{ricketson2016sparse}
L.~F. Ricketson, A.~J. Cerfon, Sparse grid techniques for particle-in-cell
  schemes, Plasma Physics and Controlled Fusion 59~(2) (2016) 024002.

\bibitem{spitkovsky2005simulations}
A.~Spitkovsky, Simulations of relativistic collisionless shocks: shock
  structure and particle acceleration, in: AIP Conference Proceedings, Vol.
  801, American Institute of Physics, 2005, pp. 345--350.

\bibitem{buneman1993computer}
O.~Buneman, Computer space plasma physics, simulation techniques and softwares,
  ed, H. Matsumoto and Y. Omura (Terra Scientific, Tokyo, 1993) p 67.

\bibitem{jolliet2007global}
S.~Jolliet, A.~Bottino, P.~Angelino, R.~Hatzky, T.-M. Tran, B.~Mcmillan,
  O.~Sauter, K.~Appert, Y.~Idomura, L.~Villard, A global collisionless {PIC}
  code in magnetic coordinates, Computer Physics Communications 177~(5) (2007)
  409--425.

\bibitem{chang2008spontaneous}
C.-S. Chang, S.~Ku, Spontaneous rotation sources in a quiescent tokamak edge
  plasma, Physics of Plasmas 15~(6) (2008) 062510.

\bibitem{fonseca2002osiris}
R.~A. Fonseca, L.~O. Silva, F.~S. Tsung, V.~K. Decyk, W.~Lu, C.~Ren, W.~B.
  Mori, S.~Deng, S.~Lee, T.~Katsouleas, et~al., Osiris: A three-dimensional,
  fully relativistic particle in cell code for modeling plasma based
  accelerators, in: International Conference on Computational Science,
  Springer, 2002, pp. 342--351.

\bibitem{qiang2006three}
J.~Qiang, S.~Lidia, R.~D. Ryne, C.~Limborg-Deprey, Three-dimensional
  quasistatic model for high brightness beam dynamics simulation, Physical
  Review Special Topics-Accelerators and Beams 9~(4) (2006) 044204.

\bibitem{adelmann2019opal}
A.~Adelmann, P.~Calvo, M.~Frey, A.~Gsell, U.~Locans, C.~Metzger-Kraus,
  N.~Neveu, C.~Rogers, S.~Russell, S.~Sheehy, J.~Snuvernik, D.~Winklehner,
  {OPAL} a versatile tool for charged particle accelerator simulations, arXiv
  preprint arXiv:1905.06654.

\bibitem{vay2018warp}
J.-L. Vay, A.~Almgren, J.~Bell, L.~Ge, D.~Grote, M.~Hogan, O.~Kononenko,
  R.~Lehe, A.~Myers, C.~Ng, et~al., Warp-x: A new exascale computing platform
  for beam--plasma simulations, Nuclear Instruments and Methods in Physics
  Research Section A: Accelerators, Spectrometers, Detectors and Associated
  Equipment 909 (2018) 476--479.

\bibitem{mniszewski2021enabling}
S.~M. Mniszewski, J.~Belak, J.-L. Fattebert, C.~F. Negre, S.~R. Slattery, A.~A.
  Adedoyin, R.~F. Bird, C.~Chang, G.~Chen, S.~Ethier, et~al., Enabling particle
  applications for exascale computing platforms, The International Journal of
  High Performance Computing Applications 35~(6) (2021) 572--597.

\bibitem{myers2021porting}
A.~Myers, A.~Almgren, L.~Amorim, J.~Bell, L.~Fedeli, L.~Ge, K.~Gott, D.~P.
  Grote, M.~Hogan, A.~Huebl, et~al., Porting warpx to gpu-accelerated
  platforms, Parallel Computing 108 (2021) 102833.

\bibitem{heroux2009improving}
M.~A. Heroux, D.~W. Doerfler, P.~S. Crozier, J.~M. Willenbring, H.~C. Edwards,
  A.~Williams, M.~Rajan, E.~R. Keiter, H.~K. Thornquist, R.~W. Numrich,
  Improving performance via mini-applications, Sandia National Laboratories,
  Tech. Rep. SAND2009-5574 3.

\bibitem{bird2021vpic}
R.~Bird, N.~Tan, S.~V. Luedtke, S.~L. Harrell, M.~Taufer, B.~Albright, Vpic
  2.0: Next generation particle-in-cell simulations, IEEE Transactions on
  Parallel and Distributed Systems 33~(4) (2021) 952--963.

\bibitem{doi:10.1063/1.168723}
J.~V.~W. Reynders, J.~Cummings, P.~F. Dubois, {The POOMA Framework}, Computers
  in Physics 12~(5) (1998) 453--459.

\bibitem{veldhuizen1995expression}
T.~Veldhuizen, Expression templates, C++ Report 7~(5) (1995) 26--31.

\bibitem{CarterEdwards20143202}
H.~C. Edwards, C.~R. Trott, D.~Sunderland, Kokkos: Enabling manycore
  performance portability through polymorphic memory access patterns, Journal
  of Parallel and Distributed Computing 74~(12) (2014) 3202 -- 3216,
  domain-Specific Languages and High-Level Frameworks for High-Performance
  Computing.

\bibitem{trott2021kokkos}
C.~R. Trott, D.~Lebrun-Grandie, D.~Arndt, J.~Ciesko, V.~Dang, N.~Ellingwood,
  R.~Gayatri, E.~Harvey, D.~S. Hollman, D.~Ibanez, et~al., Kokkos 3:
  Programming model extensions for the exascale era, IEEE Transactions on
  Parallel and Distributed Systems 33~(4) (2021) 805--817.

\bibitem{quinlan1997mlb}
D.~J. Quinlan, M.~Berndt, {MLB}: Multilevel load balancing for structured grid
  applications, Tech. rep., Los Alamos National Lab.(LANL), Los Alamos, NM
  (United States) (1997).

\bibitem{ayala2020heffte}
A.~Ayala, S.~Tomov, A.~Haidar, J.~Dongarra, heffte: {H}ighly efficient fft for
  exascale, in: International Conference on Computational Science, Springer,
  2020, pp. 262--275.

\bibitem{rodriguez2020implementation}
D.~Rodr{\'\i}guez-Pati{\~n}o, S.~Ram{\'\i}rez, J.~Salcedo-Gallo, J.~Hoyos,
  E.~Restrepo-Parra, Implementation of the two-dimensional electrostatic
  particle-in-cell method, American Journal of Physics 88~(2) (2020) 159--167.

\bibitem{devroye2006nonuniform}
L.~Devroye, Nonuniform random variate generation, Handbooks in operations
  research and management science 13 (2006) 83--121.

\bibitem{tretiak2019arbitrary}
K.~Tretiak, D.~Ruprecht, An arbitrary order time-stepping algorithm for
  tracking particles in inhomogeneous magnetic fields, Journal of Computational
  Physics: X 4 (2019) 100036.

\bibitem{ho2018physics}
A.~Ho, I.~A.~M. Datta, U.~Shumlak, Physics-based-adaptive plasma model for
  high-fidelity numerical simulations, Frontiers in Physics 6 (2018) 105.

\bibitem{myers20174th}
A.~Myers, P.~Colella, B.~V. Straalen, A 4th-order particle-in-cell method with
  phase-space remapping for the {V}lasov--{P}oisson equation, SIAM Journal on
  Scientific Computing 39~(3) (2017) B467--B485.

\bibitem{kormann2015semi}
K.~Kormann, A semi-lagrangian vlasov solver in tensor train format, SIAM
  Journal on Scientific Computing 37~(4) (2015) B613--B632.

\bibitem{chen2011energy}
G.~Chen, L.~Chac{\'o}n, D.~C. Barnes, An energy-and charge-conserving,
  implicit, electrostatic particle-in-cell algorithm, Journal of Computational
  Physics 230~(18) (2011) 7018--7036.

\bibitem{sarkar2015bump}
S.~Sarkar, S.~Paul, R.~Denra, Bump-on-tail instability in space plasmas,
  Physics of Plasmas 22~(10) (2015) 102109.

\bibitem{muralikrishnan2021sparse}
S.~Muralikrishnan, A.~J. Cerfon, M.~Frey, L.~F. Ricketson, A.~Adelmann, Sparse
  grid-based adaptive noise reduction strategy for particle-in-cell schemes,
  Journal of Computational Physics: X (2021) 100094.

\bibitem{adam1995space}
S.~Adam, Space charge effects in cyclotrons-from simulations to insights, in:
  Proc. of the 14th Int. Conf. on Cyclotrons and their Applications,(World
  Scientific, Singapore, 1996), Vol. 446, 1995.

\bibitem{yang2010beam}
J.~Yang, A.~Adelmann, M.~Humbel, M.~Seidel, T.~Zhang, et~al., Beam dynamics in
  high intensity cyclotrons including neighboring bunch effects: Model,
  implementation, and application, Physical Review Special Topics-Accelerators
  and Beams 13~(6) (2010) 064201.

\end{thebibliography}

\end{document}